\documentclass[%
 aip,
 amsmath,amssymb,
 reprint,%
nofootinbib]{revtex4-2}

\usepackage{graphicx}% Include figure files
\usepackage{dcolumn}% Align table columns on decimal point
\usepackage{bm}% bold math
\usepackage{adjustbox}
\usepackage[utf8]{inputenc}
\usepackage[T1]{fontenc}
\usepackage{mathptmx}
\usepackage{xcolor}
\usepackage{ulem}
\usepackage{hyperref}
\hypersetup{
	colorlinks   = true,
	citecolor    = blue,
	linkcolor = blue,
    urlcolor = blue
}

\newcommand{\mv}[1]{\mathbf{#1}} %alternative for vectors

\begin{document}

\preprint{AIP/123-QED}

\title{Toward improved property prediction of 2D materials using many-body quantum Monte Carlo methods }
% Force line breaks with \\

\author{Daniel Wines}
\email{daniel.wines@nist.gov}
\affiliation{Materials Science and Engineering
Division, National Institute of Standards and Technology (NIST),
Gaithersburg, MD 20899, USA}

\author{Jeonghwan Ahn}%

\affiliation{ 
Materials Science and Technology Division, Oak Ridge National Laboratory, Oak Ridge, TN 37831, USA%\\This line break forced with \textbackslash\textbackslash
}%

\author{Anouar Benali}%

\affiliation{ 
Computational Science Division, Argonne National Laboratory, Argonne, IL 60439, USA}
\affiliation{Qubit Pharmaceuticals, Incubateur Paris Biotech Santé, 24 rue du Faubourg Saint Jacques, 75014 Paris, France
}%

\author{Paul R. C. Kent}

\affiliation{ 
Computational Sciences and Engineering Division, Oak Ridge National Laboratory, Oak Ridge, TN 37831, USA
}%

\author{Jaron T. Krogel}%

\affiliation{ 
Materials Science and Technology Division, Oak Ridge National Laboratory, Oak Ridge, TN 37831, USA%\\This line break forced with \textbackslash\textbackslash
}%

\author{Yongkyung Kwon}%

\affiliation{ 
Department of Physics, Konkuk University, Seoul 05029, Korea
}%

\author{Lubos Mitas}%

\affiliation{ 
Department of Physics, North Carolina State University, Raleigh, NC 27695-8202, USA
}%

\author{Fernando A. Reboredo}%

\affiliation{ 
Material Science and Technology Division, Oak Ridge National Laboratory, Oak Ridge, TN 37831, USA%\\This line break forced with \textbackslash\textbackslash
}%

\author{Brenda Rubenstein}%

\affiliation{ 
Department of Chemistry, Brown University, Providence, RI 02912, USA
}%
\affiliation{ 
Department of Physics, Brown University, Providence, RI 02912, USA
}

\author{Kayahan Saritas}%

\affiliation{ 
Materials Science and Technology Division, Oak Ridge National Laboratory, Oak Ridge, TN 37831, USA%\\This line break forced with
}%

\author{Hyeondeok Shin}%

\affiliation{ 
Computational Science Division, Argonne National Laboratory, Argonne, IL 60439, USA
}%

\author{Ivan {\v S}tich}%

\affiliation{ 
Institute of Informatics, Slovak Academy of Sciences, 845 07 Bratislava, Slovakia
}%

\author{Can Ataca}
 \email{ataca@umbc.edu}
\affiliation{%
Department of Physics, University of Maryland Baltimore County, Baltimore MD 21250%\\This line break forced% with \\
}%

\date{\today}% It is always \today, today,

\begin{abstract}

The field of 2D materials has grown dramatically in the past two decades. 2D materials can be utilized for a variety of next-generation optoelectronic, spintronic, clean energy, and quantum computing applications. These 2D structures, which are often exfoliated from layered van der Waals materials, possess highly inhomogeneous electron densities and can possess short- and long-range electron correlations. The complexities of 2D materials make them challenging to study with standard mean-field electronic structure methods such as density functional theory (DFT), which relies on approximations for the unknown exchange-correlation functional. To overcome the limitations of DFT, highly accurate many-body electronic structure approaches such as diffusion Monte Carlo (DMC) can be utilized. In the past decade, DMC has been used to calculate accurate magnetic, electronic, excitonic, and topological properties in addition to accurately capturing interlayer interactions and cohesion and adsorption energetics of 2D materials. This approach has been applied to 2D systems of wide interest, including graphene, phosphorene, MoS$_2$, CrI$_3$, VSe$_2$, GaSe, GeSe, borophene, and several others. In this review article, we highlight some successful recent applications of DMC to 2D systems for improved property predictions beyond standard DFT.

\end{abstract}
\maketitle

\footnote{Notice: This manuscript has been authored by UT-Battelle LLC under contract DE-AC05-00OR22725 with the US Department of Energy (DOE). The US government retains and the publisher, by accepting the article for publication, acknowledges that the US government retains a nonexclusive, paid-up, irrevocable, worldwide license to publish or reproduce the published form of this manuscript, or allow others to do so, for US government purposes. DOE will provide public access to these results of federally sponsored research in accordance with the DOE Public Access Plan (\url{https://www.energy.gov/doe-public-access-plan}).}

\section{Introduction}\label{intro}

Since the synthesis of graphene in 2004 by Geim and Novoselov that led to the 2010 Nobel Prize,\cite{doi:10.1073/pnas.0502848102,doi:10.1126/science.1102896,graphene-rise} there has been an overwhelming interest in the field of 2D materials. 2D materials are single-layer crystalline structures that have a large lateral dimension compared to their thickness. Oftentimes, these monolayers are exfoliated from layered materials that are held together by weak van der Waals (vdW) bonds. Due to their lack of surface groups or dangling bonds and large surface-to-volume ratio, 2D materials can possess interesting properties that are substantially different from those of their bulk counterparts.\cite{doi:10.1073/pnas.0502848102,doi:10.1126/science.1102896,graphene-rise,PhysRevLett.105.136805,mos2-transistor,mos2-photo,sunlight-absorption-mos2,BernardiAtacaPalummoGrossman+2017+479+493} In addition, 2D materials can possess enhanced quantum confinement and significantly reduced dielectric screening.\cite{sunlight-absorption-mos2,BernardiAtacaPalummoGrossman+2017+479+493,2dexciton} These materials also present interesting phenomena such as enhanced carrier mobility,\cite{doi:10.1073/pnas.0502848102,doi:10.1126/science.1102896,graphene-rise} reduction of charge carrier scattering,\cite{doi:10.1073/pnas.0502848102,doi:10.1126/science.1102896,graphene-rise,PhysRevLett.105.136805,mos2-transistor,mos2-photo,2dexciton} superior mechanical properties,\cite{PAPAGEORGIOU201775} direct-to-indirect band gap transitions,\cite{PhysRevLett.105.136805} nontrivial topological states,\cite{rucl3,2d-top-jarvis,wte2} and superconductivity \cite{2dsupercon,2djarvis-supercon} and magnetism in 2D.\cite{cri3} These physical phenomena can be exploited for future applications in optoelectronics, spintronics, quantum computing, and clean energy.\cite{applications,https://doi.org/10.1002/adma.201904302}

In addition to graphene, there exist several other synthesized monoelemental 2D materials such as germanene,
\cite{Davila_2014,Acun_2015} borophene, \cite{BOUSTANI1997355,borophene-2,https://doi.org/10.1002/anie.201505425,doi:10.1126/science.aad1080} silicene, \cite{silicene,PhysRevLett.102.236804} antimonene, \cite{https://doi.org/10.1002/anie.201411246,antimonene-1,https://doi.org/10.1002/adma.201602128} phosphorene, \cite{phosphorene,phosphorene-2,10.1063/1.4868132} and tellurene. \cite{PhysRevLett.119.106101,tellurene-2} A popular class of 2D materials for optoelectronic and magnetic applications is transition metal dichalcogenides (TMDs) \cite{PhysRevLett.105.136805,mos2-transistor,mos2-photo,ataca-mx2} such as monolayer MoS$_2$, \cite{PhysRevLett.105.136805,mos2-transistor,mos2-photo} MoSe$_2$, \cite{mose2-1,mose2-2} WSe$_2$, \cite{mose2-2,wse2,wse2-2} and WTe$_2$. \cite{wte2,wte2-other} Similarly to TMDs, there exist post-transition metal chalcogenides (PTMCs) \cite{doi:10.1002/adma.201601184,C5NR08692B} such as 2D GaSe, \cite{PhysRevB.96.035407,doi:10.1063/1.4973918,GaSe-optical,C8NR01065J,GaSe-nature-synthesis,Rahaman_2018,Rahaman_2017,doi:10.1002/adma.201601184,doi:10.1063/1.5094663} GeSe, \cite{https://doi.org/10.1002/adfm.201704855,10.1063/1.4931459} and InSe \cite{inse,Zhou_2018} that have also shown promise for optoelectronic applications. Another class of 2D materials with applications in energy storage is MXenes, \cite{mxene,mxene2,C6NR01333C} which are monolayer structures of transition metal carbides, nitrides, or carbonitrides. Monolayer halide-based materials such as CrI$_3$, \cite{cri3} CrBr$_3$, \cite{crbr-exp} and NiI$_2$ \cite{nii2} have also been synthesized as 2D magnets, and RuCl$_3$ has been shown to have topological properties. \cite{rucl3}

The electronic density in 2D materials is highly inhomogeneous. 2D materials may possess a combination of short-range, strong correlations (if {\it d} orbitals are involved) with long-range, weak correlations. These complicated electronic properties of 2D materials make them difficult to model using methods such as density functional theory (DFT).\cite{PhysRev.136.B864,PhysRev.140.A1133} DFT calculations are,  in principle, exact but rely on approximations for the unknown exchange-correlation (XC) functional that models the many-body quantum interactions.\cite{PhysRev.136.B864,PhysRev.140.A1133} Although modern approximations for density functionals have advanced throughout the past few decades, progressing from local (local density approximation [LDA] \cite{PhysRevB.23.5048,PhysRevLett.45.566}), to semilocal (generalized gradient approximation [GGA] \cite{PhysRevLett.77.3865,PhysRevB.46.6671} and meta-GGA \cite{PhysRevLett.115.036402,r2scan,PhysRevB.84.035117,10.1063/1.2213970,PhysRevLett.102.226401}), to hybrid functionals (i.e., HSE, PBE0, and B3LYP\cite{10.1063/1.2404663,hybrid,10.1063/1.464913,10.1063/1.472933,10.1063/1.478522,PhysRevB.37.785,doi:10.1139/p80-159,hybrid}), certain shortcomings remain. The most obvious limitation of DFT is that its results can strongly depend on the XC functional, and there is no a priori way to know the significance of the approximations in the chosen functional. In addition, many DFT functionals require explicit corrections to handle vdW interactions.\cite{PhysRevLett.92.246401,Klimes_2010,PhysRevX.6.041005,PhysRevB.82.081101,https://doi.org/10.1002/jcc.20495,10.1063/1.3382344,https://doi.org/10.1002/jcc.21759,PhysRevLett.102.073005} The vdW interactions are crucial for low-dimensional layered materials, whose properties are significantly impacted by which vdW correction or functional is used in DFT.\cite{PhysRevB.98.014107,jarvis-vdw}

%FAR Comments
%Strongly correlated electronic systems are difficult to model with DFT. This failure can be due to standard exchange-correlation functionals (LDA, GGA) over-delocalizing valence electrons \cite{https://doi.org/10.1002/qua.24521},  which is why DFT fails for strongly localized systems, such as transition metal-based materials that contain $d$ and $f$ electrons. To correct this delocalization error, the Hubbard (U) correction \cite{PhysRevB.57.1505}, which is an additional onsite Coulomb interaction, can be applied to correlated electronic states. \frrem{Although this is a valid approach,}
%FAR the use of the word valid can be missunderstood by accurate.
%\fradd{ Although self-consistent approaches for obtaining U have been developed \cite{PhysRevB.103.045141,10.1063/1.2987444},} the choice of U value can be arbitrary and is often either fit to experiments or selected from previous literature values \cite{Tolba18},\frrem{ although self-consistent approaches for obtaining U have been developed \cite{PhysRevB.103.045141,10.1063/1.2987444}.} \cite{Tolba18}, \frrem{A} \fradd{The} vast majority of interesting 2D materials are electronic, magnetic, and topological materials that contain transition metals with correlated $d$ and $f$ electrons and DFT+U has been previously applied to handle such correlations \cite{PhysRevB.102.165157,Torelli_2019}. 

%DW Corrections

Modeling strongly correlated electronic systems poses a challenge within DFT frameworks due to the tendency of standard XC functionals (like LDA and GGA) to overly delocalize valence electrons.\cite{https://doi.org/10.1002/qua.24521} This issue predominantly stems from the self-interaction error,\cite{PhysRevB.23.5048} rendering DFT inadequate for accurately describing strongly localized systems, such as those found in transition metal--based materials containing $d$ and $f$ electrons. To correct this delocalization error, the Hubbard (U) correction, \cite{PhysRevB.57.1505} which is an additional on-site Coulomb interaction, can be applied to correlated electronic states. Although self-consistent approaches for obtaining U have been developed, \cite{PhysRevB.103.045141,10.1063/1.2987444} the choice of U value can be arbitrary and is often either fit to experiments or selected from previous literature values.\cite{Tolba18} The vast majority of interesting 2D materials are electronic, magnetic, and topological materials that contain transition metals with correlated $d$ and $f$ electrons, and DFT+U has been previously applied to handle such correlations.\cite{PhysRevB.102.165157,Torelli_2019} Hybrid/exact-exchange functionals and some meta-GGAs also reduce the degree of self-interaction error, but a consistent high-accuracy treatment remains elusive. The problem of self-interaction error was the main topic of the original LDA DFT paper,\cite{PhysRevB.23.5048} and the problem persists more than 40 years later.

Another major shortcoming of DFT is its underestimation of the band gap for semiconductors and insulators, which results in significant disagreements with experiments. This underestimation is clearly apparent for local and semilocal functionals such as LDA and GGA, with noted improvements for meta-GGA functionals.\cite{https://doi.org/10.1002/qua.560280846,bandgap-benchmark,bandgap-benchmark2} Significant improvements to the accuracy of DFT electronic structure predictions can be achieved through the use of hybrid functionals, but there are fundamental theoretical issues that can limit DFT's accuracy.\cite{https://doi.org/10.1002/qua.560280846,bandgap-benchmark,bandgap-benchmark2} DFT is a ground-state theory that maps the electronic structure problem to a fictitious Kohn-Sham system of single electrons occupying energy bands. Therefore, the band gap computed with DFT can be understood as the Kohn-Sham energy gap. For an infinite solid calculated with the exact DFT functional, the DFT gap would be the exact quasiparticle gap.\cite{doi:10.1073/pnas.1621352114} In real materials, electrons feel the effects of surrounding electrons and holes, a phenomenon that is not captured at the quasiparticle level. To accurately predict optical excitations (e.g., the promotion gap, excitonic effects), these quasi-electron and quasi-hole interactions must be taken into account.\cite{RevModPhys.74.601} These interactions can be obtained by using post-DFT methods such as many-body perturbation theory (i.e., the GW approximation) \cite{RevModPhys.74.601,PhysRev.139.A796,PhysRevB.34.5390} and the Bethe-Salpeter equation (BSE) \cite{RevModPhys.74.601,PhysRev.78.382} for electron-hole interactions or by exciting the system within a many-body stochastic theory such as diffusion Monte Carlo (DMC).\cite{RevModPhys.73.33} The quasiparticle picture and excitonic effects are extremely important for understanding 2D materials.\cite{sunlight-absorption-mos2,2dexciton} The reduced dielectric screening in low-dimensional systems can result in strong excitonic effects (exciton binding energies in the 10--100 meV range).\cite{sunlight-absorption-mos2,BernardiAtacaPalummoGrossman+2017+479+493,2dexciton} 

In addition to reduced screening, spin-orbit coupling (SOC) can play an important role in  understanding the electronic structure of 2D materials.\cite{soc,PhysRevB.100.125422,Premasiri_2019,valley,2d-top-jarvis} Several TMDs have SOC-induced band splitting in the valence and conduction bands, which can be exploited for spintronic and valleytronic applications. In addition, SOC can induce band inversions and other topological properties in 2D materials.\cite{soc,PhysRevB.100.125422,Premasiri_2019,valley,2d-top-jarvis} Modeling spin-orbit in 2D materials can be challenging with standard DFT methods. There have been several computational studies that have shown the considerable impact of SOC on the electronic structure and have demonstrated that results can significantly vary depending on the choice of methodology.\cite{soc-paper,2d-top-jarvis,Staros_CrI3,wines-crx3,soc-paper2,D0RA08279A,D2TA05255E}

2D materials are promising and complicated structures that have significant vdW interactions, correlated electronic and magnetic properties, and unique excitonic and topological properties that make them the perfect test bed for more accurate computational methodologies. The strong dependence of the results on the density functional and Hubbard correction can severely limit the reliability of DFT for these 2D systems. In addition, methods that can correct bands gaps and other DFT predictions such as GW and BSE still significantly rely on the Kohn-Sham eigenvalues obtained from a particular DFT functional. For this reason, DMC \cite{RevModPhys.73.33} is an ideal method to accurately describe the properties of 2D materials. Although DMC is orders of magnitude more computationally intensive than DFT, it is extremely scalable on modern computers and scales as $N^{3-4}$, where $N$ is the number of electrons in the simulation.\cite{RevModPhys.73.33} In addition, the stochastic many-body nature of DMC and its controllable approximations allow for results that have a much weaker dependence on the starting wavefunction and density functional.\cite{RevModPhys.73.33} 

Although there are major shortcomings of DFT that can be overcome by using many-body methods such as DMC, DFT is an absolutely essential tool in the electronic structure community. For a majority of problems in physics, chemistry, and materials science, standard DFT is suitable to obtain important qualitative trends in predicted properties. In fact, for closed-shell systems and scenarios without strong correlations, DFT does an adequate job of describing electronic properties, and systematic trends are well described. The low computational expense also allows more systems to be studied than with many-body methods, facilitating the construction of materials databases. However, fundamentally DFT and many-body approaches such as QMC have different properties, so we consider it best to ask when, where and how each method may be most productively applied. Unlike DFT, the majority of QMC methods offer variational properties enabling the few approximations in them to be systematically tested and potentially reduced in small enough systems. QMC methods (such as DMC) can describe strong electron correlations and van der Waals interactions more accurately than DFT, making them very suitable for 2D materials. They are therefore ideal for as-yet unsynthesized novel materials, where there is no empirical data to guide DFT approximation selection or where there are known issues with DFT approximations for the particular materials class. For the materials covered in this review, QMC can be applied directly, but can also be used to validate the selection of exchange-correlation functional.

The systematic trends from DFT functionals can be used to our advantage when attempting to understand complex phenomena. For example, the underbinding and subsequent lattice constant errors of GGA were used as a method to identify new exfoliable vdW materials. \cite{jarvis-vdw} One might ask to what extent DFT is successful and when it should be trusted as a complement to DMC. In turn, one might ask in what situations is DMC absolutely necessary. Due to the fact that DFT is significantly less computationally expensive than DMC allows us to perform DFT calculations at a much larger scale, in terms of number of calculations and system sizes. Fundamentally, however, the methods have different properties, so it is better to ask when and where each method may be best applied. Most importantly, we can use our DMC results as a metric to compare various DFT calculations to, and find which DFT functional best matches the DMC results. This can aid in the design of new DFT functionals (i.e., improving existing vdW functionals). It also allows us to make an informed decision about which DFT functional (or value of Hubbard U, amount of HF mixing in hybrid functionals, which vdW correction, etc.) to use for a system of interest in order to obtain more complicated properties that are difficult to obtain with DMC (i.e., band structure, magnetic anisotropy, simulations at larger system sizes). This is especially relevant for systems that do not have an existing experimental benchmark to compare DFT to or if we are focusing on properties that are not easily obtained by experiments (i.e., interlayer binding energy, magnetic exchange). DFT and DMC generally focus on ground state properties, since they are both based on a procedure of energy minimization. Despite this fact, time-dependent DFT \cite{PhysRevLett.52.997} can be applicable to a broad range of excited states while many-body approaches such as DMC can yield accurate excited state results (i.e., excitonic effects). While modern QMC rests on the foundations of the development work done in the 1980s and 1990s, critical advances made since then include the development of twist averaging (enabling practical calculation of metallic materials), more accurate trial wavefunction optimization methods, including the critical breakthrough of the linear method/energy minimization, and the development of better methods for pseudopotentials, which are critical for the heavier elements in dichalcogenides. These are cited at several points in the review \cite{PhysRevLett.94.150201,PhysRevLett.98.110201,atomic-energies,PhysRevB.106.075127,PhysRevB.102.155151,10.1063/1.4907589,PhysRevB.93.075143,10.1063/1.4984046,10.1063/1.4995643,10.1063/1.5038135,10.1063/1.5040472,10.1063/1.5121006,10.1063/5.0087300,10.1063/5.0109098,zhou2023new,10.1063/1.3380831,10.1063/1.460849,PhysRevB.78.125106,PhysRevLett.73.1959,PhysRevB.51.10591,enhanced-twist,PhysRevA.93.042502}.

In this review article, we provide an overview of how the DMC method has recently been applied to a variety of low-dimensional systems. Section \ref{theory} provides a brief overview of the theory, Section \ref{magnetic} describes how DMC has been applied to correlated 2D magnetic systems, Section \ref{electronic} details how DMC has been used for improved predictions of electronic properties (including excitonic, spin-orbit, and topological effects), Section \ref{binding} focuses on how DMC has been used to improve the prediction of interlayer interactions, Section \ref{energetics} describes how DMC has been used for the  cohesion and adsorption energetics of low-dimensional materials, and Section \ref{conc} provides concluding remarks
and offers future perspectives.

\section{Theory}\label{theory}

In this review article, we survey the application of the DMC \cite{RevModPhys.73.33} method, a real-space quantum Monte Carlo (QMC) method, to low-dimensional systems. Although the exact specifications of the calculations may vary among the works presented in this manuscript, we provide a broad overview of the theory and approximations used throughout these calculations.

%FAR Comments
%DMC is a projector-based method that can be used to obtain the ground state energy of a many-body system. In this method, the time-dependent Schr{\"o}dinger equation is recast into the imaginary-time ($\tau$) Schr{\"o}dinger equation:  
%\begin{equation}\label{imag-schro}
%-\frac{\partial_{\tau} \Psi(\mv{R},\tau)}{\partial \tau} = (\hat{H}-E)\Psi(\mv{R},\tau),
%\end{equation}
%where $\hat{H}$ is the Hamiltonian operator which consists of kinetic and potential energy contributions, $\tau$ measures the progress in imaginary time, $\Psi(\mv{R},\tau)$ is the \frrem{ground state}  wavefunction \fradd{at imaginary time $\tau$}, and $E$ is the offset of the ground state energy. As \frrem{the offset is adjusted} %FAR this is not correct
%\fradd{$\tau \rightarrow \infty$}, the weight of the DMC configurations with higher energy are damped exponentially \frrem{with their propagation in imaginary time}. Therefore, the ground state wavefunction is projected out once a steady state has been reached.    

%DW Corrections
DMC is a projector-based method that can be used to obtain the ground-state energy of a many-body system. In this method, the time-dependent Schr{\"o}dinger equation is recast into the imaginary-time ($\tau$) Schr{\"o}dinger equation:  
\begin{equation}\label{imag-schro}
-\frac{\partial_{\tau} \Psi(\mv{R},\tau)}{\partial \tau} = (\hat{H}-E)\Psi(\mv{R},\tau),
\end{equation}
where $\hat{H}$ is the Hamiltonian operator, which consists of kinetic and potential energy contributions; $\tau$ measures the progress in imaginary time; $\Psi(\mv{R},\tau)$ is the wavefunction at imaginary time $\tau$; and $E$ is the offset of the ground-state energy. As $\tau \rightarrow \infty$, the weight of the DMC configurations with higher energy is damped exponentially. Therefore, the ground-state wavefunction is projected out once a steady state has been reached.  

The fluctuations in the reweighting process (adjusting the statistical weight of sampled configurations to account for differences between the sampled and targeted distributions) have been shown to be reduced significantly by using a trial (guiding) wavefunction ($\Psi_T (\mv{R})$), which transforms the wavefunction in Equation \ref{imag-schro} into $f(R,t)=\Psi_T (\mv{R})\Psi(\mv{R},\tau)$. The quality of the trial wavefunction can be improved by a set of Jastrow factors: \cite{PhysRev.98.1479} \begin{equation}\label{Slater-jastrow} 
\Psi_T (\mv{R}) = e^{J(\mv{R})} \sum_{k} \alpha_k D^{\uparrow}_k(\mv{R})D^{\downarrow}_k(\mv{R}),
\end{equation}
where $J(\mv{R})$ represents the Jastrow factors; $D^{\uparrow}_k$, $D^{\downarrow}_k$ represent the $k$-th Slater determinants of up and down spins in the multideterminant expansion; and $\alpha_k$ represents the weight of the $k$-th determinant configuration. The Slater determinants in the trial wavefunction usually come from a DFT or Hartree-Fock (HF) calculation. For most of the work presented in this review, a single-determinant trial wavefunction is used. More general representations constructed from multideterminant expansions are also possible. In fact, the application of millions of determinants to solids is now possible.\cite{BenaliJCP2020,MalonePRB2020} The Jastrow factor explicitly includes electron correlations such as parameterized electron-ion, electron-electron, and electron-electron-ion terms. These terms are found through optimization of the wavefunction and the ground-state energy and/or variance.\cite{PhysRevLett.94.150201,PhysRevLett.98.110201} The increased accuracy of the trial wavefunctions---being closer to an exact eigenstate---also serves to reduce the number of statistical samples needed to reach a specific error bar, usually reducing computational costs.

DMC formally treats the many-body electron-electron interaction exactly; however, it utilizes several approximations 
%FAR the fixed-node is sometimes denoted as uncontrolled same as the locality approximation
to reduce the cost of performing the calculations in the presence of core electrons and the fermion sign problem.  
Fixed-node 
%fixed-node is also used for multi determinants.
DMC produces the ground-state energy with the constraint $\Psi_T (\mv{R})=0 \implies \Psi_0(\mv{R})=0$, such that the DMC wavefunction shares the same nodes or phase as the trial wavefunction.\cite{PhysRevLett.45.566,10.1063/1.447637,10.1063/1.443766,RevModPhys.73.33} 
This fixed-node approximation is enforced to maintain the antisymmetry of the wavefunction. Given the exact nodal surface of a system from an exact trial wavefunction, the fixed-node DMC method will yield the exact energy. For approximate nodal surfaces, the error introduced is variational (positive) in the energy.  Fortunately for approximate trial wavefunctions, this error is typically small, and because the error is variational, different choices of input wavefunction can be tested and the most accurate selected. There have been some guided efforts to estimate the fixed-node error \cite{vdw-benali,10.1063/5.0026275,PhysRevB.103.205206,PhysRevB.93.094111,atomic-energies}, and particularly for low-dimensional materials \cite{Shin_PRM_2021,Tb-soc,PhysRevB.106.075127,PhysRevB.101.205115}, but the number of results is small enough that it is premature to infer general trends. One approach to estimate the error involves performing higher accuracy quantum chemistry calculations such as selected Configuration Interaction (sCI). These can be used either directly or as a source of improved trial wavefunction in QMC, enabling a sensitivity and convergence analysis of the fixed-node error provided a sufficient number of determinants can be considered. It has been found that the magnitude of this error is entirely system dependent. It is important to note that sCI cannot always serve as a direct benchmark, unless the relevant space of determinants is not very large. However, sCI provides a way to estimate DMC nodal errors in systems where the sCI energy can be sufficiently converged in terms of basis set and system size (e.g., in primitive cells of solids and in small molecules). For large enough systems, sCI becomes impractically expensive to converge, and the DMC errors cannot be directly estimated via sCI. For example, the fixed-node error of Si systems was found to be extremely negligible for the ground state (1.3(2)$\%$), \cite{PhysRevB.103.205206} using the recovery of the nearly-exact correlation energy as an error metric. This fixed-node error has been estimated to be similar in magnitude for systems containing Sn (2$\%$), \cite{Tb-soc} Ru (4$\%$), \cite{PhysRevB.106.075127} and Cl (3$\%$) \cite{PhysRevB.106.075127} but significantly higher for systems containing Tb (12$\%$) \cite{Tb-soc} and Mn (9$\%$).\cite{Tb-soc} However, it has been demonstrated that these large biases can cancel when calculating quantities such as molecular binding energies.\cite{Tb-soc} It is important to note that systems for which one cannot compute well converged sCI energies (due to system size), it may still be possible to compute well converged CC energies. CC (ideally, CC calculations with single, double, and perturbative triple excitations (CCSD(T)) in the complete basis set limit) may be used to estimate the fixed-node error in systems where the CC ansatz works well, such as in closed-shell systems and systems that are largely single-reference. For example, the fixed-node bias was estimated for phosphorene from a fully converged CCSD(T) calculation of a P$_4$ cluster \cite{PhysRevX.9.011018}. It has been demonstrated that CC and DMC have strong agreement for ground state energy difference-based quantities for large noncovalent complexes \cite{10.1063/5.0026275} and materials dominated by vdW interactions.\cite{vdw-benali} There have been some assessments of the fixed node errors for single determinant DMC band gaps, where it has been reported that for certain systems, the fixed node error for the ground state and the excited state are different, leading to a slight bias in the gap. \cite{PhysRevB.103.205206,PhysRevX.9.011018,PhysRevB.101.205115}
On the other hand, there have been instances where the fixed node error for the single determinant DMC band gap has been negligible.\cite{Shin_PRM_2021} Practical methods to minimize or test the fixed-node bias involve changing the trial wavefunction by adjusting the Hubbard parameter in DFT+U or varying the amount of exact exchange in hybrid DFT and finding the minimum DMC energy variationally.\cite{PhysRevB.102.045103,PhysRevB.106.075127,PhysRevMaterials.3.124414,PhysRevMaterials.2.085801,PhysRevX.4.031003} In cases where the fixed-node error is high, there exist more sophisticated techniques to improve the accuracy of the trial wavefunction, such as optimizing the orbitals within QMC \cite{PhysRevB.102.155151,devaux_electronic_2015,casula_improper_2013}, using multideterminants \cite{multi-det,MalonePRB2020,BenaliJCP2020} or backflow transformations.\cite{PhysRevE.74.066701,PhysRevB.93.094111} Backflow transformations can be particularly challenging and computationally expensive for solids.\cite{PhysRevB.93.094111,PhysRevB.101.205115} It has been reported that the gain in accuracy from backflow is overshadowed by errors from finite-size (FS) and nonlocal pseudopotenitals.\cite{PhysRevB.93.094111,PhysRevB.101.205115,PhysRevE.74.066701,PhysRevB.105.184114} Extensive benchmarking of backflow for molecules and solids can be found in Ref. \onlinecite{PhysRevB.93.094111}, and benchmarking of backflow for a 2D system (hexagonal boron nitride [hBN]) can be found in Ref. \onlinecite{PhysRevB.101.205115}. Accuracy can also be improved by optimizing the orbitals entirely within the QMC calculation. This has yet to be done systematically in solid-state materials for production-level calculations (there have yet to be any examples for 2D materials), but preliminary applications indicate that the approach should be practical \cite{PhysRevB.102.155151,devaux_electronic_2015,casula_improper_2013} and it has been used in studies of cuprates \cite{PhysRevX.4.031003,PhysRevB.90.125129,PhysRevB.92.161116}, the Ce volume collapse \cite{PhysRevB.91.081101}, and for phonon studies. \cite{PhysRevB.103.L121110,PhysRevB.109.205151}

%More detailed backflow paragraph, cutting down at the suggestion of PRCK
%It has been reported that the gain in accuracy from backflow is overshadowed by errors from finite-size and nonlocal pseudopotenitals \cite{PhysRevB.93.094111,PhysRevB.101.205115,PhysRevE.74.066701,PhysRevB.105.184114}. In addition, the gain in accuracy from backflow is not controllable, which can result in not being able to take advantage of cancellation of errors for energy differences of certain systems. Extensive benchmarking of backflow for molecules and solids can be found in Ref. \onlinecite{PhysRevB.93.094111} and benchmarking of backflow for a 2D system (hBN) can be found in Ref. \onlinecite{PhysRevB.101.205115}. 

In addition to variational Monte Carlo (VMC) and DMC, other robust techniques such as Coupled Cluster (CC) and Auxiliary Field QMC (AFQMC) have been refined and applied to difficult problems throughout the last decade. Although these methods have not yet been extensively applied to 2D or quasi-2D systems (as is the case for VMC and DMC), CC has been used to compute the electronic spectrum of TMDs such as monolayer MoS$_2$, MoSe$_2$, WS$_2$, and WSe$_2$,\cite{PhysRevB.101.241113} and AFQMC has been applied to study Moir\'e systems, including a metal-insulator transition in a semiconductor heterobilayer model.\cite{PhysRevLett.132.076503} Aside from low-dimensional materials, CC has been used to obtain the quasiparticle band structure of simple bulk solids \cite{cc-diamond,PhysRevLett.131.186402} and AFQMC has been used for a wider range of materials including oxides, diamond, LiF and Al. \cite{10.1063/1.5040900,PhysRevB.102.161104,10.1063/5.0031024} We expect the expanded application to 2D materials to follow in the coming years.

Pseudopotentials are an essential aspect of DMC calculations. Pseudopotentials are required in DMC to avoid the large cost of performing all-electron calculations. High-quality pseudopotentials are essential for ensuring the accuracy of DMC simulations while simultaneously removing chemically inert core electrons. Luckily, in the past decade, there has been a strong effort to develop DMC-specific norm-conserving pseudopotentials with suitable accuracy.\cite{10.1063/1.4907589,PhysRevB.93.075143,10.1063/1.4984046,10.1063/1.4995643,10.1063/1.5038135,10.1063/1.5040472,10.1063/1.5121006,10.1063/5.0087300,10.1063/5.0109098,zhou2023new} Although the original method for evaluating pseudopotentials in DMC,  the locality approximation, \cite{10.1063/1.3380831} was nonvariational,  newer methods such as the T-moves approach \cite{10.1063/1.460849} restore the variational properties of the overall method.

The accuracy of DMC results is also impacted by FS errors.\cite{RevModPhys.73.33} Specifically, there are two types of FS errors that can impact the DMC simulations. The first type, one-body FS errors, arises from insufficient sampling of the Brillouin zone.\cite{RevModPhys.73.33} This error can be significantly reduced by using twist-averaged boundary conditions. Various twist-averaging (TA) schemes have been applied to solids throughout the years, ranging from the use of the computationally inexpensive Baldereschi-point grids \cite{PhysRevB.7.5212,PhysRevB.78.125106,PhysRevLett.73.1959,PhysRevB.51.10591} to more recent enhanced TA techniques for magnetic metals.\cite{enhanced-twist} In most common production-level calculations of 2D materials, the total energy is converged at the DFT level with respect to $k$-point grid size, and this grid is used as the subsequent number of twists in the DMC calculations. Ideally, one would wish to converge the number of twists at the DMC level, but this is a computationally expensive endeavor. The second type, two-body FS errors, arises from the artificial interaction of periodic XC holes.\cite{RevModPhys.73.33} This can be minimized by performing the simulation at multiple supercell sizes and then extrapolating to the thermodynamic (infinite-size) limit. This extrapolation is essential for achieving accurate observable quantities, and careful calculations at supercells large enough to eliminate this bias must be conducted. In addition to FS errors, errors can arise from the choice of time step $\delta\tau$. In DMC, the walkers are propagated in imaginary time by statistically sampling a Green's function. This relies on the Green's function projector within the short time approximation, which becomes exact as $\delta\tau$ approaches 0. For this reason, finite time steps can introduce errors in the DMC energies, and careful analysis and testing must be performed when choosing a time step to minimize the target uncertainty.\cite{RevModPhys.73.33} In most of 2D material works presented in this review, generously small time steps on the order of 0.01 Ha$^{-1}$ (0.27 eV$^{-1}$) were chosen to minimize time-step error.\cite{PhysRevX.9.011018,PhysRevResearch.5.033223,10.1063/5.0030952,Staros_CrI3, vse2-wines,wines-crx3,mno2-qmc,Shin_PRM_2021}

For systems with significant spin-orbit effects, the QMC methods have been generalized to handle electron spins as quantum variables so that any spin state, including noncollinear spins, can be described. The first key difference from spinless Hamiltonians is that one-particle
spatial-only orbitals are replaced by two-component spinors. The spinors can be written as 
$\chi(\mathbf{r},s)=\alpha\varphi^{\uparrow}({\bf r})\chi^{\uparrow}(s)
 +\beta\varphi^{\downarrow}({\bf r})\chi^{\downarrow}(s) $,
where orbitals $\varphi^{\uparrow}, \varphi_{\downarrow}$ are calculated in spinor-based self-consistent approaches (DFT, Dirac-Fock, etc.). The spinors also include spin variables that enter the spin functions $\chi^{\uparrow}(s),\chi^{\downarrow}(s)$, while $\alpha,\beta$ are complex constants such that $\alpha^2+\beta^2=1$.
The  fixed-node Slater-Jastrow trial functions are generalized to accommodate the spinors as given by 
\begin{equation}
\Psi_T=e^{J({\bf R})}\sum_k c_k {\rm det}_k[\chi_i({\bf r}_j,s_j)].
\end{equation}
For example, we can see that a singlet wavefunction in Eq. (\ref{Slater-jastrow}) is built from  determinants of $(N/2)\times (N/2)$ matrices,  while in the spinor wavefunction, there are  determinants of $N\times N$ matrices (because spins vary, the position space does not factorize into up- and down-spin subspaces).
Correspondingly, the fixed-node 
approximation is generalized to the fixed-phase approximation because the trial function is inherently complex. We can write $\Psi(\mathbf{R},\tau) = \rho(\mathbf{R},\tau)e^{i \Phi(\mathbf{R},\tau)}$ and substitute it into the imaginary-time Schr\"{o}dinger equation.   Decomposition into real and imaginary parts leads to an exact equation for the amplitude $\rho(\mathbf{R},\tau)\geq 0$  given by% and for the phase $\Phi(\mathbf{R},\tau)$ as given by
\begin{equation}
 -\frac{\partial \rho(\mathbf{R},\tau)}{\partial \tau} = \left[ T_{kin} + V(\mathbf{R}) + \frac{1}{2} \left| \nabla \Phi(\mathbf{R},\tau)\right|^2\right] \rho(\mathbf{R},\tau). 
    \end{equation}
 The fixed-phase approximation
 %\cite{ortiz1993}
    imposes
 the unknown phase $\Phi$ to be equal to the phase of the trial state $\Psi_T({\bf R})=\rho_T({\bf R})e^{i\Phi_T({\bf R})}$ as given by
 %that is independent of $\tau$ 
 \begin{equation}
 \Phi(\mathbf{R},\tau) 
 \stackrel{!}{=}
  \Phi_T(\mathbf{R}).
 \end{equation}
 The fixed-phase approximation is variational because the repulsive potential $V_{ph}  = \frac{1}{2} \left| \nabla \Phi_T\right|^2$
can only raise the energy for any approximate  $\rho$.  The spin-orbit terms are typically included in the pseudopotentials with the formalism generalized accordingly; see further details in Refs.
\onlinecite{PhysRevA.93.042502,10.1063/1.4954726}.

Although various real-space DMC codes have been developed over the past few decades and applied to 2D materials (e.g., CASINO, \cite{Needs_2010} PyQMC, \cite{10.1063/5.0139024} QWalk \cite{WAGNER20093390}), most of the work in this review article utilizes the QMCPACK code.\cite{Kim_2018,10.1063/5.0004860} The QMCPACK code has been routinely updated and maintained to track the latest algorithmic and hardware developments and utilize computational resources more efficiently (\url{https://github.com/QMCPACK/qmcpack}). The QMCPACK code is equipped with the Nexus \cite{nexus} workflow automation software, which helps create and monitor DFT/DMC workflows that 
can significantly reduce the user time by tracking computational dependencies. More detailed information about these theoretical frameworks and codes can be found in Refs. \onlinecite{RevModPhys.73.33,PhysRevLett.45.566,10.1063/1.447637,10.1063/1.443766,Kim_2018,10.1063/5.0004860,Needs_2010,10.1063/5.0139024,WAGNER20093390,nakano_turbogenius_2023}. For this review article, it is important to note that the error bars in each figure and the uncertainty noted in $\pm$ in concise notation (i.e., 5.3 $\pm$ 0.1 can be expressed as 5.3(1)) refer to the standard error about the mean for each observable quantity calculated with QMC (e.g., energy, band gap, magnetic moment, lattice constant).

\section{DMC Properties of 2D Materials}\label{results}

\subsection{Magnetic Properties}\label{magnetic} 

After the experimental synthesis of monolayer CrI$_3$, which was measured to have a Curie temperature ($T_c$) of 45 K, \cite{cri3} interest in identifying and physically understanding similar magnetic 2D materials has significantly increased. It has been experimentally reported that monolayer VSe$_2$ is ferromagnetic (FM) on a vdW substrate \cite{vse2} and that FM order exists in the bilayer limit for Cr$_2$Ge$_2$Te$_6$.\cite{crgete} Some other examples of low-dimensional magnetic systems include FePS$_3$, \cite{feps3} NiPS$_3$, \cite{nips3} MnPS$_3$, \cite{PhysRevLett.124.027601} and Fe$_3$GeTe$_2$.\cite{fe3gete2} Several computational works have also predicted magnetism in a variety of 2D systems.\cite{PhysRevB.88.201402,PhysRevB.94.184428,C7CP07953B,ERSAN2019111} 

The Mermin-Wagner theorem \cite{PhysRevLett.17.1133} states that magnetic order in a 2D material cannot persist unless magnetic anisotropy is present. Moreover, it requires anisotropy that breaks continuous symmetries. Whether the anisotropy is perpendicular to the plane (easy axis) or parallel to the plane (easy plane) determines the type of transition temperature. For easy-axis anisotropy, the system can be described by the Heisenberg model with a finite Curie temperature $T_c$. For easy-plane anisotropy, there is no explicit transition between the ordered and unordered states but instead a Kosterlitz-Thouless transition at finite temperature $T_{KT}$, where the system has quasi-long-range magnetic order below $T_{KT}$. To obtain accurate estimates for observables such as $T_c$ and $T_{KT}$, an accurate computation of the magnetic exchange and magnetic anisotropy energies is essential. %\frcom{In order to obtain magnetic anisotropy one must include spin-orbit
%coupling effects. Since it is essentially a single particle effect one just need to get the total moment per atom correct. I would
%remove this paragraph unless some one is doing DMC calculations with spin-orbit.} \textcolor{red}{DW: Section A1 and A2 deal with
%DMC-informed DFT calculations or spin-orbit/magnetic anisotropy, so it might be a good idea to leave this paragraph in} 
Oftentimes, these results are extremely sensitive to the choice of density functional and Hubbard U parameter, resulting in inconclusive results with DFT. Additionally, these 2D magnetic materials can have a strong interdependence between structural parameters and magnetic properties, making it even more difficult to make accurate predictions. Due to this, using a method such as DMC to understand the electron correlation effects that drive magnetic ordering in 2D is necessary. These accurately computed ab initio magnetic parameters can then be used in conjunction with either classical Monte Carlo methods or other analytical models \cite{Torelli_2018} to estimate observable quantities such as transition temperatures. Next, we discuss a few of our works on modeling magnetic 2D structures.

\subsubsection{Monolayer CrI$_3$}\label{cri3}

%{\bf{\textit{CrI$_3$: Structural and Magnetic Properties}}}

\begin{figure}
\begin{center}
\includegraphics[width=0.5\textwidth]{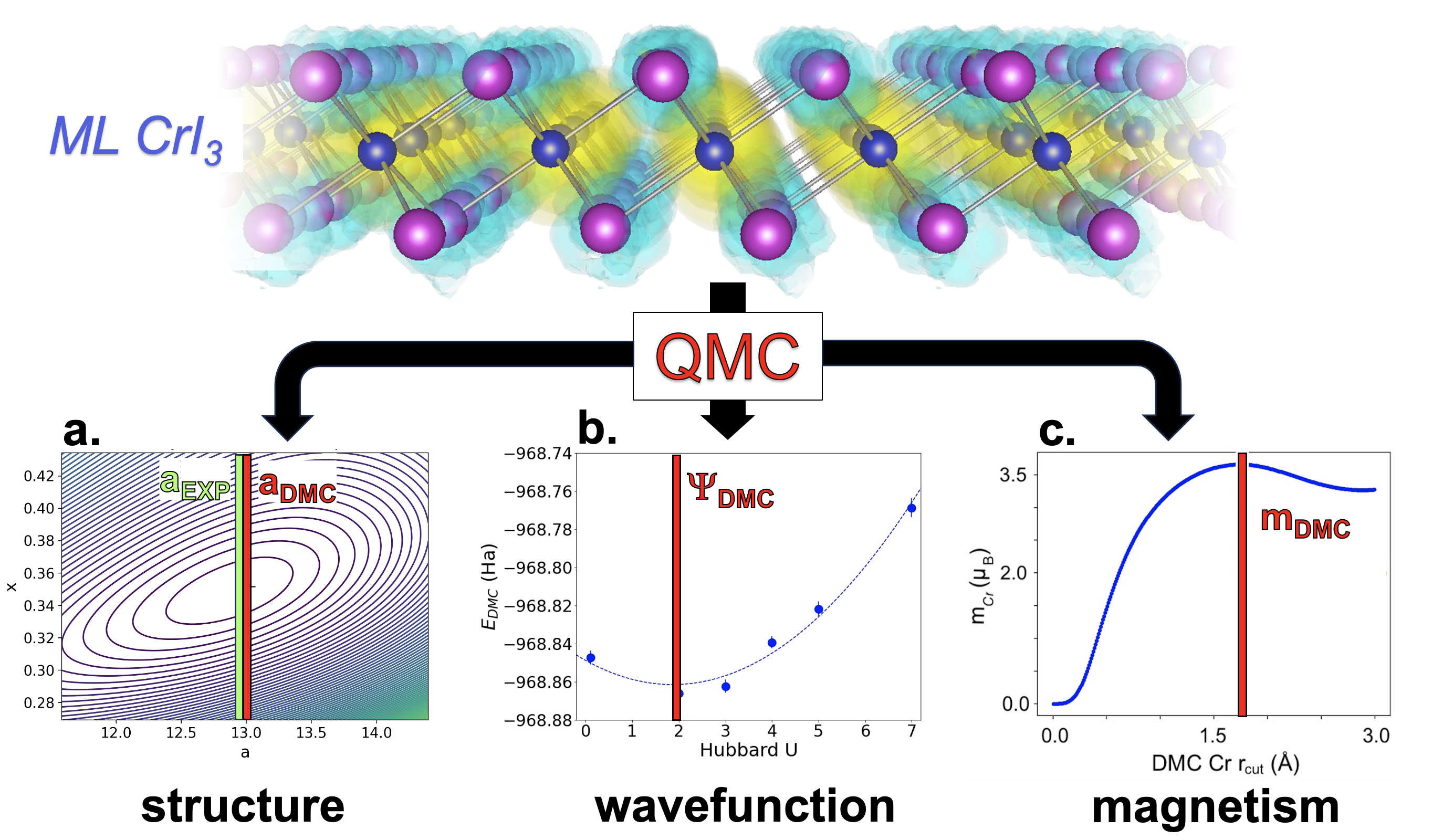}
\caption{To resolve the origin of monolayer CrI$_3$'s magnetism, we modeled this material in three key steps: We (a) first employed the surrogate Hessian line-search method to determine its high-accuracy geometry;
(b) then determined an optimal U at $U=2$ to employ in our PBE+U trial wavefunctions; and (c) finally, integrated the resulting DMC spin density to obtain estimates of CrI$_3$'s magnetic moment of 3.62 $\mu_{B}$. Reproduced from Staros et al., J. Chem. Phys. 156, 014707 (2022), with the permission of AIP Publishing.}
\label{fig:crx3-hessian}
\end{center}
\end{figure}

One of the first monolayer magnets discovered,\cite{cri3} monolayer CrI$_3$ is a Mott insulator \cite{PhysRevB.105.165127} that consists of chromium atoms octahedrally coordinated by six iodine atoms (see Fig.~\ref{fig:crx3-hessian}). Because of the negative exchange interactions among them, the spins on the Cr atoms arrange themselves perpendicular to the plane of the material,\cite{Lado_2017,PhysRevLett.124.017201} resulting in a monolayer ferromagnet with a critical temperature of 45 K.\cite{cri3, Gibertini_NatureNano} Upon stacking, CrI$_3$ heterostructures oscillate between exhibiting ferromagnetism for odd numbers of layers and antiferromagnetism for even numbers of layers.\cite{Sivadas_NanoLett} Moreover, upon the application of pressure that decreases the interlayer distance, bilayer CrI$_3$ transitions from an antiferromagnet to a ferromagnet.\cite{li_pressure-controlled_2019,song_switching_2019}

These phenomena suggest that electron-electron and electron-phonon interactions play a key role in determining CrI$_3$'s magnetic and structural properties. Yet, although significant research has gone into characterizing and modifying the macroscopic properties of this material, far less research has focused on the atomistic determinants of those properties in a manner that accurately accounts for many-body interactions. As a result, only approximate explanations for the source of this material's magnetism had been put forth before the DMC research described below was conducted. In particular, based on a rudimentary analysis of oxidation states, it was originally believed that each iodine atom in CrI$_3$ would have an oxidation state of $-$1 and therefore that the chromium should have an oxidation state of $+$3. Given that neutral Cr has a [Ar]4s$^1$3d$^5$ electron configuration, this suggests that Cr$^{3+}$ should have three valence electrons remaining, giving rise to a magnetic moment of 3 $\mu_B$. More accurate yet still single-reference DFT calculations corroborated this picture---on average---but often differed significantly from one another in detail.\cite{li_single-layer_2020,Yang_JPCC,Wu_PCCP,Lado_2017,Zhang_JMaterChemC} Previous PBE+U ($U = 2$ eV) and HSE calculations, for example, predicted Cr's magnetic moment to be as large as 3.3 $\mu_{B}$,\cite{Yang_JPCC,Zhang_JMaterChemC} while PBE\cite{Zhang_JMaterChemC} and LDA+U \cite{Lado_2017} calculations predicted moments as small as 3.1 $\mu_{B}$ and 3 $\mu_{B}$, respectively---a 10\% difference among moments overall. Even larger differences could be observed among DFT predictions of the in-plane lattice parameter: PBE calculations predicted a CrI$_3$ lattice parameter of 7.008 \AA,\cite{Zhang_JMaterChemC} while LDA+U calculations predicted a lattice parameter of 6.686 \AA,\cite{Lado_2017} a 5 \% difference. These significant DFT discrepancies were further corroborated by our own DFT calculations, which demonstrated not only that the magnetic moments and lattice parameters predicted monotonically increased with the U employed in PBE+U calculations (suggesting that it would be difficult to determine a meaningful U value from DFT calculations alone) but also that the U that most closely matched experiments for one property significantly differed from those needed to match experimental predictions of other properties.\cite{Staros_CrI3} Thus, although these predictions provide general insights into CrI$_3$'s magnetism, they lack the accuracy---and precision---needed to make definitive statements. 

To bring clarity to this picture, we thus performed DMC simulations on monolayer CrI$_3$, one of the first performed on a magnetic monolayer material.\cite{Staros_CrI3} Given the known sensitivity of the electronic structure of 2D materials to strain\cite{Webster_PhysRevB,Webster_PhysChemChemPhys} and the lack of experimental results for the monolayer geometry, however, we first set out to determine a high-accuracy, \textit{many-body} monolayer geometry (see Fig. \ref{fig:crx3-hessian}). To do so, we leveraged a cutting-edge, many-body geometry optimization method: the surrogate Hessian line-search method.\cite{doi:10.1063/5.0079046,Shin_PRM_2021,vse2-wines} In this approach, a Hessian generated using DFT is used to determine conjugate directions along which to locate the minimum-energy structure. Differing from standard DFT--based conjugate gradient minimization, the surrogate Hessian line-search method then computes DMC energies in parallel along each of the conjugate directions and subsequently refines the conjugate directions until the search converges to the DMC energy minimum.\cite{doi:10.1063/5.0079046,Shin_PRM_2021,vse2-wines} This method is particularly advantageous because it makes the most efficient use of both DFT and DMC data: Instead of relying on more-expensive DMC energy gradients, it uses relatively cheap DFT gradients and then employs DMC energy calculations to refine the search, thus providing a substantially more efficient way of determining fully many-body geometries. Applying this approach to CrI$_3$, we obtained a ground-state geometry with a lattice parameter of $a_{0}=6.87$ \AA\space and a Cr-I bond distance of $d_{Cr-I}=2.73$ \AA\space.\cite{Staros_CrI3} Interestingly, while we were conducting this research, some of the first scanning tunneling microscopy--based measurements of the monolayer geometry were performed, yielding a lattice constant of $a_{0}=6.84$ \AA.\cite{li_single-layer_2020} Such strong agreement ($\le$0.5\% error) highlights the accuracy of our structure and the value of the surrogate Hessian line-search approach, despite CrI$_3$'s (and many monolayer materials') shallow potential energy surface (PES) around its minimum. This $\le$0.5\% error is quite remarkable considering that the DFT lattice constants had a 6$\%$ variation in answer (depending on the choice of DFT method).

With this high-accuracy structure in hand, we then performed DMC calculations on the monolayer. To generate a high-quality trial wavefunction from which to construct our DMC nodal surface, we first performed PBE+U calculations. Unlike previous works, we were sure to determine our U in a variational fashion by determining the U in our PBE+U trial wavefunctions that minimized our resultant DMC energies (see Fig. \ref{fig:crx3-hessian}). This resulted in an optimized U value of 2 eV.\cite{Staros_CrI3} Employing this U in our Slater-Jastrow trial wavefunctions with one- and two-body Jastrow terms, we obtained high-accuracy DMC charge and spin densities. We finally obtained DMC estimates of the Cr magnetic moment by integrating the spin density from the center of the Cr atom to the distance from the nucleus at which the spin density changed sign (the zero-recrossing radius of the sign of the spin density). This yielded a DMC magnetic moment of 3.62 $\mu_B$ for Cr atom, a moment substantially larger than that previously predicted using single-reference techniques. Importantly, PBE+U $=$ 2 eV calculations performed on the same optimized geometries yielded a similarly large Cr moment of 3.57 $\mu_B$, demonstrating the strong coupling between the moment and the geometry,\cite{Webster_PhysChemChemPhys,Webster_PhysRevB} as also observed in previous DFT simulations  and our subsequent spin-phonon and spin-lattice coupling calculations.\cite{Staros_CrI3} Although these moments seem unexpectedly large, they are counterbalanced by negative moments of $-$0.145 $\mu_B$ on the iodine atoms, leading to the anticipated \textit{averaged} moment of 3 $\mu_B$ over the entire unit cell. These simulations have therefore shed a clarifying light on the atomic-scale origin of CrI$_3$'s magnetism and produced a valuable workflow that can be readily extended to a wide range of other 2D materials.\cite{staros2023firstprinciples} Additional details of this work can be found in Ref. \onlinecite{Staros_CrI3}.

%\begin{figure*}
%\begin{center}
%\includegraphics[width=0.8\textwidth]{crx3-toc.eps}
%\caption{(a) The workflow used to extract the accurate magnetic properties of a 2D magnetic system using a combined DFT+U and QMC approach and b) top and side view of the structure of 2D CrX$_3$ (X = I, Br, Cl, F). Adapted from Ref. \onlinecite{wines-crx3}.}
%\label{fig:crx3}
%\end{center}
%\end{figure*}

\subsubsection{Monolayer CrX$_3$}\label{cri3}

%{\bf{\textit{CrX$_3$: QMC and DFT Workflow for $T_c$}}}

To expand upon this work, we created a workflow that combines DFT+U and DMC methods to calculate magnetic properties of 2D CrX$_3$ (X = I, Br, Cl, F) with improved accuracy.\cite{wines-crx3} These structures were chosen as a case study because they have been extensively investigated with DFT, \cite{olsen-data} have been experimentally synthesized, and have a finite transition temperature.\cite{cri3,olsen-data} We can map our ab initio quantities to a 2D model spin Hamiltonian to obtain properties such as $T_c$.\cite{Torelli_2018,Lado_2017} The model Hamiltonian contains easy-axis single-ion anisotropy ($D$), Heisenberg isotropic exchange ($J$), and anisotropic exchange ($\lambda$),  
\begin{equation}
{\cal {H}}= - \left( \sum_{i} D(S_i^z)^2 +\frac{J}{2}\sum_{i,i'}\vec{S}_i\cdot \vec{S}_{i'} + \frac{\lambda}{2}\sum_{i,i'}S_i^z S_{i'}^z \right).
\label{hamiltonian}
\end{equation}
The sum over $i$ runs over the lattice of Cr atoms (magnetic atoms), and $i'$ runs over $i$'s nearest-neighbor Cr atoms; only nearest neighbors are needed because the magnetic moments are strongly localized on the Cr atoms.  $D>0$ prefers off-plane easy axis, $J>0$ prefers FM interactions, and $\lambda=0$ results in fully isotropic exchange. Note that in the following sections, $J$ refers to the Heisenberg exchange and not the J in the DFT+U+J formalism.\cite{PhysRevResearch.5.013160} Two magnetic orientations can be constructed from the CrX$_3$ unit cell: the FM (two spin-up Cr atoms) and antiferromagnetic (AFM) (one spin-up Cr atom and one spin-down Cr atom) orientations. $J$, $\lambda$, and $D$ can be extracted from noncollinear (spin-orbit) DFT calculations. Specifically, the energy differences between configurations that have an easy-axis rotation of 90$^{\circ}$ and nonrotated configurations for FM and AFM magnetic orientations can be computed.

These results can be systematically improved with DMC. First, the nodal surface of the trial wavefunction can be optimized by changing the U value (variationally determining which U value yields the lowest energy). Second, by performing DMC for the FM and AFM configurations, $J$ can be estimated. In our case, these DMC energies are collinear (without SOC). Although SOC has been implemented in DMC, the energy differences needed to compute magnetic anisotropy parameters are smaller in magnitude than the DMC uncertainty. For this reason, we used the optimal U value determined from the nodal surface optimization (in our case, it was $U = 2$ eV for CrI$_3$ and CrBr$_3$, similar to Ref. \onlinecite{Staros_CrI3}) to perform DFT+U calculations to obtain $D$ and $\lambda$. We performed this DMC variational optimization of U for both the FM and AFM configurations of 2D CrI$_3$ and CrBr$_3$, where both magnetic configurations yielded an optimal U value of 2 eV. The starting geometry for our DMC calculations was obtained from the vdW-DF-optB88 \cite{Klimes_2010} functional (vdW-DF-optB88 produces a geometry for CrI$_3$ in near-perfect agreement with experiment \cite{li_single-layer_2020} and Ref. \onlinecite{Staros_CrI3}). After extracting our parameters from our combined DMC/DFT+U approach, we can plug those parameters into analytical models such as the one developed by Torelli and Olsen \cite{Torelli_2018} to estimate the Curie temperature. We determined a maximum $T_c$ of 43.56 K for CrI$_3$ and 20.78 K for CrBr$_3$, which is in excellent agreement with the measured values of 45 K \cite{cri3} and 27 K, \cite{doi:10.1073/pnas.1902100116,crbr3-exp} respectively. More details of this work can be found in Ref. \onlinecite{wines-crx3}.

\subsubsection{Monolayer MnO$_2$}\label{mno2}

\begin{figure}
\begin{center}
\includegraphics[width=0.45\textwidth]{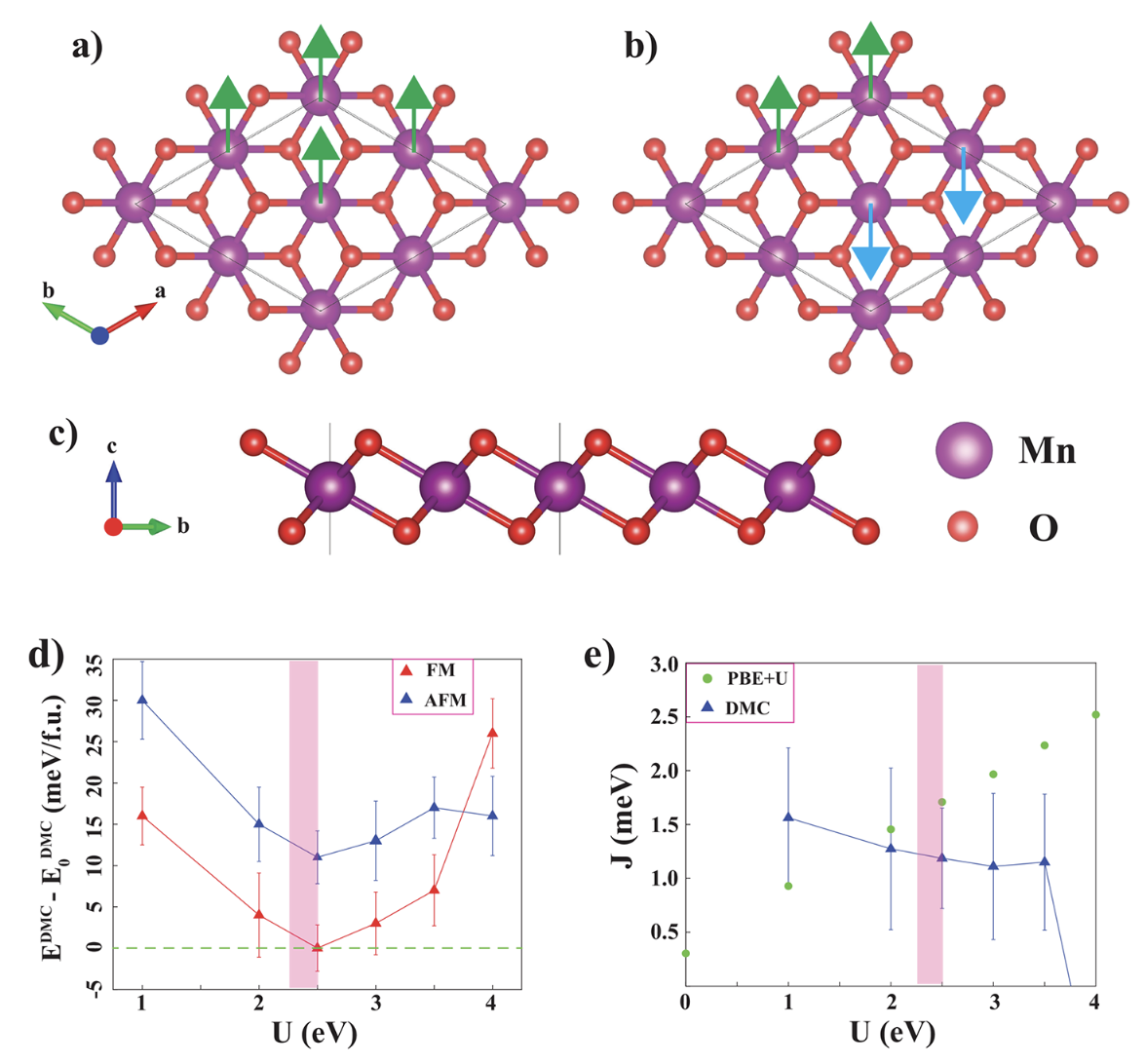}
\caption{(a, b) Top and (c) side depictions of monolayer MnO$_2$, where FM ordering is shown in (a) and AFM ordering is shown in (b) by green arrows. (d) Shows the total relative DMC energy as a function of U parameter for the FM and AFM orientations, and (e) shows the nearest-neighbor Heisenberg exchange ($J$) as a function of U calculated with DMC (blue) and PBE+U (green). The magenta rectangle represents the fitted optimal U value of 2.4(1) eV. Reproduced with permission from Wines et al., J. Phys. Chem. C, 126 (13), 5813-5821 (2022). Copyright 2022 American Chemical Society. }
\label{fig:mno2}
\end{center}
\end{figure}

The workflow from Section \ref{cri3} is general and can be extended beyond CrX$_3$ materials. In this work, we applied the same workflow to monolayer MnO$_2$.\cite{mno2-qmc} Single-layer MnO$_2$ is a commercially available transition metal oxide semiconductor that has been synthesized and studied with DFT.\cite{mno2-exp,PhysRevB.93.045132,Rong_2019,irradiation,C1CP20634F,https://doi.org/10.1002/adma.201602281,ataca-mx2,fm-mno2} Kan et al. \cite{fm-mno2} predicted the FM ordering to be more favorable than the AFM ordering with PBE+U (using a U correction of 3.9 eV obtained from previous literature \cite{PhysRevB.73.195107}). $J$ was extracted from these calculations, and a magnetic coupling Hamiltonian based on the Ising model was used to perform classical Monte Carlo simulations to obtain a Curie temperature of 140 K.\cite{fm-mno2} Although these PBE+U results are promising, there are aspects that can be revisited with more accurate methods such as DMC.

First, previous results and our own DFT+U benchmarking results show that the energy difference between the FM and AFM orientations is heavily dependent on the functional and U value.\cite{fm-mno2, mno2-qmc} In fact, our DFT benchmarking shows an FM-AFM energy difference range of $-$55 meV to 18 meV, indicating that there is a discrepancy regardless of whether the FM or AFM orientation is more energetically favorable. Using the same workflow from Section \ref{cri3}, we determined an optimal U value of 2.5 eV and a $J$ value of 1.2(5) meV with DMC (see Fig. \ref{fig:mno2}). The starting geometry for our DMC calculations was obtained from the strongly constrained appropriately normed (SCAN) \cite{PhysRevLett.115.036402} meta-GGA functional (because the SCAN-computed lattice constant is in near-perfect agreement with experiment \cite{mno2-exp}). Using this optimal U value in subsequent PBE+U calculations, we find monolayer MnO$_2$ to have out-of-plane magnetic anisotropy. From these magnetic constants, we estimated the maximum value of $T_c$ to be 28.8 K. In addition, we analyzed the spin density obtained from DMC for Mn and O and compared these with those obtained from PBE+U. Specifically, we found that DMC predicts that the spin density of O atoms is polarized antiparallel with respect to the Mn atoms, which is in agreement with our PBE+U results for $U = 2.5$ eV and $U = 3.5$ eV (the O spin density is polarized parallel with respect to Mn for $U = 0$ eV and $U = 1$ eV). More details of this work can be found in Ref. \onlinecite{mno2-qmc}.

\subsubsection{Monolayer 2H- and 1T-VSe$_2$}\label{vse2}
\begin{figure}
\begin{center}
\includegraphics[width=0.4\textwidth]{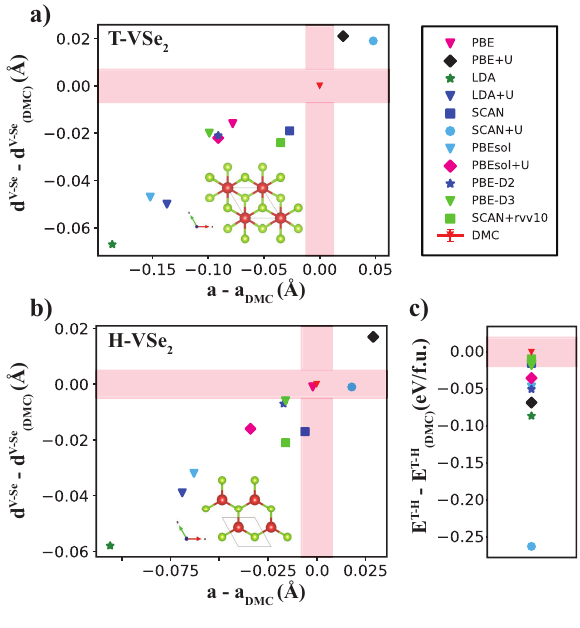}
\caption{The deviation of structural parameters (lattice constant ($a$) and V-Se distance ($d^{\textrm{V}-\textrm{Se}}$)) compared to the DMC results for (a) T-VSe$_2$ and (b) H-VSe$_2$. (c) The deviation of the T $-$ H energy from several different DFT functionals ($U = 2$ eV) relative to the DMC calculated T $-$ H energy (E$^{\textrm{T} - \textrm{H}}$). The atomic structures are depicted in the insets. Reproduced with permission from Wines et al., J. Phys. Chem. Lett. 14 (14), 3553-3560 (2023). Copyright 2023 American Chemical Society.}
\label{fig:vse2}
\end{center}
\end{figure}

Monolayer VSe$_2$ has been a source of controversy throughout the theoretical and experimental literature, with substantial claims of near--room temperature ferromagnetism (from 291 K to 470 K \cite{vse2,https://doi.org/10.1002/adma.201903779,vse2-exp,vse2-bkt}).The T (octahedral phase (1T)-centered honeycombs) phase and the H (the trigonal prismatic phase (2H)-hexagonal honeycombs) phase (shown in Fig. \ref{fig:vse2}) of 2D VSe$_2$ have a close lattice match and similar total energies, which makes it difficult to discern which phase is being observed experimentally.\cite{struc-phase,C9CP03726H,https://doi.org/10.1002/adma.201903779,vse2} The difficulty in discerning the relative stability of the phases and related structural uncertainty undoubtedly imply that the structural parameters are coupled to the magnetic properties.\cite{vse2,https://doi.org/10.1002/adma.201903779,vse2-exp,vse2-bkt,vse2-moment-exp} Similarly to Refs. \onlinecite{Staros_CrI3} and \onlinecite{Shin_PRM_2021}, we used a combination of DMC, DFT, and the surrogate Hessian line-search structural optimization technique \cite{doi:10.1063/5.0079046} to clarify the discrepancies in structural parameters and relative phase stability of the T and H phases of monolayer VSe$_2$.

Fig. \ref{fig:vse2} depicts a summary of DFT results (multiple functionals with and without U) alongside DMC (red bars) for the lattice constant ($a$), V-Se bond distance ($d^{\textrm{V}-\textrm{Se}}$), and the relative phase energy (E$^{\textrm{T} -\textrm{H}}$). As expected, there is a large disagreement among DFT functionals for $a$, $d^{\textrm{V}-\textrm{Se}}$, and E$^{\textrm{T} - \textrm{H}}$, indicating the need for a theory such as DMC. With DMC and the Hessian line-search method, we computed the lattice constant and V-Se distance to be 3.414(12) \AA\space and 2.505(7) \AA, respectively, for T-VSe$_2$ and 3.335(8) \AA\space and 2.503(5) \AA, respectively, for H-VSe$_2$. The relative energy between the T and H phases was found to be 0.06(2) eV, with the H phase being lower in energy than the T phase in freestanding form. We then constructed a phase diagram between the T and H phases with DMC accuracy and found that an H-to-T phase transition can be induced by applying small amounts of strain ($\approx$3$\%$). More details of this study can be found in Ref. \onlinecite{vse2-wines}. As a follow-up to this work, the authors intend to focus on studying the magnetic properties of 2D T-VSe$_2$ with DMC methods, specifically running DMC for various magnetic configurations and predicting the transition temperature.

\subsection{Electronic Properties}\label{electronic}

%{\bf Electronic structure and geometries}

{\bf Electronic structure:} It is now established that 2D systems such as graphene, phosphorene, and TMDs belong to a class of materials that is exceptionally interesting for both fundamental research and future technological use. The electronic structure of 2D systems is astonishingly rich because it provides an unusual combination of periodicity and free boundaries, the possibility of stacking with new variational freedom (twistronics), significant effects related to the presence of substrates, and further potential modifications with doping, straining (straintronics), and related processing. Due to the presence of unique electronic phases, these systems pose a number of challenges for any theoretical or computational method.

%The simulatenous presence of several 
%i) possibilities for remarkable coincidence of variety of bonding mechanisms that includes valence, ionic, metallic but also non-covalent (van der Waals/dispersive) interactions

%ii) free boundary(ies) open possibilities for conformal and geometry relaxation as well as reconstruction notrivial reconstructions as well as several metastable conformations; additional effects from substrates provide another interesting angle that needs to be taken into a consideration

%iii) the usual charge screening in solids is significantly diminished, in fact it has a non-monotonous behaviour
%with maximum at short distances while rapidly fading beyond (\cite{Louie})

%iv) stacking of multiple layers opens new possibilities for structural variations 
%and indeed, geometries are quite challenging (As somebody said, crystals are from God, while surfaces are from Evil (or something of that sort). :-)) 

From the perspective of electronic structure theory,  the 2D materials in this review are not strictly systems with electrons in a 2D plane but rather 2D slabs with atomic thickness. Consequently, the electronic structure, one-particle orbitals, and other properties  indeed depend on all three spatial coordinates. (Previous theoretical work has shown that calculations of slabs have their intricacies, as can be seen for the homogeneous electron gas in a slab geometry; see Ref. \onlinecite{PhysRevB.76.035403}.) This is important because many properties that apply to strictly 2D geometries might not apply or might be significantly modified. Clearly, electrons have a nonzero probability to be on either side of the material in the orthogonal $z$-direction, and they are not confined within a 2D $(x,y)$ plane. This leads to the possibility of Janus materials,\cite{janus-ML} in which the opposing surfaces have different properties due to different functionalizations, doping, and symmetry-breaking. An important consequence of this may be seen in their electronic structure, as most clearly revealed by the dependence of their dielectric constant on the interelectronic distance. In  3D solids, the dielectric constant grows monotonically until it saturates at large distances to the bulk value. In 2D materials, it initially grows; however, at the distance of a few atomic bonds, it exhibits a rather sharp maximum and then rapidly falls off to a small value. This small value  corresponds to long-range-distance vdW-like behavior in which distant regions appear electrically neutral with typical dipole or higher-order dispersive fluctuations. This implies that at short distances,  the system's bonds look locally like those in a 3D material, while at large distances, the system exhibits dispersive effects resembling effective interactions with noncovalent character.

\begin{figure}
\begin{centering}
\includegraphics[clip,width=1.0\columnwidth]{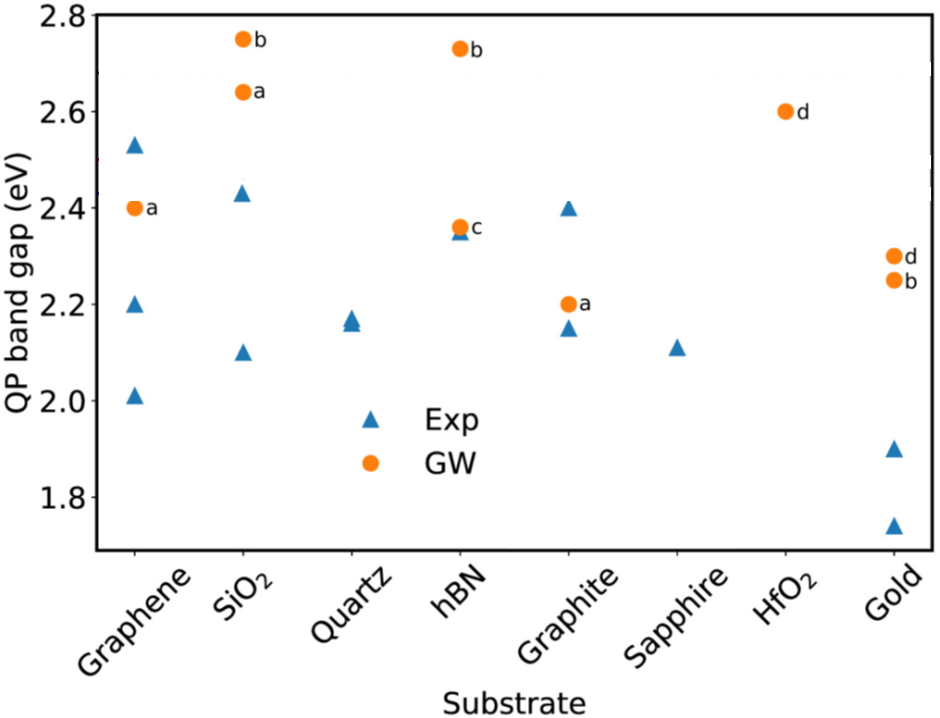}
\par\end{centering}
\caption{
Quasiparticle band gap of an MoS$_{2}$ monolayer on different substrates reported in the literature. The experimental results
shown by blue triangles were obtained with scanning tunneling microscopy spectroscopy, absorbance, and angle-resolved (inverse) photoemission spectroscopy. The GW band gaps are shown by the orange disks. Reproduced with permission from Zibouche et al., Phys. Rev. B 103, 125401 (2021). Copyright 2021 American Physical Society.
\label{fig:exp-th_compar}
}
\end{figure}

\begin{figure}
\begin{centering}
\includegraphics[clip,width=0.75\columnwidth]{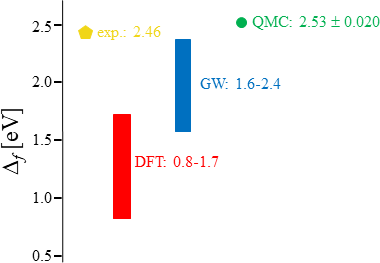}
\par\end{centering}
\caption{
Comparison of the quasiparticle band gaps of a freestanding phosphorene monolayer as calculated by the DFT, GW, and QMC (DMC) methods against the experimental value measured on a freestanding sample.\cite{acs.nanolett.9b03928} Values taken from Ref. \onlinecite{PhysRevX.9.011018,PhysRevResearch.5.033223}.
\label{fig:DFT-GW-QMC-exp_comp}
}
\end{figure}

The DFT and GW methods have been widely used to study the electronic structure of 2D systems. The accuracy of the former is determined by the XC functional, whereas that of the latter is determined by the unperturbed state and the way the perturbation expansion is terminated. This introduces significant biases between the two approaches as well as within each method. Quasiparticle band gaps typically differ between these methods by up to $\approx$1 eV. As a result, experiments are expected to provide the final answer. However, 2D materials typically must be supported by substrates (quartz, sapphire, etc.), which compromises their 2D character by providing unintentional dielectric embedding. As shown in Fig.~\ref{fig:exp-th_compar}, the experimental effect of the tuning by the substrate is also of the order of $\approx$1 eV, complicating the comparison with calculations. In such a situation, DMC may provide the final answers and even challenge the achievable experimental accuracy. The situation is illustrated in Fig.~\ref{fig:DFT-GW-QMC-exp_comp}, where the quasiparticle gap of monolayer phosphorene calculated by the DFT (PBE and B3LYP), GW (G$_0$W$_0$ and GW$_0$ using both PBE and B3LYP wavefunctions), and DMC methods is compared to the rare result where the experimental band gap is measured on a freestanding sample.\cite{acs.nanolett.9b03928} The DMC result agrees with the experimental result almost with chemical accuracy, given that no vibronic effects were considered in the DMC calculation, while both the DFT and GW methods tend to underestimate the band gap and can feature a very wide spread in their calculated values.

%{\bf Cohesions.}

{\bf Excited states:} One of the key characteristics of a 2D material important for many applications is the value of the band gap and its sensitivity to variations caused by doping, substrates, strain, multilayer stacking, and both internal (e.g., spin-orbit) and external influences. In DMC calculations of band gaps, the key complications come from  two main sources of error---namely, FS and fixed-node biases. The fixed-node error is determined by the quality of the antisymmetric part of the trial function, which is often just single reference because robust methodologies for multireference wavefunctions for periodic systems are under development. In QMC, the band gaps can be calculated essentially by two approaches as outlined in the following.

\textit{Fundamental band gap:} One way is to use the definition of the fundamental gap, which implies charging the system with $N$ electrons by an additional electron $(N+1)$ and then by an additional hole $(N-1)$ and calculating the response $\Delta_{f}=E(N+1)+E(N-1)-2E(N)$ as the difference of total energies for charged and neutral states. This has been routinely applied in 3D and often also to a number of 2D systems.\cite{10.1063/5.0030952,10.1063/5.0023223,Shin_PRM_2021,PhysRevB.101.205115} Some cases showed a minor (0.1 to 0.2) eV upward bias that can be qualitatively understood by the fact that the cation typically relaxes more than the anion because the anionic state with its conduction state occupation should typically involve more than a single configuration. The reason is that the conduction state excitation could mix other available (i.e., unoccupied) conduction states that are not too far energy wise and that can form linear combinations that produce the same symmetry as in the excitation. It is possible to have either increases or decreases in energy from mixing of excitations and each case requires careful analysis. It is even more pertinent when energy differences (band gaps) are involved, since each of the corresponding states can be impacted differently. In practice, the conduction states are often less accurate and less well converged than the ground state calculation.
This explains the resulting minor bias, which is surprisingly small considering just a single-reference approximation of the nodal hypersurfaces for both charged and neutral systems. 

\begin{figure}
\begin{centering}
\includegraphics[clip,width=1.0\columnwidth]{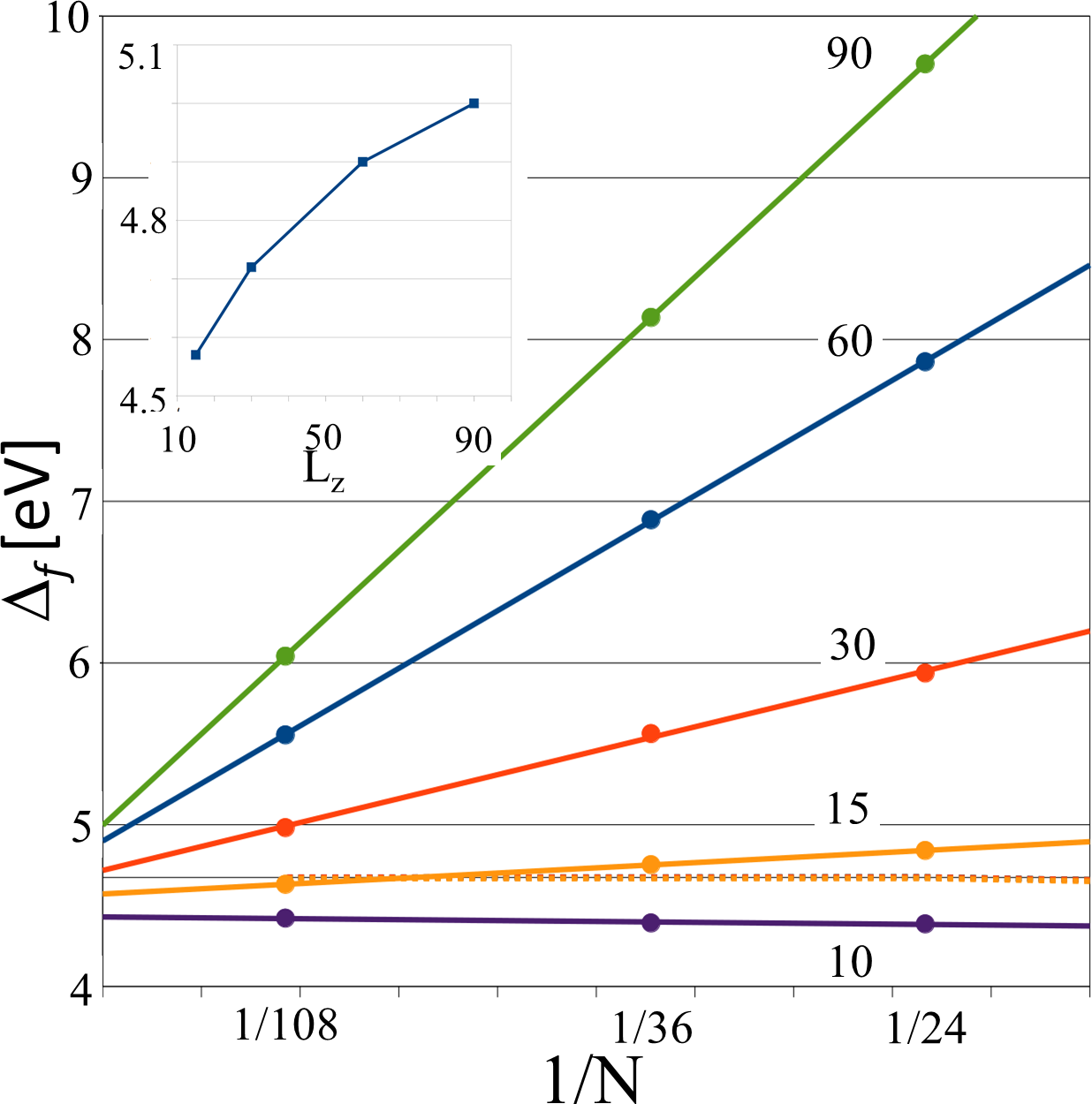}
\par\end{centering}
\caption{
Behavior of the band gap in a 2D supercell of hBN, computed from charged systems using uniform neutralizing background. In-plane FS scaling with box height L$_{z}$ (simulation cell vector length orthogonal to 2D material)  ranging from 10 \AA~to 90 \AA~as parameter. The inset shows the L$_{z}$ convergence for in-plane converged gaps. Example from a DFT-PBE study. The broken horizontal orange line shows the HOMO (highest occupied molecular orbital)--LUMO (lowest unoccupied molecular orbital) gap, which is converged at L$_{z}$= 15 \AA. $N$ represents the number of primitive unit cells.  
\label{fig:Gap_charged_scaled}
}
\end{figure}

 Additional bias could come from FS scaling because the charged states must be compensated to make the periodic supercell neutral. The commonly used compensation by the opposite sign constant background charge eliminates the leading (monopole) diverging term. However, it does not perfectly compensate the subleading terms, and although the impact mostly appears to be small,\cite{PhysRevB.103.205206} this need not be valid.  There are also subtleties in using charge compensation and Ewald sums when treating 2D systems and increasing the size of the box in the orthogonal direction. An example for band gap in hBN\cite{charged_gap} is shown in Fig.~\ref{fig:Gap_charged_scaled}, where first the in-plane FS scaling with the number of primitive unit cells, $N$, is shown for various lengths of the simulation cell vector orthogonal to 2D material (L$_{z}$). The inset shows the L$_{z}$ convergence of the in-plane scaled band gaps. As the neutralizing background dilutes with increasing L$_{z}$, the band gap appears to diverge with L$_{z}$ due to artificial charge decompensation. In fact, proper compensation for slab geometries in real systems is still a matter of current studies.\cite{charged_gap} In addition to our own benchmarking calculations displayed in Fig.~\ref{fig:Gap_charged_scaled}, Hunt et al. \cite{PhysRevB.101.205115} previously studied the electronic and excitonic properties of 2D hBN with DMC, carefully taking FS errors and vibrational renormalization into account (discussed in more detail in Section \ref{hbn}). 

 \textit{Promotion band gap:} The second method is closer to  experiments that provide the optical spectrum---namely, promotion of an electron from valence into the conduction band. This excitation can be  direct or indirect (i.e., approximately modeling a phonon-assisted optical excitation). Because the optical spectrum typically involves excitonic states, it is good to distinguish the two possibilities that can occur:

 {\em Wannier excitons:} In systems with strong covalent bonds and large dielectric constants, one typically encounters Wannier excitons with exciton binding energies from meVs to a few tens of meV, so the effect on the electronic bands is very small. This is what is observed in many such calculations\cite{PhysRevB.103.205206} because any relaxation into the excitonic state provides only very minor effects, which are at the level of statistical error bars in DMC.

{\em Frenkel excitons:} For systems with a possible presence of excitons with binding energies larger than 0.5 eV (Frenkel excitons), the situation is more complicated. 

 (i) In covalent or mainly covalent materials, the one-particle orbitals from band structure calculations, such as DFT, HF, and DFT+U,  provide Bloch orbitals that enter the trial wavefunction. This determines the symmetry and periodicity of the resulting trial wavefunction. In most cases, this constraint ``locks'' the overall band structure picture also in QMC calculations. Typical QMC effects come as band gap shifts; however, the overall structure of the bands remains nearly the same.  Therefore, due to the nodal constraint, the relaxation to a more-localized, more strongly bonded non-Bloch excitonic state is strongly hindered.  This is  easy to understand because the localized exciton wavefunction in the Bloch state basis necessarily requires large multireference expansion.  Therefore, any relaxation to an excitonic state is mostly absent, \cite{PhysRevX.9.011018,10.1063/5.0030952} which shows that the overlap of the exciton and Bloch state promotion is negligible or small. In the noninteracting limit, the overlap identically vanishes, thus showing the dominant one-particle nature of this behavior.
 
 (ii) However, in some cases, the nodal constraint might be less  restricting. One such possibility could occur if the excitonic state and Bloch excitation share a subgroup of symmetries. Possible mixing of such two states would have to be carried out explicitly. Another possibility could occur in environments with rapidly changing charge densities (e.g., ionic bonds or 2D layered materials with large space between the layers) because absence of proper tails in the trial function can hinder the needed charge redistribution. Therefore, the  charge relaxation and/or localization could be compromised with a resulting energy decrease toward the excitonic state. This could be observed  in some systems, \cite{Shin_PRM_2021} although the signal is comparable to the error bar range due to complications with thermodynamic limit extrapolation. This requires a new leap in the quality of the trial function to  quantify the  involved effects. There is no  fundamental obstacle in carrying out such analysis; the issues are mainly technical due to limited availability of appropriate software tools. Clearly, one would need to rebuild the trial function with localized orbitals and possibly orthogonalize it to the Bloch excited state that corresponds to the fundamental gap excitation to distinguish the fundamental vs. excitonic state explicitly. This is routinely done in molecular studies because many quantum chemical codes enable corresponding calculations.
 %what is the true value of the Frenkel exciton since that requires building the correct many-reference wave function since it is likely that there is a non-dynamical contribution to the exciton that is not captured by single-reference trial function.
 Another such case could occur for vdW molecular crystals, where the localized state(s) on a single molecule could dominate the low-lying excitation spectrum because the periodicity causes only minor energetic shifts and results in very flat low-lying bands. More dispersive bands appear in significantly higher scattering states in the conduction part of the spectrum. 

Several QMC works have computed the fundamental and promotion band gap to estimate the exciton binding energy of 2D systems.\cite{10.1063/5.0030952,10.1063/5.0023223,Shin_PRM_2021,PhysRevB.101.205115} Additionally, there have been studies by Szyniszewski et al. \cite{PhysRevB.96.075431} and Mostaani et al. \cite{PhysRevB.95.081301} that have utilized DMC coupled with other theoretical approaches to successfully estimate biexciton and trion energies in 2D semiconductors.

%QMC calculations have been applied %to basic parameters such as %cohesive energies \cite{XXXX}, %binding curves of stacked layers 
%\cite{XXXX} 

\subsubsection{Monolayer Phosphorene and MoS$_2$}\label{phos}

Alongside graphene, phosphorene and TMDs belong to the most studied 2D semiconductors. Their band gap is often direct. In phosphorene, the band gap is direct at $\Gamma$ in a monolayer and few-layers and direct even in bulk black phosphorus, albeit at the Z point. In MoS$_2$, it is direct at the K point only in the monolayer. In addition, 2D semiconductors usually possess ultrahigh carrier mobility and field-effect switching ratios, which make them ideal materials for field-effect and digital logic transistors. 2D semiconductors with an appropriate and tunable direct band gap have achieved extremely efficient photon absorption, emission, and photoelectric conversion and have been widely used in the field of optoelectronic devices. To meet more demands, various techniques have been exploited to modulate their properties, including doping, alloying, forming vdW heterostructures, and strain engineering or straintronics.\cite{iop.31.313201,inf2.12177} Due to their atomic thickness, 2D materials are highly sensitive to external perturbations, such as strain. Their resilience to mechanical deformations allows application of strains well in excess of 10\%. Applying strain enables the lattice and electronic structure to be modulated as well as their various properties, such as the carrier mobility. MoS$_2$ has traditionally been considered the quintessential straintronic material for which many straintronic experiments have been performed.\cite{nn405938z}

\begin{figure}
\begin{centering}
\includegraphics[clip,width=0.8\columnwidth]{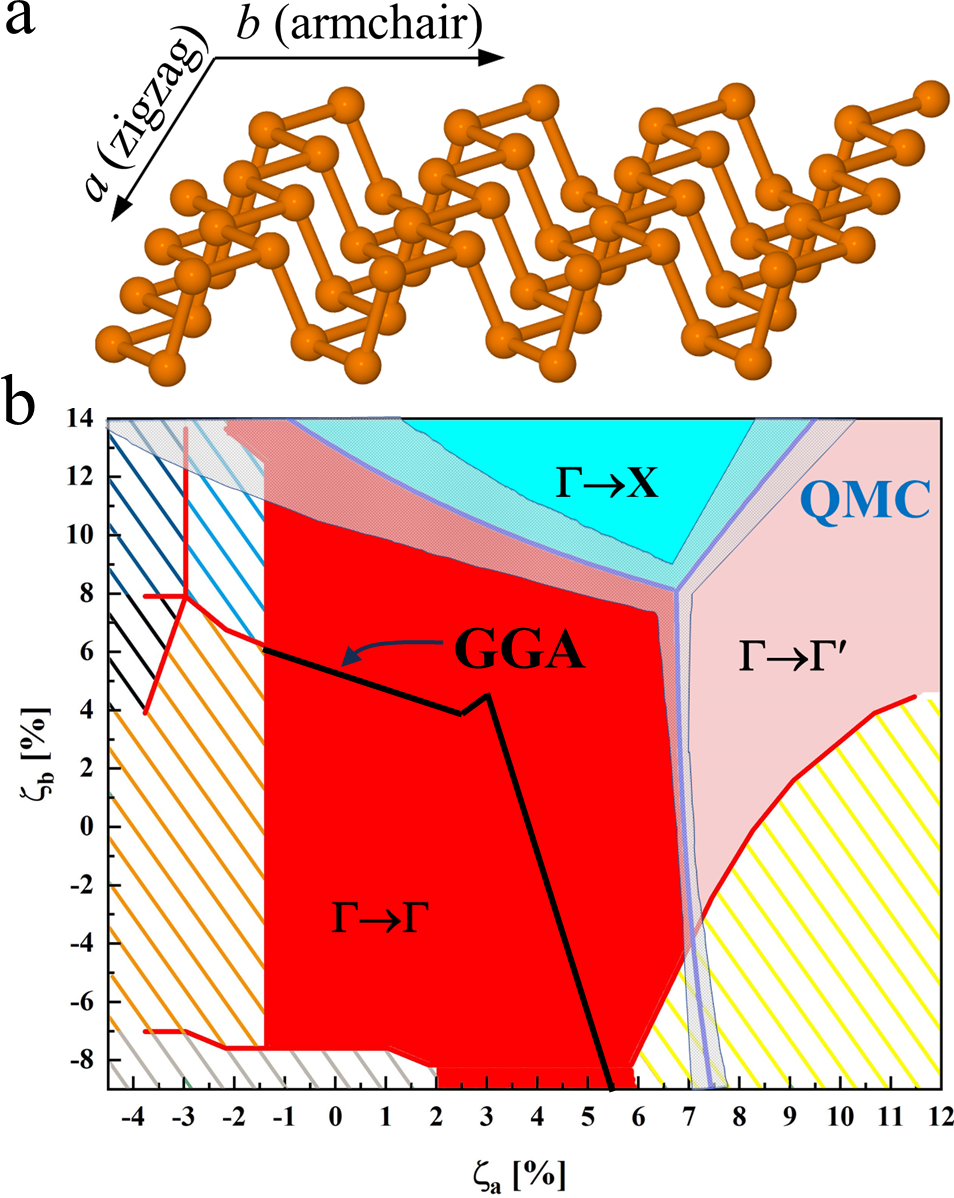}
\par\end{centering}
\caption{
(a) Model of the atomic structure of monolayer phosphorene with the lattice parameters \textit{a} and \textit{b} indicated. (b) Phase diagram of the various excitations in the $\zeta_{a}$/$\zeta_{b}$ plane. Blue lines correspond to boundaries determined by fixed-node DMC with the overlays outlining the $\pm$1$\sigma$ (standard deviation) error bar. Hatched regions correspond to DFT-PBE results with boundaries outlined by red lines except for the $\Gamma$ $\to$ $\Gamma$ region where the DFT-PBE boundary is shown in solid black line for contrast reasons and labeled GGA. Reproduced from Huang et al., Phys. Rev. Research 5, 033223 (2023), licensed under a Creative Commons Attribution (CC BY) license. 
\label{fig:phospho_FD}
}
\end{figure}

Recently, the straintronic response of two 2D semiconductors, monolayer phosphorene and MoS$_2$, has been studied using fixed-node DMC methods.\cite{PhysRevResearch.5.033223,PhysRevResearch.6.013007} In phosphorene, the strain was applied in both armchair and zigzag directions by adjusting the \textit{a} and \textit{b} lattice parameters (Fig.~\ref{fig:phospho_FD}), considering deformations of up to $\approx$10\%. These deformations determine the in-plane $\xi_{a}$, $\xi_{b}$ strains. Determination of strained properties was treated as a full optimization problem in the space of four structural variables: lattice parameters \textit{a}, \textit{b} and two internal parameters \textit{x}, \textit{y}.\cite{PhysRevResearch.5.033223} Around the minima, the ground-state PES $E_{0}$ was fitted by 4D paraboloid functions used to find the lowest point on the \textit{x}, \textit{y} subspace, leaving to further minimize only bivariate functions $E_{0} = f (a, b)$. The excited state $E_{1}$ was computed only at the minimum for the internal parameters \textit{x}, \textit{y}, and the quasiparticle band gap for any applied strain was computed as $\Delta_{f} = E_{1} - E_{0}$, subject to FS scaling. In MoS$_{2}$, only diagonal strain was considered, and a scaled-down version to two parameters (one lattice parameter \textit{a} and one internal parameter \textit{x}) was used, leaving a single variable function $E_{0} = f (a)$ to optimize.\cite{PhysRevResearch.6.013007} This approach allows researchers to study the 2D materials both in and out of equilibrium.

For phosphorene in equilibrium, we obtain (\textit{a} = 3.30 $\pm$ 0.003 \AA), (\textit{b} = 4.61 $\pm$ 0.006 \AA), and ($\Delta_{f}$ = 2.53 $\pm$ 0.020 eV); see also Fig.~\ref{fig:DFT-GW-QMC-exp_comp}. This value of the quasiparticle gap is in excellent agreement with the experimental value for freestanding monolayer phosphorene of 2.46 eV;\cite{acs.nanolett.9b03928} see also Fig.~\ref{fig:DFT-GW-QMC-exp_comp} (keeping in mind that neglecting adiabatic, vibronic, and zero-point vibrational energy tends to increase the gap value compared to experiments). The DMC optimized structure exhibits noticeable differences with respect to the structure derived from the 3D black phosphorus crystal especially in the \textit{b} parameter (4.376\AA). This variation of the \textit{b} parameter in 3D crystal is partially due to chemical interlayer interaction that reduces it by $\approx$5\% and is almost completely absent from MoS$_2$, where the interlayer interaction in the 3D crystal is fully vdW.\cite{PhysRevResearch.6.013007} These trends are fairly well described by the DFT-PBE functional and not as well by the hybrid functionals.\cite{PhysRevResearch.5.033223} All gaps in the DFT treatment are appreciably smaller than the DMC value. As expected, the smallest value by about 2 eV is obtained by the DFT-PBE functional. The hybrid functionals yield larger values but fail in predicting the equilibrium geometries.

The calculated band gap phase diagram for phosphorene is shown in Fig.~\ref{fig:phospho_FD}. 
%The diagram was constructed by calculating the boundaries between the $\Gamma$$\to$$\Gamma$ and a band gap at $\Gamma$ formed by interchange of LUMO (Least unoccupied molecular orbital) and LUMO + 1 states ($\Gamma\to\Gamma^{\prime}$) and the boundary between the $\Gamma$$\to$$\Gamma$ and $\Gamma$$\to$X due to applied strain from the intersections of the respective PESs.
%rephrased
The diagram was constructed from intersections of PESs corresponding to conditions under which the system retains the zero-strain $\Gamma\to\Gamma$ direct gap, and where it forms a direct gap at the $\Gamma$ point but with a level crossing between the lowest unoccupied molecular orbital (LUMO) and LUMO+1 states, labeled $\Gamma\to\Gamma^{\prime}$, and a third situation with a $\Gamma\to$X indirect gap. The three different PESs outline three different ``phase'' boundaries. No attempt was made to determine the other phase boundaries in Fig.~\ref{fig:phospho_FD}; hence, they correspond to DFT-PBE boundaries.
%rephrased
As for the first boundary, while both gaps are direct at $\Gamma$, the nature of the excited state is different if the order of unoccupied states is interchanged. The new LUMO state has a differing curvature; hence, the transport properties in the conduction band are expected to be significantly modified, which goes in line with straintronic as a tool for electron effective mass modification. The DMC boundaries outline a strain tuning area for the direct $\Gamma$$\to$$\Gamma$ band gap more than twice as large as that determined by DFT-PBE (and similarly by hybrid functionals\cite{PhysRevResearch.5.033223}). This area extends mostly into the region of tensile strain, which is important to prevent the material from wrinkling at higher compressive loads. The DFT results are qualitatively similar (Fig.~\ref{fig:phospho_FD}) regardless of the DFT functional. All PESs (e.g., QMC, DFT) are rather parallel, albeit strongly shifted vertically in energy, and their intersections all feature similar curvatures. The primary differences consist of relative band offsets (magnitude of band gaps) and the location of the minima. These two factors play havoc with the form of the phase diagrams. In phosphorene (Fig.~\ref{fig:phospho_FD}), the offsets reduced the $\Gamma\to\Gamma^{\prime}$ section of the diagram by $\approx$50\% in phosphorene.\cite{PhysRevResearch.5.033223} In MoS$_2$, where the strain tuning areas are more compressed, the different treatments were predicting qualitatively different tunings in the K$\to$K section: DMC mainly in the compressive region, GW and B3LYP exclusively in compressive, and HSE mainly into tensile.\cite{PhysRevResearch.6.013007} Phosphorene was also found to feature high values of the gauge factor of the order of 100 meV per \%, which is comparable to MoS$_2$. Combined with the huge tuning area of several percent leaving the direct nature of the band gap intact, this provides a colossal straintronic tunability of phosphorene with the tuning limit likely determined only by the mechanical breakdown of the material. The range of band-gap tuning by applied strain while maintaining the direct band gap at $\Gamma$ is truly huge with achievable values of the band gap in the range 2.1 eV to 3.8 eV. Surprisingly, this is at complete variance with the straintronic response of the quintessential straintronic material, monolayer MoS$_2$, which features similar a gauge factor but an order-of-magnitude smaller tuning area.\cite{PhysRevResearch.6.013007}

\subsubsection{Fluorographene}\label{flur}

Fluorographene (FG) is a 2D  stoichiometric graphene derivative (C$_1$F$_1$) material that exhibits a large band gap due to the complete out-of-plane bonding of carbon orbitals with fluorine (C-F sp$^3$ bonds). Some of the key properties of interest involve its large band gap, dielectric characteristics, and potential surface physics and chemistry applications. Interestingly, the fundamental and optical band gaps as well as related derived properties have not been fully settled until recently. In particular, from experiments, the onset of FG optical absorption spectrum has been estimated to lie between 3 eV and 5 eV. The fundamental gap from GW for the electron self-energy suggests a range of 7 eV to 8 eV. In addition, their excitonic effects appear to be very strong, and the BSE confirmed the exciton binding energy of FG close to 2 eV, regardless of technicalities such as orbital sets or related parameters. In our study, \cite{10.1063/5.0030952} %\onlinecite{10.1063/5.0030952} 
we have employed both the highly accurate GW-BSE approach and DMC to ascertain these excitation characteristics.   We found that the fundamental band gap from thoroughly converged GW and consistently extrapolated DMC methods agree within the error bars on value of 7.1(1) eV, establishing the reference for this material from two independent many-body approaches. Careful analysis of the BSE results confirmed $\approx$1.9 eV exciton binding and provided additional insights into the structure of the excitonic state based on the projection of the BSE exciton on the Bloch states.

\subsubsection{Monolayer hBN}\label{hbn}

\begin{figure}
\begin{center}
\includegraphics[width=7.5cm]{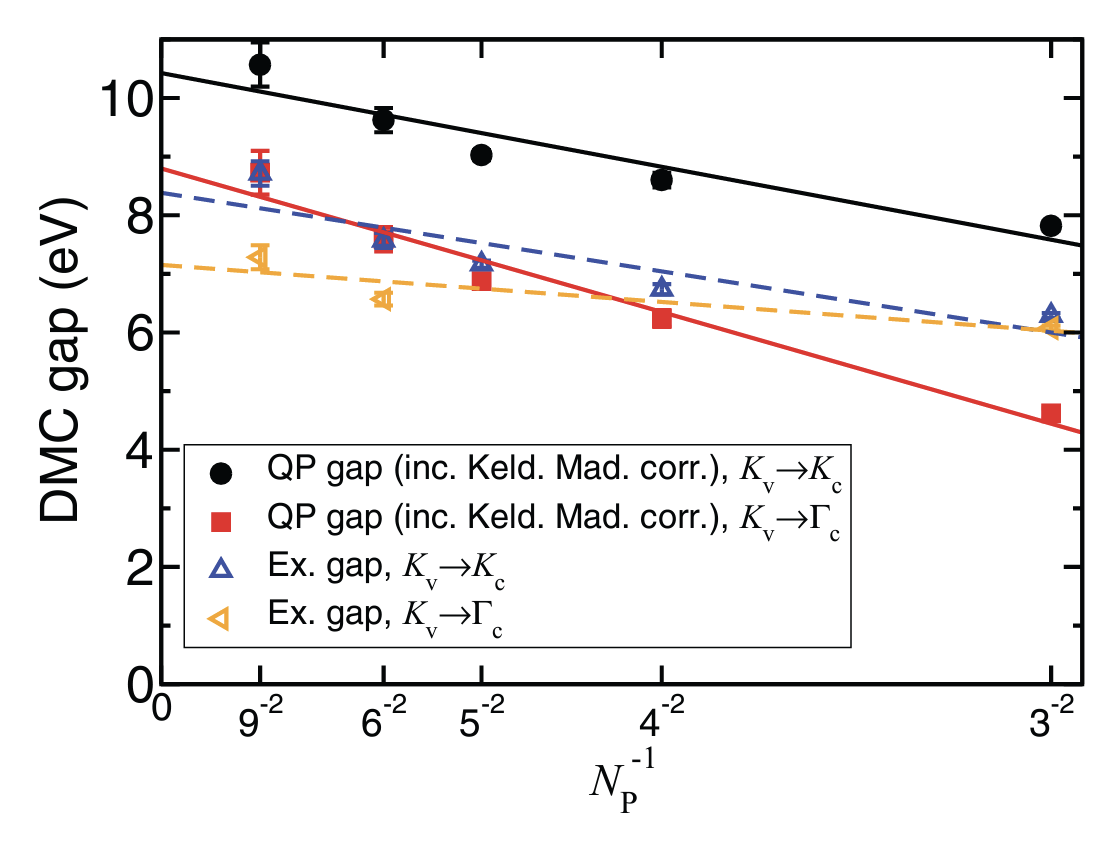}
\caption{The DMC quasiparticle (QP) and excitonic (Ex.) gaps of monolayer hBN as a function of the inverse number of primitive cells in the supercell ($N_p$). QP gaps  include the Madelung correction. Reproduced with permission from Hunt et al., Phys. Rev. B 101, 205115 (2020). Copyright 2020 American Physical Society.}
\label{hbnfig}
\end{center}
\end{figure}

Monolayer hBN is a promising insulating 2D material with potential applications in electronic devices such as vertical tunneling diodes \cite{doi:10.1126/science.1218461,hbn-tunnel-2} and supercapacitors.\cite{bn-supercap} In addition, hBN is an ideal substrate material for graphene-based electronics due to its similar lattice constant.\cite{hbn-hetero,hbn-hetero-2,hbn-hetero-3,hbn-hetero-4} Recently, Hunt et al. \cite{PhysRevB.101.205115} studied monolayer hBN through the lens of many-body techniques (DMC and GW). The DMC quasiparticle gap of monolayer hBN was found to be 8.8(3) eV and indirect (from $K_v$ - $\Gamma_c$). This value was significantly higher than those predicted by G$_0$W$_0$ and GW$_0$ (ranging from 7.43 eV to 7.72 eV). Fig. \ref{hbnfig} depicts the FS convergence of the DMC quasiparticle and excitonic gaps of monolayer hBN. It was demonstrated that the DMC quasiparticle gap falls off as a function of $N_p^{-1}$ but can be corrected by subtracting a screened Madelung constant \cite{madelung1918} from the gap. From DFT calculations, a large $-$0.73 eV vibrational correction was found for monolayer hBN. With regard to excitonic effects, an exciton binding energy of 1.9(4) eV was found for the indirect exciton ($K_v$ - $\Gamma_c$), and 1.8(4) eV was found for the direct ($K_v$ - $K_c$), where the results for the direct exciton are in excellent agreement with the GW-BSE results performed in this study. A possible major source of error for this system (overestimation of the gap) could be unaccounted fixed-node errors from using a single-determinant wavefunction. For example, it is entirely possible that the fixed-node error does not equally cancel when computing energy differences for the ground and excited states. It is also possible that the vibrational renormalization of the gap is underestimated. Further information can be found in Ref. \onlinecite{PhysRevB.101.205115}.

\subsubsection{Monolayer GeSe}\label{gese}

%\textcolor{red}{Paul, do you want to add a short paragraph and a figure?}

%\textcolor{blue}{-Ref. \onlinecite{Shin_PRM_2021}: Optimized structure and electronic band gap of monolayer GeSe from quantum Monte Carlo methods, Hyeondeok Shin, Jaron T. Krogel, Kevin Gasperich, Paul R. C. Kent, Anouar Benali, and Olle Heinonen (2021): https://journals.aps.org/prmaterials/abstract/Shin_PRM_2021}
\begin{figure}
\begin{center}
\includegraphics[width=8.5cm]{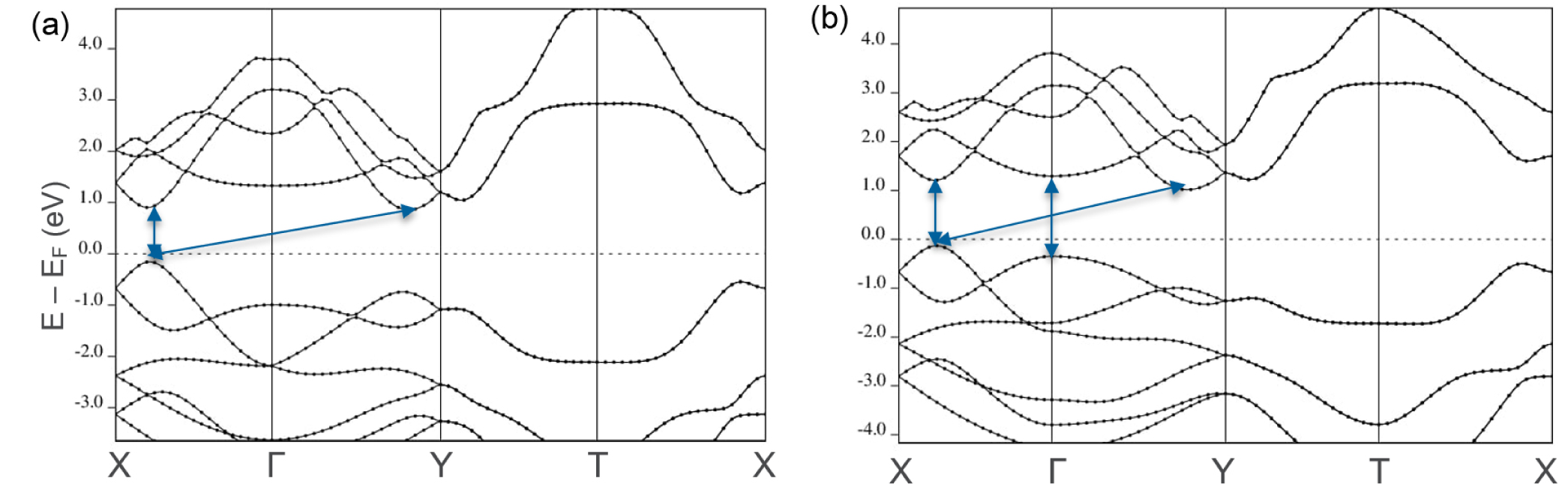}
\caption{Monolayer GeSe band structure for (a) PBE and (b) DMC geometry calculated by PBE functional. Blue lines represent candidates for direct and indirect gap. Reproduced with permission from Shin et al., Phys. Rev. Materials 5, 024002 (2021). Copyright 2021 American Physical Society. }
\label{GeSe_fig}
\end{center}
\end{figure}

GeSe is a p-type semiconductor that has been widely studied because of its unique optical properties. Its bulk structure is generally known to possess a measured indirect gap of 1.07 eV to 1.29 eV at room temperature.\cite{PhysRevB.16.1616,gese-exp,Mishra_2015} However, it was recently reported that a direct gap minimum of 1.3 eV was measured via optical spectroscopy.\cite{gese-exp2} This controversy is mainly due to the small difference between the direct and indirect gaps in bulk GeSe, and its high sensitivity in the method used for the measurement. In addition, DFT studies on GeSe showed strongly varied band gap energy and optimized lattice parameters depending on choice of XC functionals.\cite{Wang_2019,C7TA02109G,10.1063/1.4891230,10.1063/1.4931459} These limitations on computing accurate optical and structure properties from DFT lead us to confirm that the use of a more accurate method, which can describe exact optical and structural properties at the same time, is highly desired on the GeSe structure.
      
We estimated the DMC direct gap of bulk GeSe to be 1.62(16) eV, which is in excellent agreement with the experimental value of 1.53 eV.\cite{gese-qp} Based on this DMC direct gap and our DMC quasiparticle gap result of 1.95(21) eV, bulk GeSe is expected to possess a weak exciton binding energy of $\approx$0.3 eV. Because the exact geometry for the GeSe monolayer is experimentally not known yet, its geometry is fully optimized using a surrogate Hessian-based parallel line-search method \cite{doi:10.1063/5.0079046} (similar to Ref. \onlinecite{Staros_CrI3, vse2-wines}). In the optimization process, the GeSe monolayer shows a shallow PES minimum over a large range of lattice parameters, which explains the high sensitivity of optimized DFT geometries to the choice of XC functional.

In the PBE band structure for the GeSe monolayer, the DMC monolayer geometry shows a small direct gap of 1.50 eV at the $\Gamma$ point from the PBE result, which is competitive to the direct (1.24 eV) and indirect gaps (1.02 eV) located between the X and Y high symmetry point (see Fig.~\ref{GeSe_fig}). Large differences in the direct gap at the $\Gamma$-point between the PBE (2.32 eV) and DMC geometries (1.50 eV) show that the monolayer band structure is very sensitive to its geometry, suggesting that controlling strain on the monolayer can be a route to manipulate the electronic and optical properties of monolayer GeSe. Additional details can be found in Ref. \onlinecite{Shin_PRM_2021}.

\subsubsection{Monolayer GaSe and GaS$_{x}$Se$_{1-x}$ Alloys}\label{gase}

PTMCs are a class of 2D materials that have suitable band gaps for photovoltaics and transistors\cite{NI201310,PhysRevB.84.085314,PhysRevB.87.195403,Pozo_Zamudio_2015,gase-atomic-layers,C3CP50233C,doi:10.1002/adma.201201361,bulk-gase,C8CP04723E,JAPPOR2018251} and have lower exciton binding energies than TMDs,\cite{NI201310,PhysRevB.84.085314,PhysRevB.87.195403} which make them suitable for water-splitting applications.\cite{water-splitting-mx} Monolayer GaSe is a PTMC that has been reliably synthesized, and measurements of the quasiparticle gap, optical gap, and lattice constant have been performed.\cite{PhysRevB.96.035407,doi:10.1063/1.4973918,GaSe-optical,C8NR01065J,GaSe-nature-synthesis,Rahaman_2018,Rahaman_2017,doi:10.1002/adma.201601184,doi:10.1063/1.5094663} The experimental lattice constant of monolayer GaSe has a measured value of $a=b=3.74$ \AA.\cite{GaSe-nature-synthesis} In addition, 2D GaSe has an experimental indirect band gap of 3.5 eV (on a graphene substrate),\cite{PhysRevB.96.035407} which is much larger than the bulk GaSe band gap of 2.0 eV. The optical band gap has been measured to be 3.3 eV,\cite{GaSe-optical} which implies an exciton binding energy lower than 0.2 eV. Despite these well-characterized measurements, computational results can significantly vary based on which functional is used.\cite{10.1063/5.0023223}

\begin{figure}
\begin{center}
\includegraphics[width=8.5cm]{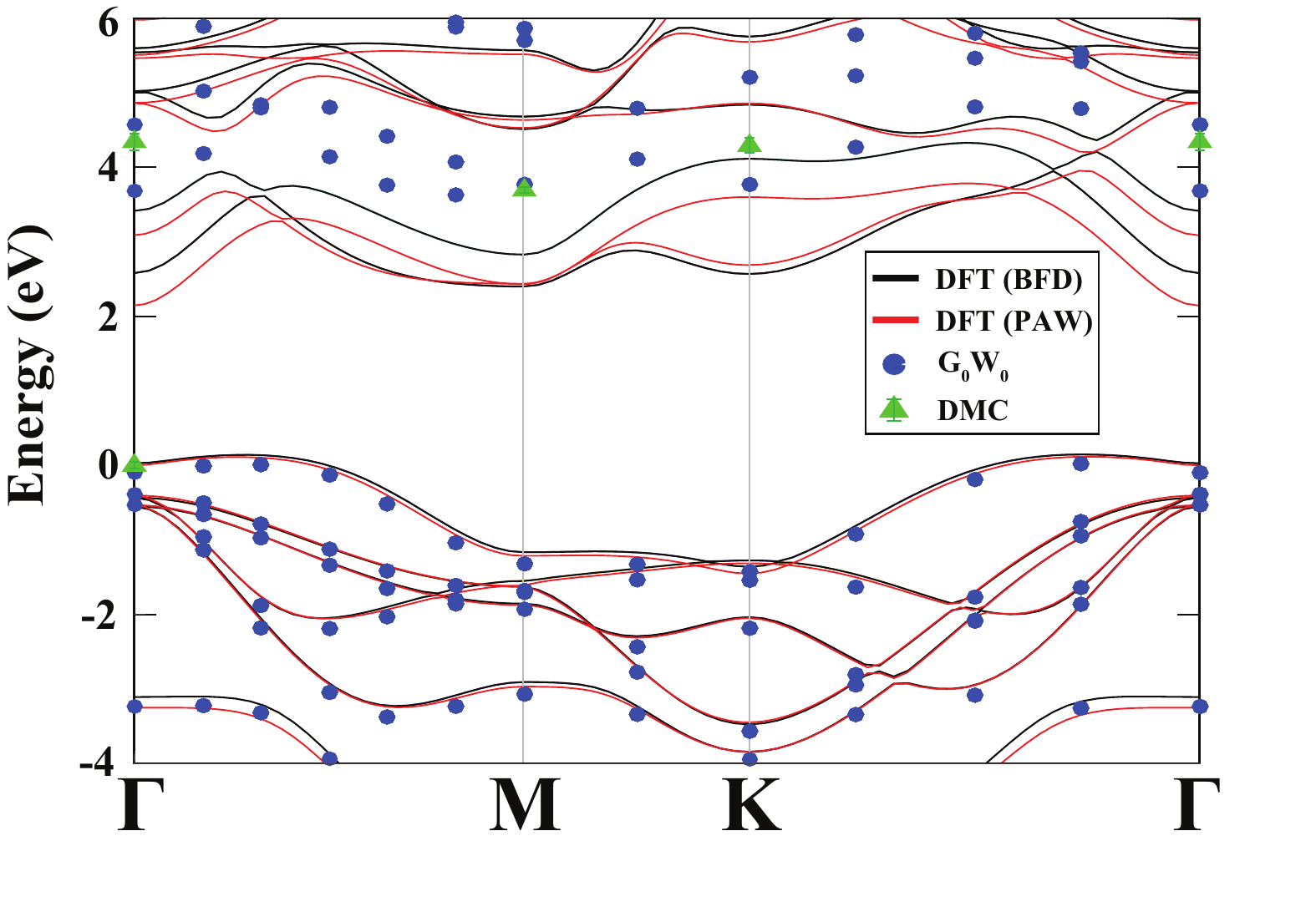}
\caption{Monolayer GaSe electronic band structure of computed with PBE using Burkatzki-Filippi-Dolg (BFD) \cite{doi:10.1063/1.2741534,doi:10.1063/1.2987872} pseudopotentials (black), projector augmented wave (PAW) \cite{PhysRevB.50.17953,PhysRevB.59.1758} pseudopotentials (red), and G$_0$W$_0$ using PBE wavefunctions and PAW potentials (blue). The DMC excitation energies and error bars (with respect to the $\Gamma$ point) are given in green at each high symmetry point. Reproduced from Wines et al., J. Chem. Phys. 153, 154704 (2020), with the permission of AIP Publishing.}
\label{band}
\end{center}
\end{figure}

The most apparent discrepancy is the location of the conduction band edge at each high symmetry $k$-point. The energy differences are so small between each high symmetry point ($\approx$0.2 eV to 0.3 eV) that different functionals/pseudopotentials can predict the gap value to have different values, and some methods even incorrectly predict the gap to be direct. Improvements to the underestimation of the gap can be achieved using methods such as GW or BSE, but these results depend on which functional is used to generate the starting wavefunction, and the indirect/direct discrepancy still exists. We used DMC to obtain the optimal lattice constant by isotropically scaling the lattice and finding the energy minima. To demonstrate the weak dependence that DMC has on the starting functional, we created the DMCtrial wavefunction with PBE, LDA, and SCAN functionals and performed the same calculation. We computed 3.74(2) \AA\space for DMC-PBE, 3.75(1) \AA\space for DMC-LDA, and 3.75(1) \AA\space for DMC-SCAN, which is in close agreement with experiment. With DMC, we confirmed that 2D GaSe is an indirect material ($\Gamma$-M) with a quasiparticle gap of 3.69(5) eV, which is in excellent agreement with the experimental value. A full comparison of the electronic properties of 2D GaSe computed with DFT, G$_0$W$_0$ and DMC is depicted in Fig. \ref{band}. From our calculation of the optical band gap, we obtained a maximum bound on the exciton binding energy to be 80 meV, which confirms that GaSe is a 2D material with a lower exciton binding energy than most TMDs,\cite{2dexciton} making it suitable for water-splitting applications. More details of this work can be found
in Ref. \onlinecite{10.1063/5.0023223}.     

We extended our approach to 2D GaS$_{x}$Se$_{1-x}$ alloys \cite{10.1063/5.0070423} since alloying is a promising technique to control the properties of single- and few-layer structures,\cite{ingaalloy,KANLI201913,TiS,C7CP06750J,Li-ads,photoresp,ersan-alloy,ataca-alloy} which includes PTMCs.\cite{Hosseini_Almadvari_2020,wines-nitride,PhysRevB.102.075414,doi:10.1002/adma.201601184,bowing} Studying alloys with accurate computational methods is a difficult endeavor because usually the local atomic ordering and stoichiometry of the different alloyed structures are unknown. To tackle this issue, ab initio methods have been coupled with methods such as cluster expansion to construct the energy hull diagram of an alloy system, which allows us to determine whether a structure is thermodynamically stable and possibly can exist in nature.\cite{atat}

\begin{figure}
\begin{center}
\includegraphics[width=7cm]{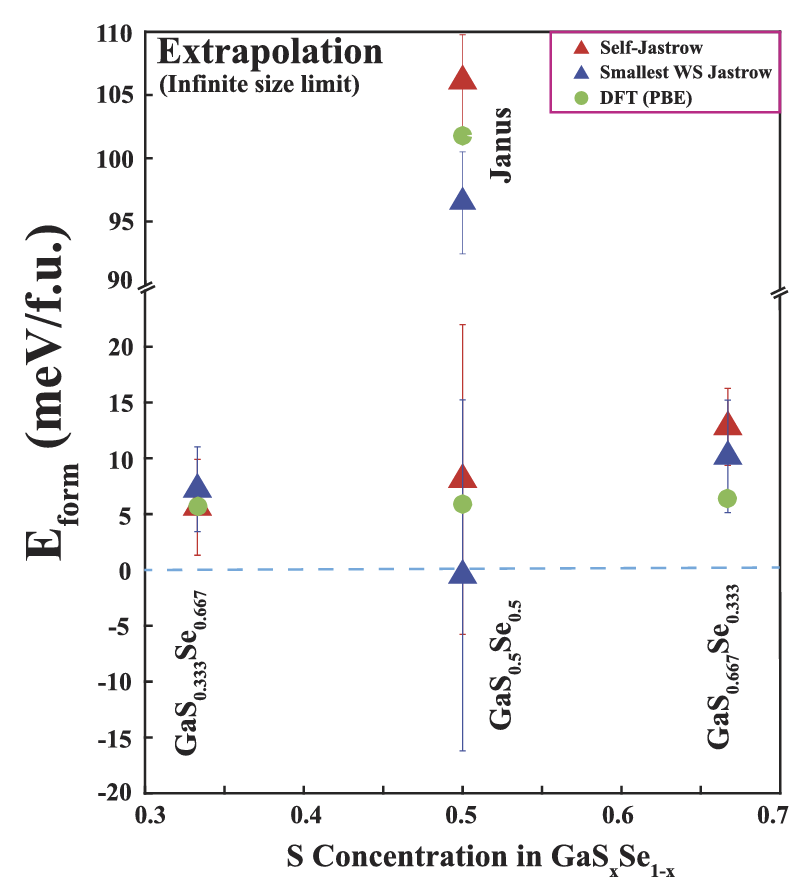}
\caption{The constructed energy hull diagram for monolayer GaS$_x$Se$_{1-x}$ obtained with PBE (green) and DMC (red using the structures' own Jastrow functions and green reusing the Jastrow functions from the alloy with the smallest Wigner-Seitz radius). Reproduced from Wines et al., J. Chem. Phys. 155, 194112 (2021), with the permission of AIP Publishing. }
\label{gasse}
\end{center}
\end{figure}

In this work, we designed a high-throughput workflow to compute the energy hull diagram of an alloy system with QMC methods, using 2D GaS$_{x}$Se$_{1-x}$ as a case study. To achieve this, we proposed a method we call \textit{Jastrow sharing}, which involves recycling the optimized Jastrow parameters between various alloys with different stoichiometries. Specifically, we optimized the Jastrow parameters of the alloy with the smallest Wigner-Seitz radius and used those parameters for other structures of interest in the alloy system. We demonstrated that this eliminates the need for unnecessary VMC Jastrow optimization simulations and can reduce the total computational time by 1/4. After testing the validity of this Jastrow sharing approach, we went on to compute the alloy formation energies with DMC (extrapolated to the thermodynamic limit) for selected points on the energy hull diagram (see Fig. \ref{gasse}). This method can easily be implemented for other 2D alloy systems where the Jastrow sensitivities of the pseudopotentials of the atoms in the system are low, which can eventually aid in accurate studies of more complex alloy systems (i.e., alloying 2D transition metal oxide materials). Additional information from this study can be found in Ref. \onlinecite{10.1063/5.0070423}.

\subsubsection{Spin-Orbit Effects, Topological States}\label{soc}

\begin{figure}
\begin{center}
\includegraphics[width=7cm]{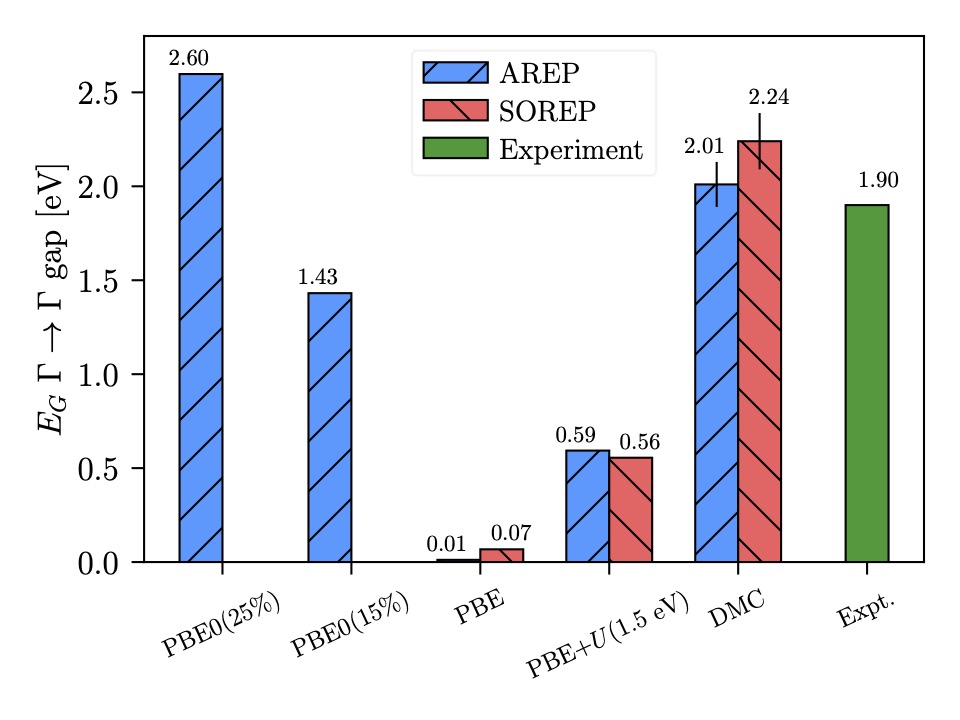}
\caption{The band gaps of RuCl$_3$ system compared with experiment. Calculations were done by several methods such as hybrid PBE0($w$), \cite{10.1063/1.472933,10.1063/1.478522} where $w$ denotes percentage of exact exchange in two ways: averaged spin-orbit (AREP) and explicit spin-orbit (SOREP). Reproduced with permission from Annaberdiyev et al., Phys. Rev. B 106, 075127 (2022). Copyright 2022 American Physical Society.  }
\label{gap_rucl3}
\end{center}
\end{figure}

\textbf{Layered RuCl$_3$ for Kitaev's spin liquid:} Recently, the QMC method has been extended to spin-orbit Hamiltonians with trial functions built upon two-component spinors and the spin-space sampled from an overcomplete set of spin configurations.\cite{PhysRevA.93.042502,10.1063/1.4954726} For the first time, we have applied this methodology to layered RuCl$_3$,\cite{PhysRevB.106.075127} where claims of sizeable
contributions of the Ru atomic spin-orbit to the gap opening have been made.\cite{PhysRevB.90.041112,rucl3-so}  Our calculations did not find support for this scenario. We found that the quasiparticle band gap opens due to expected significant Hubbard repulsion on the Ru atom. That proved to be the dominant effect, while spin-orbit provides a mild shift of approximately 0.2 eV, as also expected for the intermediate strength of the atomic spin-orbit for the Ru atom (see Fig. \ref{gap_rucl3}).

A somewhat larger impact of the spin-orbit was observed in the cohesive energy; however, this is caused almost exclusively by the larger shift in the high symmetry isolated Ru atom, not in the solid where the hybridization partially quenches and averages the corresponding splittings.  We note that RuCl$_3$ is considered to be a promising candidate for realization of Kitaev spin liquid with low-lying collective states. However, these are in the meV range (i.e., much smaller than the difference between FM and AFM states), hence outside the statistical resolution of current valence electronic structure QMC. Even if such wavefunction could be constructed, the presence of Kitaev physics would be hidden in the statistical noise, and very extensive calculations would be needed to discern such a signal. Additional information from this work can be found in Ref. \onlinecite{PhysRevB.106.075127}.

\textbf{Layered TbMn$_6$Sn$_6$ for Chern magnets:} A distinct class of materials, RMn$_6$Sn$_6$, where R denotes a rare-earth element, displays rich and intricate physical phenomena such as strong electron-electron correlations, spin-orbit effects, and possibilities of forming states with topological order. This stems from structural peculiarities with 2D Kagome layers of Mn interlaced by R and Sn layers.  In this respect, the rare-earth Tb atom is of particular interest since it is the only R element  that forms an out-of-plane spin order compound. Experiments indicated that TbMn$_6$Sn$_6$ is close to realizing a quantum-limit Chern magnet, as predicted by the Haldane model. Indeed, Kagome lattice geometry with an out-of-plane magnetization formed by the Mn atoms and the presence of strong SOC originating from the Tb and Sn atoms provide the necessary conditions for opening the Chern gap. In our very recent study (Ref. \onlinecite{Tb-soc}), we use DMC and DFT with Hubbard U (DFT+U) calculations to examine the electronic structure of TbMn$_6$Sn$_6$. We find that DFT+U and single-reference QMC calculations exhibit the same overestimation of the magnetic moments as meta-GGA and hybrid density functional approximations. Our findings point to the need for improved orbital quality and for potentially extending beyond single-determinant wavefunctions for this class of materials. We have concluded that significant multireference effects must be included to capture the static correlations necessary for an accurate prediction of magnetic properties. We have probed for the occurrence of topological order, where we have explored DFT+U with Mn magnetic moments adjusted to experiment. We have observed the Dirac crossing in bulk to be close to the Fermi level, within $\approx$120 meV, in agreement with the experiments. The possibility of crossing the Fermi level has been further enhanced in nonstoichiometric slab calculations, keeping  realization of Chern magnetism in this limit within the experimental reach. More information can be found in Ref. \onlinecite{Tb-soc}.

\textbf{Enhancing MnBi$_2$Te$_4$ stability:} MnBi$_2$Te$_4$ has an ordered layer of Mn atoms that lead to the observation of the quantum anomalous Hall effect when all the magnetic moments of its layers are aligned ferromagnetically.\cite{deng2020quantum} However, it is shown that departure from the crystalline long-range order can lead to contradictory observations with regard to the existence of a Dirac cone or a gapless topological surface state.\cite{Estyunin2020, Lee2019, Vidal2019, zeugner2019chemical, Zhang2019, Li2019a, Otrokov2019b,Chen2019, Hao2019, Li2019, Nevola2020, Swatek2020, Yan2021, Garnica2022a,ahn2023stacking} Several studies show that Mn atoms migrating to Bi sites and creating Mn$_{\rm Bi}$ defects can create a ferrimagnetic configuration and strongly modify the topological ground state.\cite{Garnica2022a, Yan2019a, Liang2020, Yuan2020} In our recent study (Ref. \onlinecite{saritas2024}), we evaluate the potential for increasing the Mn-Bi antisite formation through chemical doping. Using a comprehensive screening set using DFT and DMC calculations, we find that group III elements, such as Sc, Y, and La dopants on Bi sites, have the potential to increase the formation energy of Mn$_{\rm Bi}$ defects. None the dopants we studied on Mn sites has increased the Mn$_{\rm Bi}$ defect formation energy. We show that the low antisite formation energy of Mn$_{\rm Bi}$ defects in pristine MnBi$_2$Te$_4$ originates from internal strain between the Bi-Te and Mn-Te layers. Group III dopants on Bi sites decrease the coupling between these two layers by increasing the bond lengths between the Bi atoms and the Te atoms that are shared between the Bi and Te atoms. More information can be found in Ref. \onlinecite{saritas2024}.

%\subsection{van der Waals Interactions}\label{binding}
\subsection{Interlayer Interactions}\label{binding}

Layering 2D materials offers additional degrees of freedom, thereby enabling us to explore the correlation between interlayer coupling and physical properties. This exploration can lead to the emergence of novel electronic states in 2D materials. Interlayer coupling in these layered 2D materials varies widely, ranging from purely dispersive interactions, such as those seen in bilayer graphene,\cite{PhysRevLett.115.115501,carbon-layers} to complex interactions involving weak interlayer chemical bonding and even metallic bonding, in addition to vdW interactions.
%, as observed in 2D transition metal dichalcogenide (TMD) materials~\cite{}. 
This complexity in interlayer interaction has been verified through x-ray density investigations for a 2D TMD material.\cite{kasai2018x} It is important to note that vdW interactions among these interactions are a many-body phenomenon stemming from nonlocal electronic correlations induced by the instantaneous fluctuation of electron density. These interactions may not be adequately captured by standard DFT frameworks, highlighting the need for a more sophisticated theoretical framework that can address these interactions equally. Such a framework is essential for capturing the diverse interlayer coupling, which could vary depending on different stacking modes or twisting degrees of freedom. To illustrate this point, we present examples of DMC calculations aimed at investigating the nature of interlayer binding in 2D layered systems, providing a comprehensive understanding of interlayer coupling beyond what DFT offers.
%Layering 2D materials is one of the key features to promote additional degrees of freedom which enables us to study a correlation between interlayer coupling and physical properties, leading to emergent electronic states in 2D materials.
%Interlayer coupling in the layered 2D materials can be diverse from purely dispersive interactions like in bilayer graphene to complicated interaction composed of weak interlayer chemical bonding and even metallic interactions in addition to the vdW interactions in 2D TMD materials, as verified through the X-ray density investigation~\cite{}. 
%It is pointed out that vdW interaction among these interactions is many-body phenomena originating from non-local electronic correlations induced by instantaneous fluctuation of the electron density, which might not be captured well by standard DFT frameworks. This calls for a sophisticated theoretical framework to deal with these interactions on the equal footing, which is important to capture the interlayer coupling that could be varied depending on stacking mode or even twisting degrees of freedom.
%The followings are examples of DMC calculations for investigating the nature of the interlayer binding properties of the 2D layered system which provides a comprehensive understanding of the interlayer coupling revealed beyond the DFT pictures.

\subsubsection{Bilayer Phosphorene}\label{bi-phos}

\begin{figure*}
\begin{centering}
\includegraphics[clip,width=1.7\columnwidth]{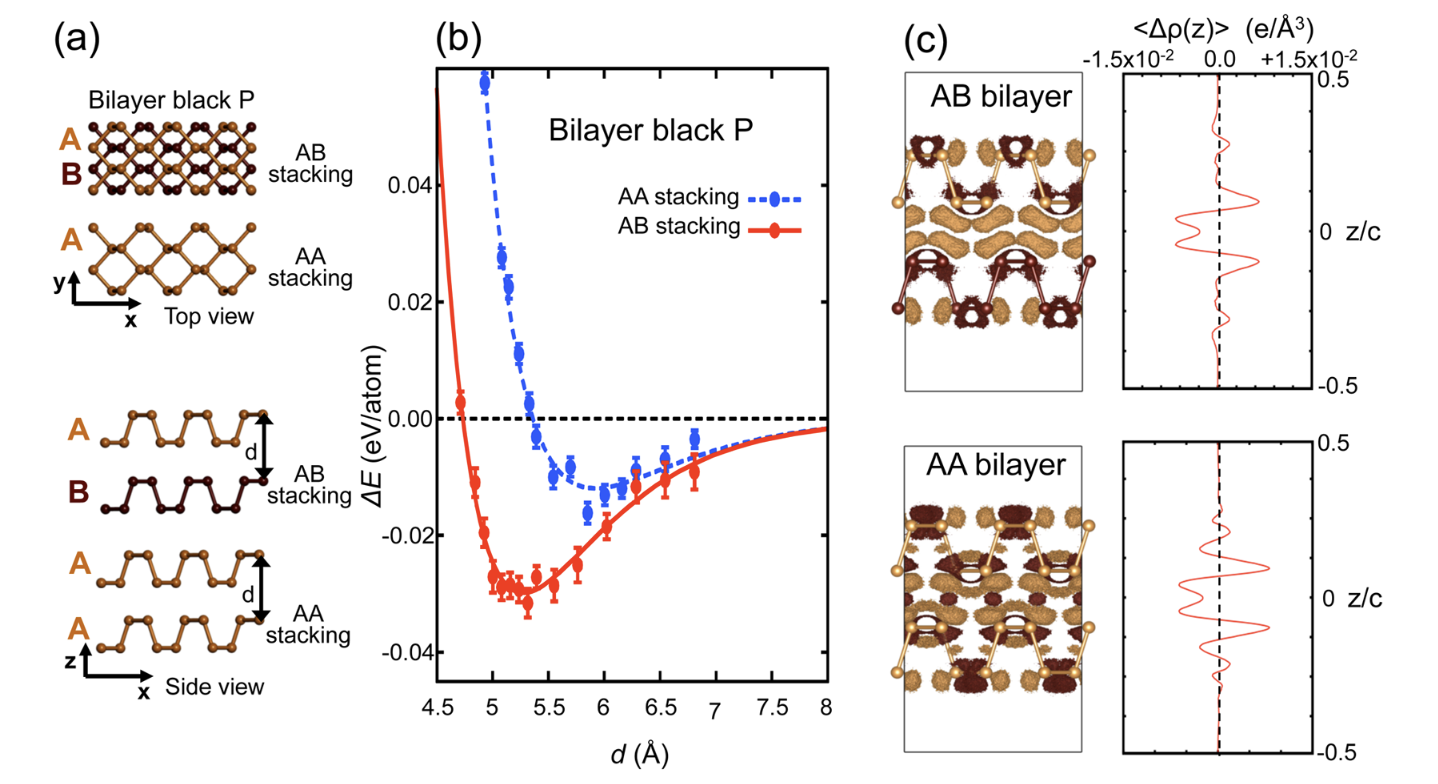}
\par\end{centering}
\caption{
DMC calculations for bonding in AA- and AB-stacked bilayer phosphorene. (a) AA- and AB-stacked bilayer geometries. (b) The relative total energy per atom $\Delta$E as a function of the interlayer spacing $d$ calculated with DMC. Lines that connect the data points are Morse fits that extrapolate to $\Delta$E = 0 as $d$ $\rightarrow \infty$. (c) The difference in electron density for the AB and AA bilayers (isosurfaces and planar averages). This electron density difference is computed by subtracting the electron density of the respective bilayer structures from the electron density of the bulk structure. The planar average, $<\rho(z)>$, is obtained by averaging across the $x$-$y$ plane of the layers, where $z/c$ indicates the relative position of the plane within the unit cell. Reproduced from Shulenburger et al., Nano Lett., 15, 8170-8175 (2015), licensed under a Creative Commons Attribution (CC BY) license. 
\label{fig:phosphorene}
}
\end{figure*}

%\textcolor{blue}{-Ref. \onlinecite{phosphorene-1st}: The Nature of the Interlayer Interaction in Bulk and Few-Layer Phosphorus, L. Shulenburger, A.D. Baczewski, Z. Zhu, J. Guan, and D. Tománek (2015): https://pubs.acs.org/doi/full/10.1021/acs.nanolett.5b03615}

%\textcolor{blue}{-Ref. \onlinecite{PhysRevB.98.085429}: Phase stability and interlayer interaction of blue phosphorene, Jeonghwan Ahn, Iuegyun Hong, Yongkyung Kwon, Raymond C. Clay, Luke Shulenburger, Hyeondeok Shin, and Anouar Benali (2018): https://journals.aps.org/prb/abstract/10.1103/PhysRevB.98.085429}

%\textcolor{blue}{-Ref. \onlinecite{blue-phos}: Metastable Metallic Phase of a Bilayer Blue Phosphorene Induced by Interlayer Bonding and Intralayer Charge Redistributions, Jeonghwan Ahn, Iuegyun Hong, Gwangyoung Lee, Hyeondeok Shin. Anouar Benali, Yongkyung Kwon (2021): https://pubs.acs.org/doi/abs/10.1021/acs.jpclett.1c03045}

Layered phosphorene allotropes have garnered a great deal of interest because of their layer-dependent properties exemplified with band gaps of black phosphorenes varying from 2.26 eV at the monolayer limit to 0.3 eV for the bulk.\cite{tran2014layer,choi2015linear,li2017direct} This is understood to be driven by hybridization of valence p$_{z}$ orbitals between the adjacent layers, which causes interlayer interactions of layered phosphorenes to be not purely dispersive compared to those of bilayer graphene interpreted as a typical vdW system. 

The distinct interlayer interaction of phosphorenes from the work of Shulenburger et al. \cite{phosphorene-1st} is shown in Fig.~\ref{fig:phosphorene}, presenting DMC interlayer binding energy curves for AA- and AB-stacked bilayer black phosphorenes. There is a significant energy difference of interlayer interactions ($\approx$18 meV per atom) and interlayer separations ($\approx$0.6~\AA) for AA- and AB-stacked bilayer black phosphorenes, not consistent with the vdW interaction between two homogeneous slabs. This is in contrast to the case of bilayer graphene showing a difference of 6 meV/atom and $\approx$0.1~\AA \space for interlayer binding energies and separations, respectively.\cite{PhysRevLett.115.115501,carbon-layers} Furthermore, the charge density redistributions computed with DMC for the two stacking modes show a delicate difference in the interlayer region where AB and AA yield the charge depletion and accumulation, respectively. This feature demonstrates that the nature of interlayer interactions in black phosphorenes is complicated, not characterized by purely
dispersive interactions.

\begin{figure}
\begin{centering}
\includegraphics[width=8.5cm]{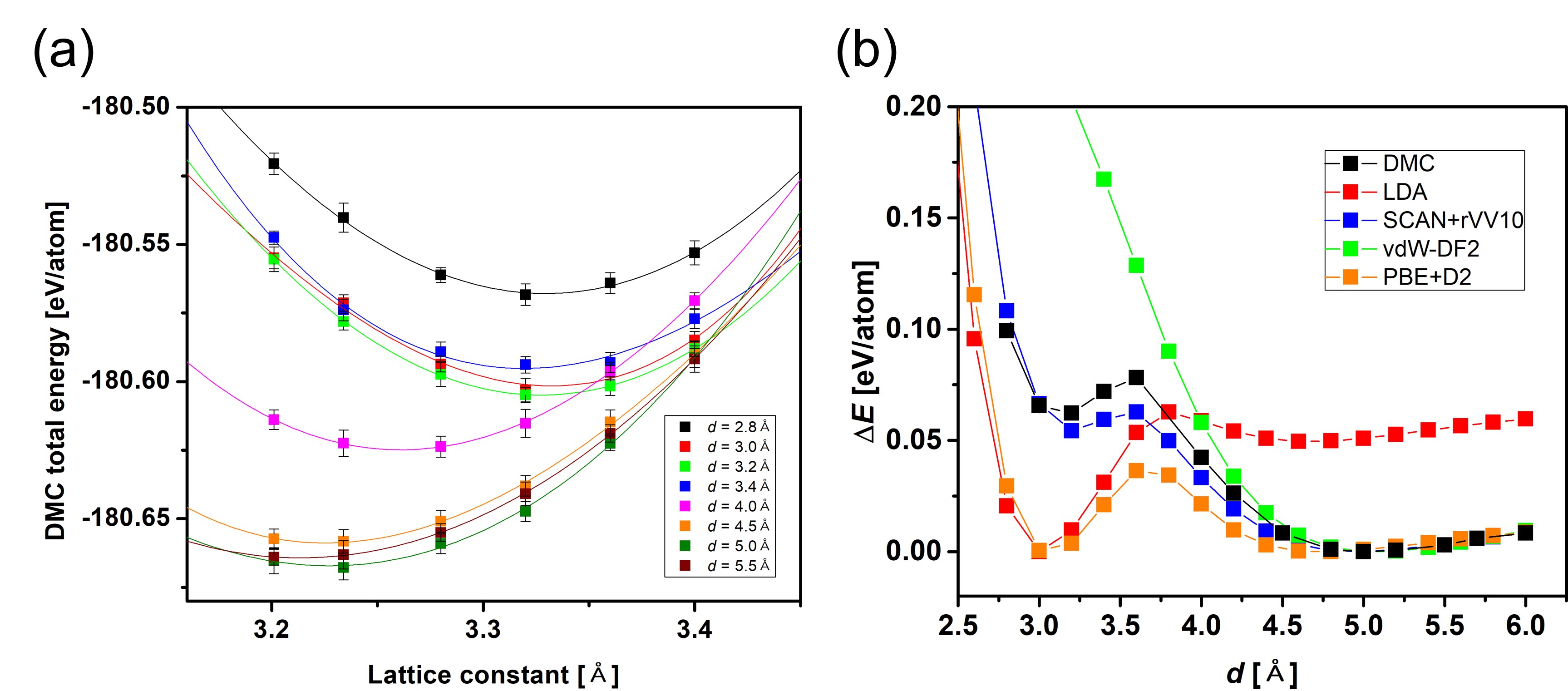}
\par\end{centering} \caption{(a) DMC total energy $E_{\text{bi}}$ of a $4 \times 4 \times 1$ supercell of bilayer blue phosphorene as a function of the lattice constant for a given interlayer distance $d$, where the solid lines represent Murnaghan fits for each $d$. (b) DMC relative energy $\Delta E=E_{\text{bi}}-E_{min}$, along with the corresponding DFT results computed with LDA, SCAN+rVV10, vdW-DF2, and PBE+D2 XC functionals, as a function of the interlayer distance $d$, where $E_{min}$ is the minimum total energy in the respective computation. Statistical errors of DMC data in (b) are smaller than the symbol sizes (less than 0.002 eV/atom). Reproduced with permission from Ahn et al., J. Phys. Chem. Lett. 2021 12 (45), 10981-10986 (2021). Copyright 2021 American Chemical Society.
\label{fig:blue_phosphorene}
}
\end{figure}

Additionally,  blue phosphorene, a 2D phosphorus allotrope with a puckered honeycomb structure, shows interlayer binding properties analogous to those of black phosphorene, according to our DMC study reported in Ref.~\onlinecite{PhysRevB.98.085429}. Furthermore, as the interlayer distance decreases, a blue phosphorene bilayer could exhibit a semiconductor-to-metal transition with the metallic state being manifested at short interlayer distances comparable to the intralayer bond length,\cite{blue-phos} which is observed for the A$_{1}$B$_{-1}$ stacking mode, one of five possible bilayer stacking modes of a puckered honeycomb structure.\cite{arcudia20} A successful capture of this feature requires a well-balanced description of long-range dispersive forces and short-range chemical bondings with DMC because DFT calculations based on different XC functionals show large inconsistency in predicting equilibrium interlayer binding energies and interlayer separations (as displayed in Fig.~\ref{fig:blue_phosphorene}.) 
The balanced description of diverse interactions is a key feature achieved through DMC, which has been proven to exhibit the accuracy of CC calculations with single, double, and perturbative triple excitations (CCSD(T)), the gold standard in quantum chemistry, in systems such as the benzene dimer\cite{dubecky2013quantum,azadi2015chemical} and methane-water.\cite{vrezavc2015extensions,gillan2015energy} This often positions DMC as a reference method for relatively large molecular systems where the application of CCSD(T) is limited.\cite{tkatchenko2012first,ambrosetti2014hard,benali2014application,hermann2017nanoscale,stohr2021coulomb}
%but also manifests local stabilization of its metallic state at the short interlayer distance comparable to intralayer bond length~\cite{blue-phos}. This feature is observed for the A$_{1}$B$_{-1}$ stacking mode, which is one of five possible bilayer stacking modes of a puckered honeycomb structures~\cite{arcudia20}, and interestingly, DFT calculations shows large inconsistency in predicting the interlayer binding energetics and interlayer separations along with even the number of energy minima as displayed in Fig.~\ref{fig:blue_phosphorene}(b). This highlights a need of the well-balanced description of the interlayer binding and intralayer bonding on the equal footing in order to correctly capture the interlayer coupling from short- to long-range regions as well as the intermediate region.

Fig.~\ref{fig:blue_phosphorene}(b) presents the DMC interlayer binding curve as a function of the interlayer distance, along with the corresponding DFT ones, for which the lattice constant optimized at a given interlayer distance was used (see Fig.~\ref{fig:blue_phosphorene}(a)).  The DMC binding curve is seen to possess two minima with their energy difference being 76(1) meV per atom.
%a local minimum at a shorter interlayer distance and the global minimum at a longer distance. 
Among the DFT functionals considered here, the SCAN+rVV10 (SCAN plus vdW correction) \cite{PhysRevX.6.041005} is found to produce the results in the best agreement with DMC in terms of the two-minima feature and the relative energetics between them. The local minimum at a short interlayer distance and the global minimum at a longer distance turn out to have a metallic and a semiconducting band structure, respectively.  The metastable metallic minimum at a short interlayer distance is understood to arise from the interlayer hybridization between p$_{z}$ orbitals of the highest occupied valence bands of the two monolayers as well as intralayer charge redistribution.
%From the optimization of the lattice constant for given interlayer distance with DMC calculations as shown in Fig.~\ref{fig:blue_phosphorene}(a), DMC yields the interlayer binding curve with possessing both local and global minima at short and large interlayer distances, respectively, with their energy difference being 76(1) meV/atom. 
%SCAN+rVV10 is found to be in agreement with predicting the two minima and their relative energetics. The local and global minima turns out to have metallic and semiconducting band structures and the metastable metallic phases at short interlayer distance is understood to arise from the interlayer hybridization between p$_{z}$ orbitals of the highest occupied valence bands of the two monolayers and intralayer charge redistributions.
Random phase approximation (RPA) calculations of Arcudia et al.\cite{arcudia20} also revealed two energy minima in the interlayer binding energy curve with different electronic phases for an A$_{1}$B$_{-1}$-stacked blue phosphorene bilayer. Contrary to the DMC results, however, the RPA study predicted that the metallic minimum at a short distance was energetically favored over the semiconducting one at a longer distance. This discrepancy is understood to be due to a well-known RPA limitation in describing short-range interactions which play crucial roles in stabilizing the metallic phase. Additional information regarding this work can be found in Ref.~\onlinecite{PhysRevB.98.085429}. 
%Remarkably, previous RPA calculations also predicts two energy minima with different electronic phase, but the metallic minimum at short distance is more stable than the semiconducting one at large distance as opposed to the DMC prediction. Although noting lack of validation of RPA calculations for the phosphorene allotropes and its well-known limitation of describing interactions at short-range regime where the metallic phase occurs~\cite{}, the discrepancy between DMC and RPA calculations calls for further experimental investigations.

\subsubsection{Bilayer Arsenene}\label{arsenene}

%Feel free to swap this figure out for another
\begin{figure}
\begin{center}
\includegraphics[width=8.5cm]{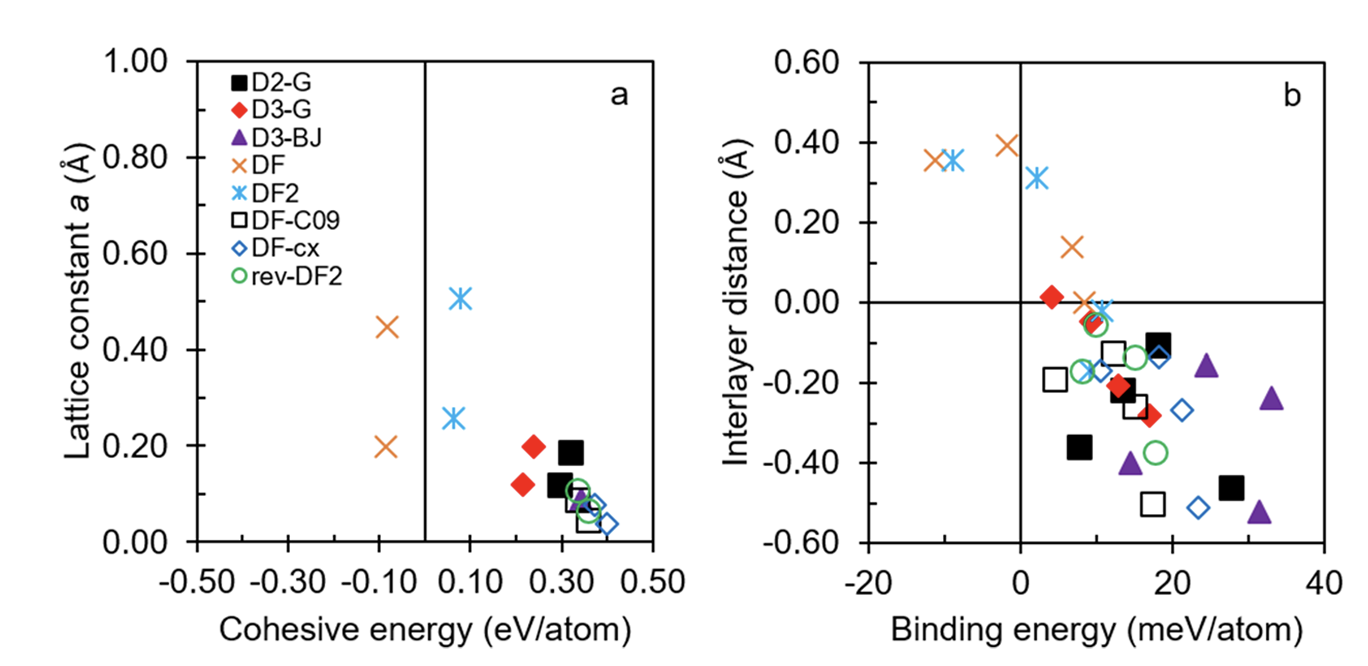}
\caption{The deviation of DFT from DMC for a) cohesive energy and lattice constant of monolayer b-As and w-As and b) binding energy and interlayer distance of bilayer b-As-AA, b-As-AB, w-As-AA, and w-As-AB. Reproduced from Kadioglu et al., J. Chem. Phys. 148, 214706 (2018), with the permission of AIP Publishing.}
\label{Arsenene}
\end{center}
\end{figure}

Similar to phosphorene, arsenene (As) is another monoelemental 2D material that displays promising electronic properties.\cite{PhysRevB.91.085423,Zhang_2015,https://doi.org/10.1002/anie.201507568,arsenene-exp} Monolayer arsenene can exist in the washboard (w-As) honeycomb structure (similar to black phosphorene) and the buckled (b-As) honeycomb structure (similar to blue phosphorene). For bilayer arsenene, the ordering of energetic stability and the electronic properties have not been clearly established, and various local, semilocal, vdW-corrected, and hybrid functionals have been used to study bilayer arsenene.\cite{CAO2015501,https://doi.org/10.1002/anie.201411246,arsenene-paper,Mi_2017,PhysRevB.94.205409} Because this system is weakly bonded, vdW functionals are a viable method to describe the complex physics of bilayer arsenene. To benchmark these vdW functionals and obtain an accurate benchmark of the stability order of possible stacking configurations of bilayer arsenene, we performed DMC calculations.\cite{10.1063/1.5026120}     

In this work, we performed DMC and DFT (using local, semilocal, and vdW density functionals and functionals with semi-empirical vdW corrections) calculations of the binding energy and interlayer distance of w-As and b-As in AA- and AB-stacking configurations (see Fig. \ref{Arsenene}). Our DMC results revealed the AA stacking to be lower in energy than the AB stacking for b-As. We also find that the layer-layer interactions are dispersive because the energy changes by 22$\%$ and the interlayer distance changes by 0.1 \AA \space when going from AA to AB stacking. We find that for b-As, the interlayer distance changes by 0.65 \AA \space when going from the AB to AA  stacking, which most likely signifies a complex layer-layer interaction. When benchmarking our DMC results along with DFT, we find that vdW density functionals (i.e., DF \cite{PhysRevLett.92.246401}) can reproduce DMC energetics, but structural parameters are better described by semi-empirically corrected functionals such as the D3-G method of Grimme.\cite{10.1063/1.3382344} Additional information regarding this work can be found in Ref. \onlinecite{10.1063/1.5026120}.

\subsubsection{Bilayer Graphene and Graphyne}\label{graph}
%\textcolor{blue}{-Ref. \onlinecite{carbon-layers}: Nature of Interlayer Binding and Stacking of sp-sp2 Hybridized Carbon Layers: A Quantum Monte Carlo Study, Hyeondeok Shin, Jeongnim Kim, Hoonkyung Lee, Olle Heinonen, Anouar Benali, Yongkyung Kwon (2017): https://pubs.acs.org/doi/full/10.1021/acs.jctc.7b00747}

\begin{figure}
\begin{center}
\includegraphics[width=8.5cm]{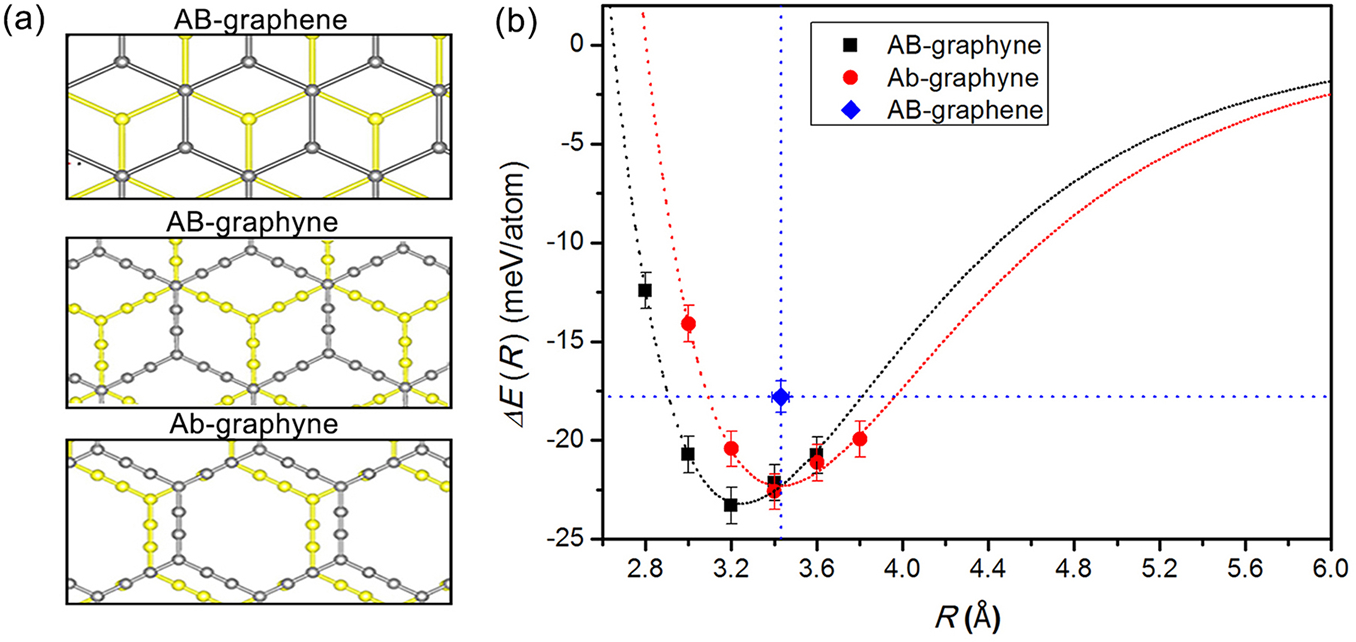}
\caption{(a) Stacking configurations of AB-stacked bilayer graphene and two stable modes (AB and Ab) of a bilayer $\alpha$-graphyne. The yellow and gray structures represent the low and upper layers of a bilayer, respectively. (b) DMC interlayer binding energies of AB- and Ab-stacked bilayer $\alpha$-graphynes as functions of an interlayer distance. The blue diamond symbol represents a DMC interlayer binding energy for an AB-bilayer graphene at an equilibrium interlayer distance reported in Ref. \onlinecite{PhysRevLett.115.115501}. Reproduced with permission from Shin et al., J. Chem. Theory and Comp. 13 (11), 5639-5646 (2017). Copyright 2017 American Chemical Society.}
\label{bilayer_graphyne_fig}
\end{center}
\end{figure}

Regarding low-dimensional carbon allotropes, $\alpha$-graphyne was predicted to possess weaker DFT binding energy for the bilayer structure than bilayer graphene but has attracted a great deal of attention due to its larger honeycomb structure than that of graphene.\cite{carbon-layers} However, DFT studies of the bindings of graphynes failed to confirm the most stable stacking mode because of the strong dependence of vdW-corrected XC functional on computed DFT binding energy. Those uncertainties in the choice of the XC functional for DFT and the weak binding energy of graphynes revealed that a more precise method is imperative for accurately predicting the stable stacking mode of graphynes and their binding energies.

Fig.~\ref{bilayer_graphyne_fig} shows our DMC binding energies of the AB- and Ab-stacking modes of bilayer $\alpha$-graphyne as a function of the layer spacing. The negligible difference in equilibrium binding energy between AB and Ab modes from DMC indicates the difficulty of synthesizing a pristine AB- or Ab-stacking $\alpha$-graphyne structure. Both AB (23.2(2) meV per atom) and Ab modes (22.3(3) meV per atom) of graphyne show larger equilibrium binding energies than AB-stacked bilayer graphene (17.8(3) meV per atom). This can be attributed to the different nature of interlayer interactions in the sp-sp$^2$ hybridized graphyne structure, as opposed to graphene whose interlayer coupling is primarily governed by weak vdW interactions. Due to this difference of interlayer binding nature between sp$^2$-bonded and sp-sp$^2$ hybridized carbon networks, it is found that vdW-corrected DFT functionals including DFT-D2, \cite{https://doi.org/10.1002/jcc.20495} vdW-DF, and rVV10 significantly underestimate the binding energy of bilayer $\alpha$-graphynes, while an overestimation of the binding energy was shown on the pristine sp$^2$-bonded graphene. Among the vdW-corrected DFT functionals, binding energies computed by the rVV10 functional show the closest result to the corresponding DMC result, while vdW-DF provides the closest charge density distributions. This inconsistent trend indicates the importance of an accurate description for dispersion and density correction in vdW-corrected DFT functionals, which can provide guidelines to improve vdW descriptions for the future Kohn-Sham scheme. More information can be found in Ref. \onlinecite{carbon-layers}. 

In addition to our work presented in this section, Mostaani et al. \cite{PhysRevLett.115.115501} computed the DMC binding energy of bilayer graphene to be 11.5(9) meV/atom and 17.7(9) meV/atom for AA and AB stacking, respectively, and Krongchon et al. \cite{PhysRevB.108.235403} utilized DMC and tight binding models to accurately describe the registry-dependent potential energy and lattice corrugation of twisted bilayer graphene, a highly correlated system.

\subsubsection{Bilayer and Bulk TiS$_2$}\label{tis2}

%Feel free to swap this figure out for another
\begin{figure}
\begin{center}
\includegraphics[width=8cm]{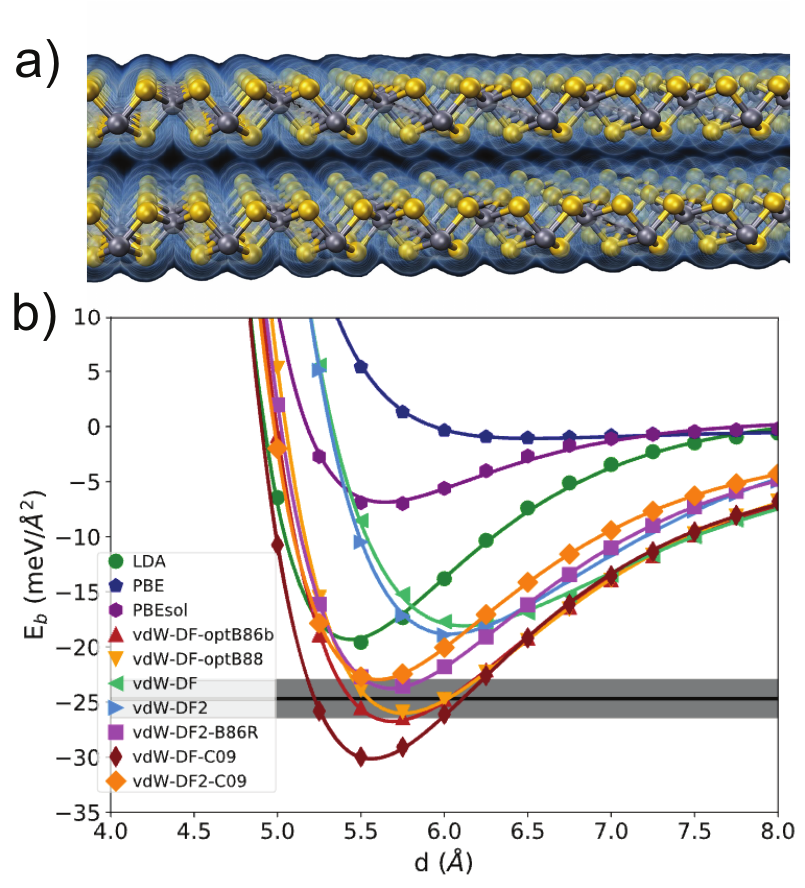}
\caption{(a) Atomic structure and charge density isosurfaces and (b) interlayer binding energy curve for bilayer TiS$_2$ calculated with a variety of DFT functionals alongside the DMC
results (black line with gray region indicating the uncertainty). Reproduced with permission from Krogel et al., J. Phys. Chem. A, 124 (47), 9867-9876 (2020). Copyright 2020 American Chemical Society.}
\label{fig:tis2}
\end{center}
\end{figure}

The family of vdW density functionals has been largely successful for a wide variety of systems, including 2D materials. Due to the success of this approach, there have been numerous improvements to the original vdW-DF functionals, where an emphasis has been placed on incorporating semilocal exchange.\cite{PhysRevB.81.161104} Despite these improvements, highly accurate benchmarking calculations, such as those performed with DMC, are crucial in paving the way to developing a general-purpose vdW density functional that can accurately reproduce fundamental properties such as the total energy and the electron density. In this specific work, we chose to benchmark the properties of bulk and bilayer TiS$_2$ with a variety of local, semilocal, and vdW DFT functionals against DMC.\cite{tis2} TiS$_2$ (part of the dichalcogenide family) is a system of interest due to its potential applications as a cathode material for Li-ion batteries \cite{doi:10.1126/science.192.4244.1126} and has been debated to be either a heavily self-doped narrow gap semiconductor or a semimetal,\cite{tis2-semimetal-semiconductor} adding complexity and novelty to the noncovalent interactions.\cite{PhysRevB.21.615} 

After performing these benchmarking calculations (depicted in Fig. \ref{fig:tis2}), we found a close relationship between the accuracy of the interlayer distance and binding energy. We find that more recently developed vdW functionals such as vdW-DF-optB88 \cite{Klimes_2010} perform well for both properties. In terms of the response of the electron density to binding, we find that functionals such as LDA and PBEsol \cite{PhysRevLett.100.136406} outperform the vdW functionals and can correctly describe the interlayer charge accumulation. The vdW functionals such as vdW-DF-C09,\cite{PhysRevB.81.161104} which was developed on purely theoretical grounds, perform the best in terms of simultaneously reproducing the DMC energy and electron density. This highlights the need for a theory-driven path forward to develop a fully predictive and consistent vdW functional, with highly accurate benchmarks from methods such as DMC paving the way. Additional information can be found in Ref. \onlinecite{tis2}.

\subsubsection{Bulk CrI$_3$}\label{cri3-layered}

In this section, we revisit the CrI$_3$ system (discussed extensively in Section \ref{cri3}) from a different perspective. As previously mentioned, CrI$_3$ exhibits long-range magnetic ordering from few-layer to monolayer.\cite{cri3} Bulk and few-layer CrI$_3$ possess long-range interlayer and short-range intralayer interactions. The long-range vdW interlayer (noncovalent) interactions, which are due to strongly correlated electrons occupying $d$ orbitals of Cr, and the competing intralayer correlations make this an extremely challenging system to model with approaches such as local or semilocal DFT. Certain vdW corrections in DFT can improve the accuracy of these interlayer forces, but the results can vary drastically depending on which vdW correction is employed.     

\begin{figure}
\begin{center}
\includegraphics[width=8cm]{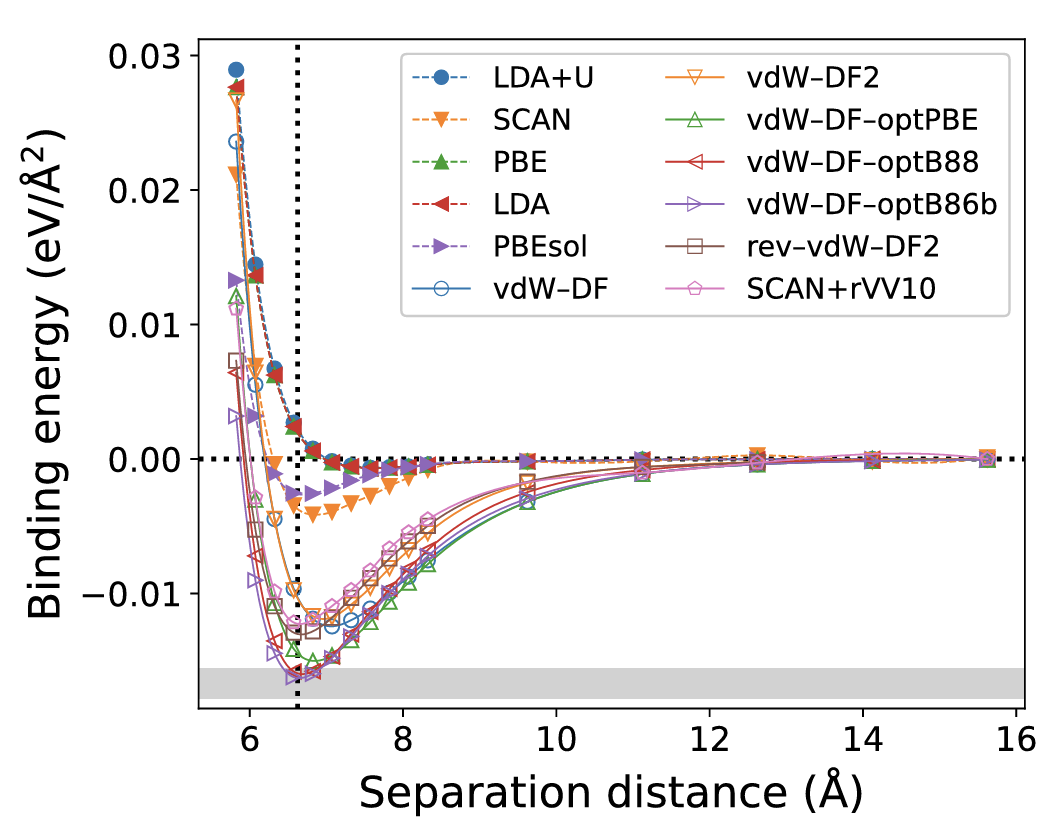}
\caption{The interlayer binding energy as a function of separation distance for monoclinic CrI$_3$ for various DFT functionals alongside DMC (gray shaded region indicates uncertainty). The vertical line indicates the experimental separation distance. Reproduced with permission from Ichibha et al., Phys. Rev. Materials 5, 064006 (2021). Copyright 2021 American Physical Society.}
\label{fig:cri3-layered}
\end{center}
\end{figure}

In this work, we studied the binding properties of bulk CrI$_3$ with DMC, with the goal of simultaneously describing the short- and long-range correlations and overcoming the shortcomings of DFT.\cite{PhysRevMaterials.5.064006} For the monoclinic bulk CrI$_3$ structure, we calculated the interlayer separation distance to be 6.749(73) \AA \space with DMC, which is in excellent agreement with the experimental value of 6.623 \AA.\cite{cri3-bulk} We also estimated the interlayer binding energy to be between 14.3 and 17.9 meV/\AA$^2$. We benchmarked several DFT functionals with and without vdW corrections against our DMC results for interlayer separation distance and binding energy (see Fig. \ref{fig:cri3-layered}), and we found that vdW-DF-optB88 \cite{Klimes_2010}
and vdW-DF-optB86b \cite{PhysRevB.83.195131} are closest at reproducing our DMC values. In addition, we studied the bulk rhombohedral structure of CrI$_3$ and found that the rhombohedral and monoclinic are within thermal energy differences of each other, which is in agreement with experiment.\cite{cri3-bulk} We believe our DMC benchmark calculations can be useful for testing other promising current and future vdW-corrected functionals. Further information about this work can be found in Ref \onlinecite{PhysRevMaterials.5.064006}.

\subsubsection{Graphene-Supported Pt Layers}\label{Ptlayer}
%\textcolor{blue}{-Structural Stability of Graphene-Supported Pt Layers: Diffusion Monte Carlo and Density Functional Theory Calculations, Jeonghwan Ahn, Iuegyun Hong, Gwangyoung Lee, Hyeondeok Shin, Anouar Benali, Jaron Krogel, Yongkyung Kwon (2023): https://pubs.acs.org/doi/abs/10.1021/acs.jpcc.3c03160}

\begin{figure}
\begin{centering}
\includegraphics[width=8.0cm]{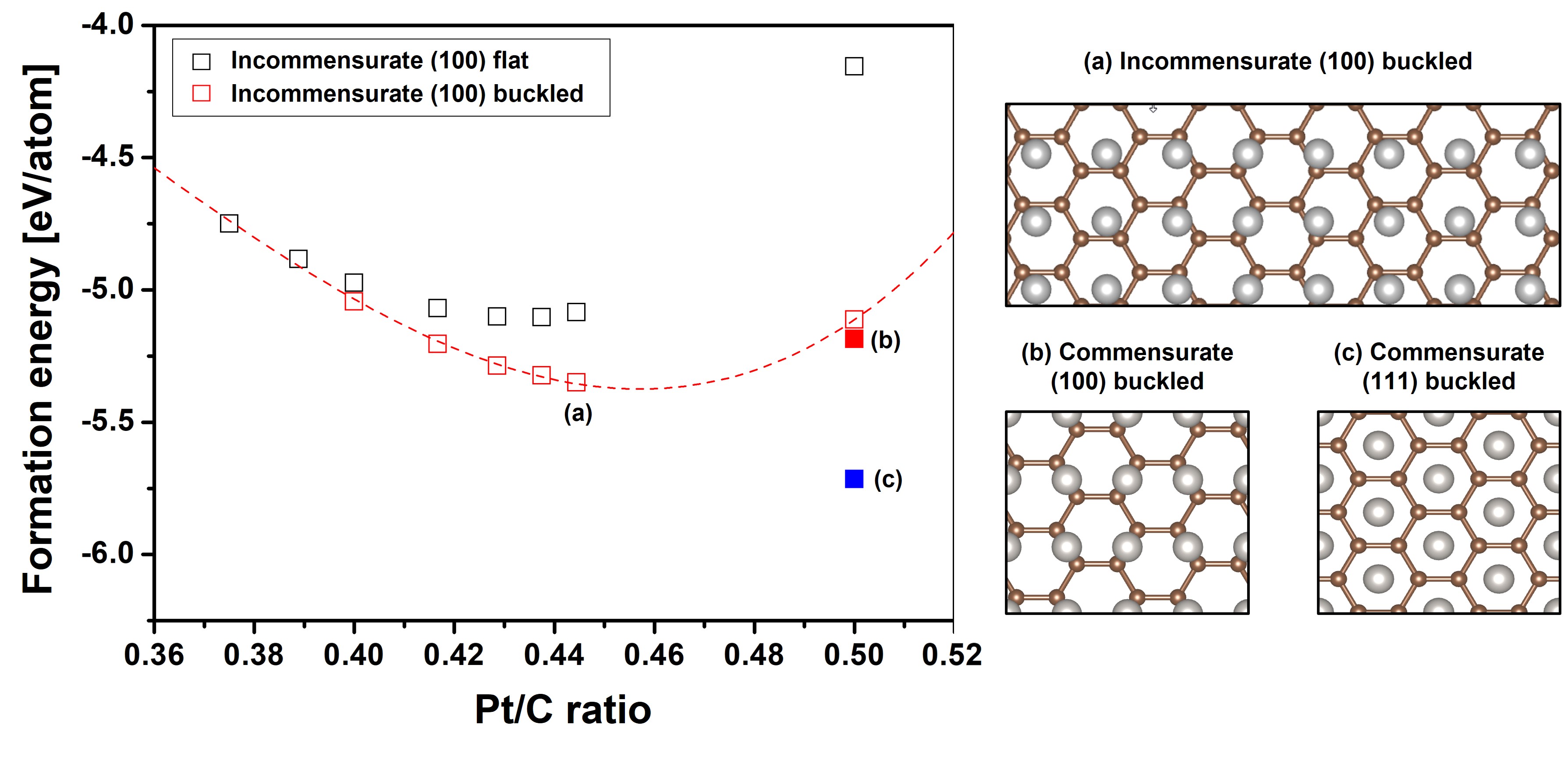}
\par\end{centering}
\caption{DFT-SCAN+rVV10 equilibrium formation energies of various Pt monolayer on graphene, as functions of the Pt/C atomic ratio,
where open black and red squares represent the formation energies of (100)-packing flat and buckled structures on pristine graphene, respectively.
The solid symbols denote the equilibrium formation energies of the respective Pt layers on graphene whose lattice is allowed to relax.
The red dotted lines represent third-order polynomial fits of the lower formation energies between flat and buckled structures at the respective Pt/C atomic ratio. SOC was not included in these calculations.  Reproduced with permission from Ahn et al., J. Phys. Chem. C, 127 (37), 18630-18640, (2023). Copyright 2023 American Chemical Society.}
\label{fig:Pt_adsorption}
\end{figure}

Recent experimental realization of 2D layered structures of Pt atoms above a graphene surface\cite{abdelhafiz2018epitaxial,choi2019contiguous,robertson2019atomic} has opened an interesting avenue for exploration of their various morphologies, as well as the bonding nature of Pt-graphene complexes. In the sense that the layer formation is a consequence of a competition between nondirectional metallic bondings among Pt atoms and their covalent and vdW bondings to graphene, DMC benchmarking is critical for accurate theoretical investigation of these layered systems with large degrees of freedom (including different Pt/C atomic ratios). As far as their geometries and relative energetics between different layered structures of Pt atoms are concerned, DFT results based on the SCAN+rVV10 functional are found to best agree with the corresponding DMC results. 
%such as an incommensruate Pt layer on the graphene surface. 
%A commensurate structure of monolayer and bilayer Pt layers with (100)- and (111)-packing order on the graphene surface were treated in this study by considering their (100)- and (111)-packing orders with the number of ratios betwene Pt and C atoms being fixed to 1/2. The incommensurate Pt structures relative to the underlying the graphene lattice

%The DMC benchmark of DFT calculations from several different DFT functionals shows the closest agreement between SCAN+rVV10 and DMC calculations for the geometries and relative energies among the commensurate graphene-supported Pt layers considered in this study which has a Pt/C atomic ratio of 1/2. 
Our DMC-benchmarked DFT calculations show that
a (111)-packing structure, where each of the Pt atoms is adsorbed at a hollow site of graphene (see Fig.~\ref{fig:Pt_adsorption}(c)), is more stable than the (100)-packing ones when forming the commensurate Pt-graphene complex at a Pt/C atomic ratio of 1/2. This can be understood by a significant lattice mismatch (>10$\%$) between pristine graphene and a freestanding (100)-packing Pt layer. 
%It is found that (111)-packing commensurate structure is more stable than the (100)-packing one, which can be understood by the significant lattice mismatch (>10$\%$) between pristine graphene and a freestanding (100)-packing Pt layer when forming the commensurate Pt-graphene complex. 
Energetic stability of incommensurate Pt layers with different Pt/C atomic ratios was also investigated. 
%Furthermore, the impact of Pt/C atomic ratio on the stability of graphene-supported layers were also investigated.
Fig.~\ref{fig:Pt_adsorption} displays DFT-SCAN+rVV10 formation energies of both flat and buckled incommensurate (100)-packing Pt monolayers (see Fig.~\ref{fig:Pt_adsorption}(a)) as functions of the Pt/C atomic ratio. The lowest-energy configuration is found at the ratio of 0.457, and its formation energy is lower than the corresponding energies for buckled (100)-packing monolayer with the atomic ratio of 1/2, which is in line with recent experimental findings.\cite{robertson2019atomic} However, the formation energy of this optimal incommensurate structure is still higher than that of the (111)-packing commensurate structure. 
%The bilayer results are analogous to the case of monolayer. 
Through its systematic investigation for various possible morphologies of layered Pt-on-graphene systems, this DMC-benchmarked study contributes to the expansion of a family of available metallic Pt layers. Additional information can be found in Ref.~\onlinecite{Ptlayer}. 
%This study has a significance in terms of expanding a family of available Pt layers and providing an opportunity to further investigate other 2D metallic layers.

\subsection{Cohesion and Adsorption Energetics}\label{energetics}

Since the successful isolation of the graphene sheet, numerous types of 2D materials have been theoretically proposed, each exhibiting intriguing electronic properties. Several of these materials have also been synthesized. In light of this, obtaining accurate assessments of their ground-state properties, including cohesive energies and their relative differences, is of paramount importance. These assessments not only guide experimental synthesis efforts but also aid in the development of sophisticated DFT XC functionals to deal with 2D materials. Quantitative agreement between DMC calculations and experimental results for some existing 2D materials has solidified DMC as the optimal approach for fulfilling this role among current available first-principles calculations for periodic solid systems. DMC calculations have effectively established the ground-state energetics for many proposed 2D materials whose DFT energy differences are sensitive to the choice of density functionals. It is worth noting that, in many instances, DFT tends to overestimate their cohesive energies compared to those derived from DMC calculations.\cite{shin14,PhysRevB.98.085429,hong2020competition,borophene} This discrepancy has been attributed to various factors, including dimensional effects and the types of bondings involved. Below, we provide several examples illustrating the application of DMC calculations in investigating the cohesion and adsorption energetics of 2D materials.

%Since the successful isolation of the graphene sheet, many kinds of 2D materials has been theoretically proposed with their intriguing electronic properties, among a few of which have been synthesized. In this regard, an accurate estimation of their ground-state properties such as the cohesive energies along with their relative difference is an integral ingredient to guide not only experimental synthesis, but also development of the sophisticated DFT exchange-correlation functionals.
%On the basis of quantitative agreement between DMC and experimental results for the existing 2D materials, DMC calculations have proven to be the best approach in taking this role among the current available first-principles calculations for the periodic solid systems and have established the ground-state energetics for many proposed 2D materials with their energy difference being sensitive to the choice of the DFT functionals~\cite{}.
%As shown in many cases, DFT cohesive energies tend to be overestimated compared to the DMC one, which has been observed to be affected by dimensional effects, types of bondings and so on.
%The followings are a few examples of application of DMC calculations to investigating the cohesion and adsorption energetics of 2D materials.

\subsubsection{Carbon Allotropes}\label{graphyne}

%\textcolor{blue}{-Cohesion energetics of carbon allotropes: quantum Monte Carlo study, Hyeondeok Shin, Sinabro Kang, Jahyun Koo, Hoonkyung Lee, Jeongnim Kim, Yongkyung Kwon (2014): https://aip.scitation.org/doi/abs/10.1063/1.4867544}
\begin{figure}
\begin{center}
\includegraphics[width=8.0cm]{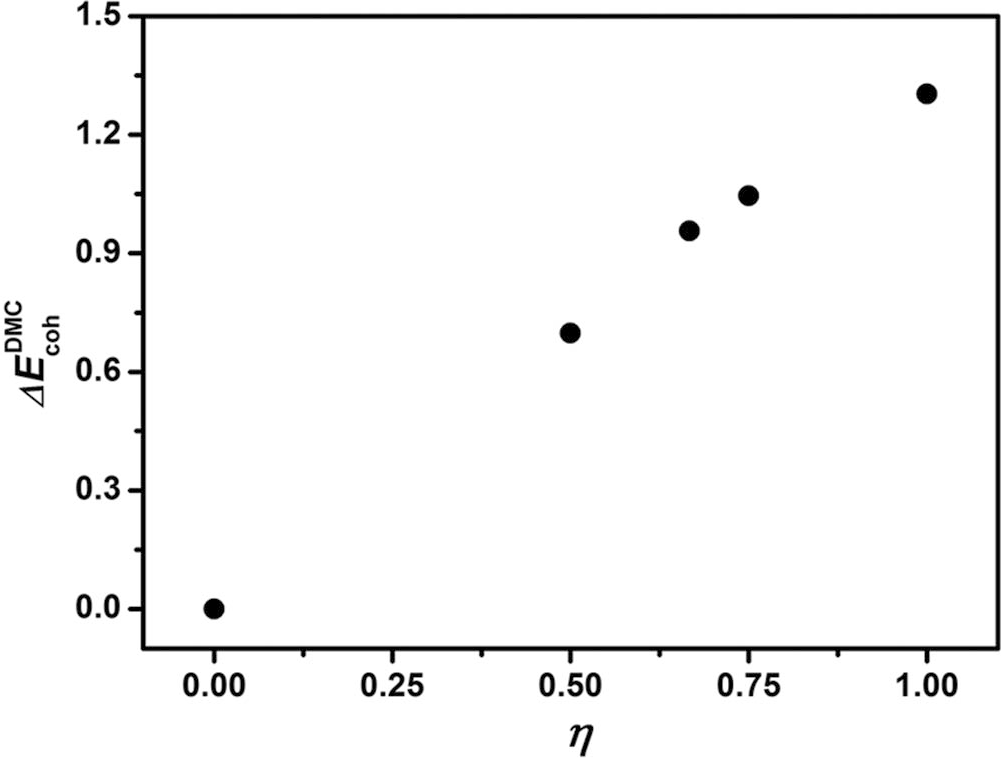}
\caption{The DMC cohesive energy difference between a low-dimensional carbon allotrope and graphene, $\Delta E^{DMC}_{coh} = E^{DMC}_{coh}(graphene) - E^{DMC}_{coh}(allotropes)$, as a function of the ratio of sp-bonded carbon atoms, $\eta$. The energies are in units of eV/atom. Reproduced from Shin et al., J. Chem. Phys. 140, 114702 (2014), with the permission of AIP Publishing.}
\label{cohesion_fig}
\end{center}
\end{figure}

Graphyne is one of the proposed sp-sp$^2$ hybridized 2D carbon allotropes expected to have exotic electronic properties, such as possessing a Dirac cone, high carrier mobilities, and a larger surface area than graphene.\cite{shin14} Since the successful synthesis of graphene, various sp- or sp-sp$^2$ hybridized low-dimensional carbon allotropes have been proposed via various first-principle studies. However, only $\gamma$-graphyne has been experimentally synthesized so far,\cite{graph-exp,Haley+2008+519+532,https://doi.org/10.1002/adma.200902623} while DFT predicted the possible existence of other stable graphyne structures. To investigate the accurate structural stability of carbon allotropes, QMC is employed on various carbon allotropes, including sp$^3$ diamond, sp$^2$ graphene, sp-sp$^2$ hybridized graphyne, and sp-bonded carbyne chain. We define a parameter $\eta$, which denotes the ratio of sp-bonded carbon atoms: $\eta= N_{\textrm{sp}}/N_{\textrm{atom}}$, where $N_{\textrm{sp}}$ is the number of sp-bonded carbon atoms, and $N_{\textrm{atom}}$ is the total number of carbon atoms in the unit cell. For the carbon allotropes in this study, $\eta$ is 0 for graphene; 0.5, 0.67, and 0.75 for $\gamma$, $\beta$, and $\alpha$-graphyne, respectively; and 1.0 for carbyne.

In the results obtained from our DMC calculations for the cohesive energy of carbon allotropes, it is found that the estimated DMC cohesive energies for diamond and graphite are in excellent agreement with corresponding experimental results.\cite{Brewer1977} On the other hand, the computed cohesive energy of graphynes from DMC demonstrates a monotonically decreasing trend with the increase of the ratio of sp-bonded atoms on their system. Based on the linear increase of cohesive energy difference between graphene and graphynes with increasing the ratio of sp-bond $\eta$ as shown in Fig.~\ref{cohesion_fig}, we expect that prediction of cohesive energy for other newly proposed graphyne structures can be achieved by utilizing computed DMC bond energies of the sp-, sp$^2$-, and sp$^3$- bond as $N_{atom}E_{coh} = \epsilon_{s}N_{s} + \epsilon_{d}N_{d} + \epsilon_{t}N_{t}$.
$E_{coh}$ and $\epsilon$ represent cohesive energy and bond energies for the single bond ($\epsilon_{s}$), double bound ($\epsilon_{d}$), and triple bond ($\epsilon_{t}$), respectively. Note that $N$ indicates the total number of the carbon atoms ($N_{atoms}$), the single bonds ($N_{s}$), the double bonds ($N_{d}$), and the triple bonds ($N_{t}$) per unit cell, respectively. More information regarding this work can be found in Ref. \onlinecite{shin14}. 

%\textcolor{blue}{-Competition between Huckel's Rule and Jahn--Teller Distortion in Small Carbon Rings: A Quantum Monte Carlo Study, Iuegyun Hong, Jeonghwan Ahn, Hyeondeok Shin, Hyeonhu Bae, Hoonkyung Lee, Anouar Benali, Yongkyung Kwon: https://pubs.acs.org/doi/abs/10.1021/acs.jpca.0c02577}

%I think we can leave the Huckel's rule and JT distortion work out since it does not directly deal with 2d or quasi-2d materials 
%YK: I agree.

\subsubsection{Boron Allotropes}\label{sec:borophene}
%\textcolor{blue}{-Energetic Stability of freestanding and Metal-Supported Borophenes: Quantum Monte Carlo and Density Functional Theory Calculations, Jeonghwan Ahn, Iuegyun Hong, Gwangyoung Lee, Hyeondeok Shin, Anouar Benali, Yongkyung Kwon (2020): https://pubs.acs.org/doi/full/10.1021/acs.jpcc.0c06883}

\begin{figure}
\begin{center}
\includegraphics[width=8.0cm]{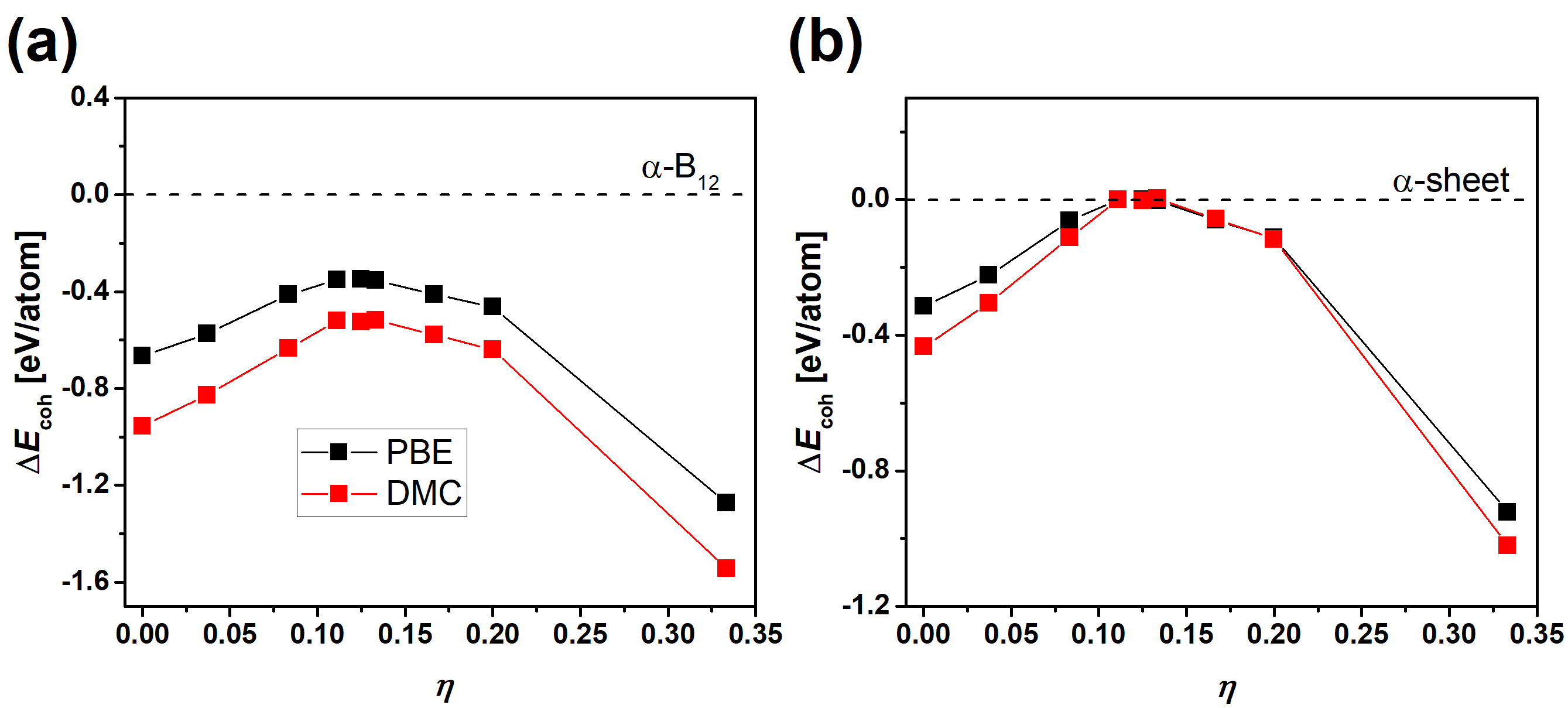}
\caption{DMC and PBE cohesive energies of freestanding boron sheets relative to those of (a) the $\alpha$-B$_{12}$ solid and (b) the $\eta=\frac{1}{9}$ $\alpha$-sheet, as a function of $\eta$. 
Statistical errors are smaller than the symbol sizes. Reproduced with permission from Ahn et al., J. Phys. Chem. C, 124 (44), 24420-24428 (2020). Copyright 2020 American Chemical Society.}
\label{fig:borophene_cohesion}
\end{center}
\end{figure}

Although no freestanding boron monolayer has been found in nature, a 2D sheet of boron, named borophene, was reported to be synthesized on a metal surface.\cite{mannix2015synthesis,borophene-2,zhang2016substrate,kiraly2019borophene,li2018experimental} The structure of borophene is characterized by a hexagonal hole density $\eta$, the ratio of the number of single-atom vacancies to the total number of atomic sites of the triangular boron sheet. 
%, called hexagonal hole density $\eta$. 
%This parameter leads to various combinations of triangular and hexagonal orderings in the borophene structures. 
With different $\eta$ values, borophene structures show various combinations of triangular and hexagonal orderings from a complete triangular lattice ($\eta=0$) to a honeycomb structure ($\eta=1/3$). On the theoretical side, a series of DFT calculations predicted the polymorphism of energetically degenerate borophene structures in the range from $\eta = \frac{1}{9}$ to $  \frac{2}{15}$.\cite{penev2012polymorphism,wu2012two} On the other hand, subsequent experiments have reported the synthesis of borophene phases only outside the polymorphic range predicted by DFT calculations,\cite{mannix2015synthesis,borophene-2,zhang2016substrate,kiraly2019borophene,li2018experimental} calling for more accurate and systematic investigation of borophene energetics. 
%This led DMC calculations to be employed to investigate the ground-state energetics among the borophene structures of various range of $\eta$.

Our DMC calculation for bulk $\alpha$-B$_{12}$ yielded a cohesive energy in very good agreement with its experimental value,\cite{Kittel2004} confirming its accuracy for cohesion energetics of boron allotropes. Subsequent DMC calculations for freestanding borophene structures showed the same polymorphism as predicted by DFT studies (see Fig.~\ref{fig:borophene_cohesion}). Fig.~\ref{fig:borophene_cohesion}(a) and (b) present DMC cohesive energies of freestanding borophenes relative to those of $\alpha$-B$_{12}$ and $\alpha$-sheet borophene ($\eta=1/9$), respectively, along with the corresponding DFT-PBE results.
%DMC calculations for boron allotropes were validated by perfect agreement between the DMC and experimental cohesive energies for a bulk $\alpha$-B$_{12}$ solid. On the basis of this validation, DMC calculations confirmed the polymorphism among the 2D borophenes as predicted by previous DFT studies as shown in Fig.~\ref{fig:borophene_cohesion}(a) and (b) presenting comparison between DMC and PBE relative cohesive energies for borophenes against $\alpha$-B$_{12}$ solid and borophene with $\eta = \frac{1}{9}$ ($\alpha$-sheet), respectively, as a function of $\eta$.
Although PBE overestimates energetic stability of borophene against $\alpha$-B$_{12}$ solid (see Fig.~\ref{fig:borophene_cohesion}(a)), PBE relative energetics between different borophene structures are in quantitative agreement with the DMC results, as seen in Fig.~\ref{fig:borophene_cohesion}(b). 
This justifies the use of PBE energies to investigate the relative energetics among various borophene structures formed on metal surfaces.
%Furthermore, it is revealed that PBE relative energetics among borophenes are in quantitative agreement with the DMC results (See Fig.~\ref{fig:borophene_cohesion}(b)) while PBE overestimates the stability of borophene against $\alpha$-B$_{12}$ solid compared to the case of DMC (See Fig.~\ref{fig:borophene_cohesion}(a)). The benchmarked PBE relative energetics to the DMC calculations among the borophene structures were further utilized to investigate the relative energetics on the metal surfaces. 
The DMC-benchmarked PBE calculations show that because of the charge transfer between a metal surface and borophene, the polymorphic range can be expanded to the $\eta$ values of experimentally synthesized borophenes on top of the Ag(111) and the Au(111) surfaces. Furthermore, the PBE calculations also predict a possible formation of bilayer borophene with $\eta = \frac{1}{12}$ on the Au(111) surface to extend a borophene family. This not only offers further insight into a mechanism of stabilizing 2D borophene structures but also opens a possibility of borophene-based electronic devices. Additional details of this work can be found in Ref. \onlinecite{borophene}.
%As a consequence of the charge transfer between metal surface and borophenes, the polymorhpic range is identified to be broadened to $\eta$ for the experimentally synthesized borophenes on top of the Ag(111) and Au(111) surfaces, which provides further insight into understanding of a mechanism of stabilizing borophenes. Finally, PBE even predicts possible formation of bilayer borophene with $\eta = \frac{1}{12}$ on Au(111) surface, which extends borophene family and opens an possibility of borophene-based electronic devices.

\subsubsection{Atomic/Molecular Adsorption}\label{borophene}
%\textcolor{blue}{-Importance of Van der Waals Interactions in Hydrogen Adsorption on a Silicon-carbide Nanotube Revisited with vdW-DFT and Quantum Monte Carlo, Genki I. Prayogo, Hyeondeok Shin, Anouar Benali, Ryo Maezono, Kenta Hongo (2021): https://pubs.acs.org/doi/full/10.1021/acsomega.1c03318}
%I think we can leave the SiC nanotube work out since it does not directly deal with 2d or quasi-2d materials 

%\textcolor{blue}{-Diffusion Monte Carlo study of O2 adsorption on single layer graphene, Hyeondeok Shin, Ye Luo, Anouar Benali, and Yongkyung Kwon (2019): https://journals.aps.org/prb/abstract/10.1103/PhysRevB.100.075430}
\begin{figure*}
\begin{centering}
\includegraphics[clip,width=1.7\columnwidth]{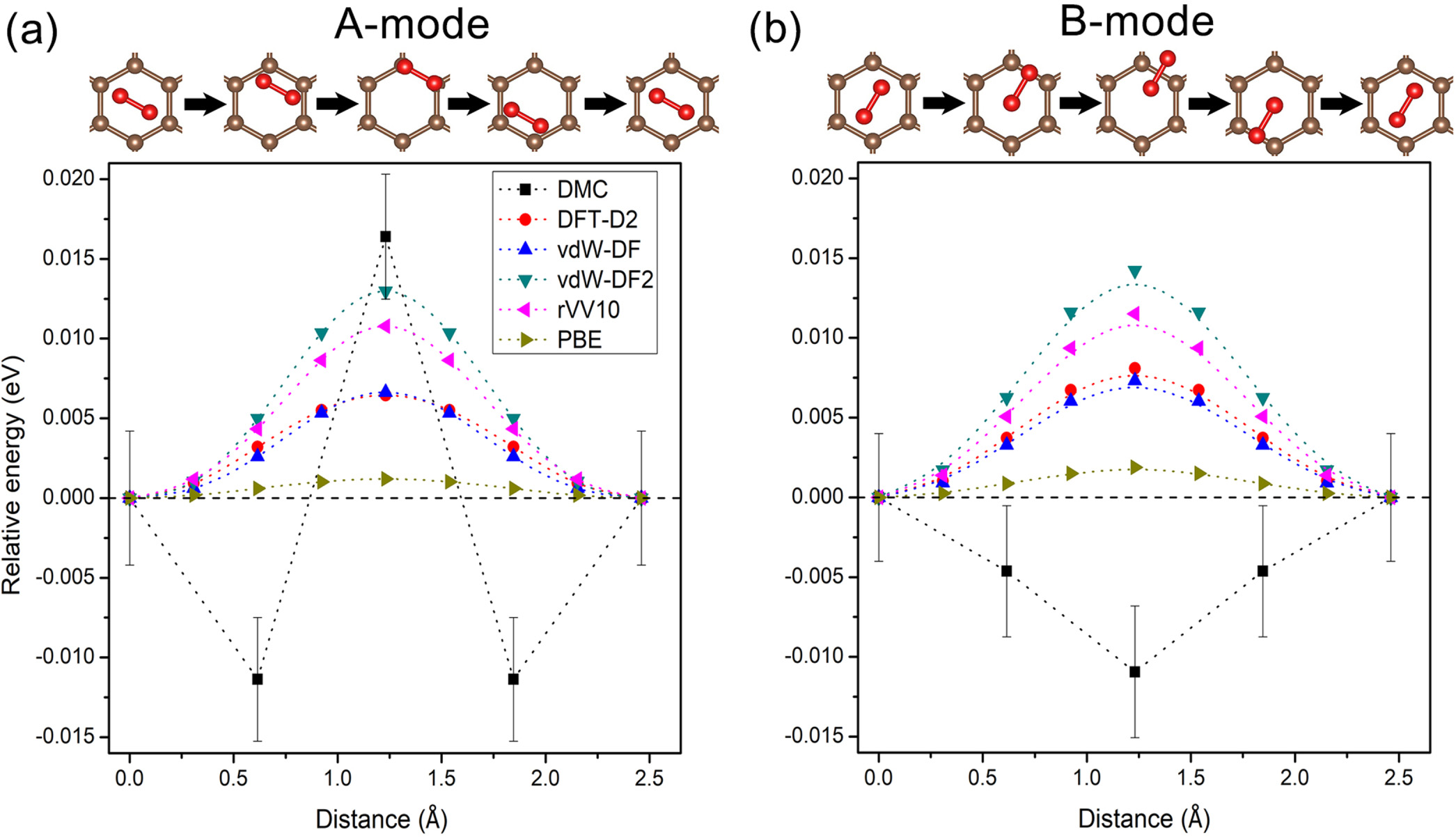}
\par\end{centering}
\caption{ DMC and DFT in-plane diffusion barrier for the hollow-bridge-hollow path for O$_2$ aligned in the (a) A and (b) B orientation modes. Note that the adsorption energies at a hollow site are set to be zero, and the dotted lines are guides for the eye. Shin et al., Phys. Rev. B 100, 075430 (2019). Copyright 2019 American Physical Society.}
\label{fig:O2_adsorption}
\end{figure*}

\textbf{O$_2$ adsorption on graphene:} Capture of oxygen molecules (O$_2$) is important for various industrial applications because O$_2$ can be used to control the rate of combustion, and its rate in the atmosphere should be controlled to avoid corrosion. Graphene has been considered as a suitable substrate for efficient O$_2$ capture, and its adsorption energy was successfully measured experimentally using temperature-programmed terahertz emission microscopy.\cite{graphene-o2} However, the stable O$_2$ adsorption site and its preferred orientation mode on the surface of graphene have not been confirmed experimentally yet.

Our DMC study of O$_2$ adsorption in triplet-state on graphene \cite{shin19} found that the O$_2$ orientation mode parallel to the graphene surface is more favorable than the vertical mode. The DMC adsorption energy for two different parallel modes of A and B (see Fig.~\ref{fig:O2_adsorption}) on the hollow site of the graphene surface is estimated as $-$0.130(4) eV and $-$0.126(4) eV, respectively. This similar adsorption energy between A and B indicates possible free planar rotation of O$_2$ at the hollow site. On the other hand, the DMC in-plane diffusion barrier, as seen in Fig.~\ref{fig:O2_adsorption}, shows that a hollow site is not the most stable adsorption site, while vdW-corrected DFT functionals show the lowest adsorption energies at the hollow sites, as seen in Fig.~\ref{fig:O2_adsorption}. These stable O$_2$ adsorptions at the bridge site of graphene can be understood by the interplay between repulsive interaction and vdW interaction. Our DMC results confirmed that the B orientation mode at the bridge site is the most stable orientation mode for O$_2$ adsorption, and the mode at the bridge site is hard to accurately describe within the Kohn-Sham framework, even with vdW correction.

%\textcolor{blue}{-Adsorption of a single Pt atom on graphene: spin crossing between physisorbed triplet and chemisorbed singlet states, Jeonghwan Ahn, Iuegyun Hong,  Gwangyoung Lee, Hyeondeok Shin, Anouar Benali, Yongkyung Kwon (2021): https://pubs.rsc.org/en/content/articlelanding/2021/cp/d1cp02473f}

\begin{figure*}
\begin{centering}
\includegraphics[clip,width=1.7\columnwidth]{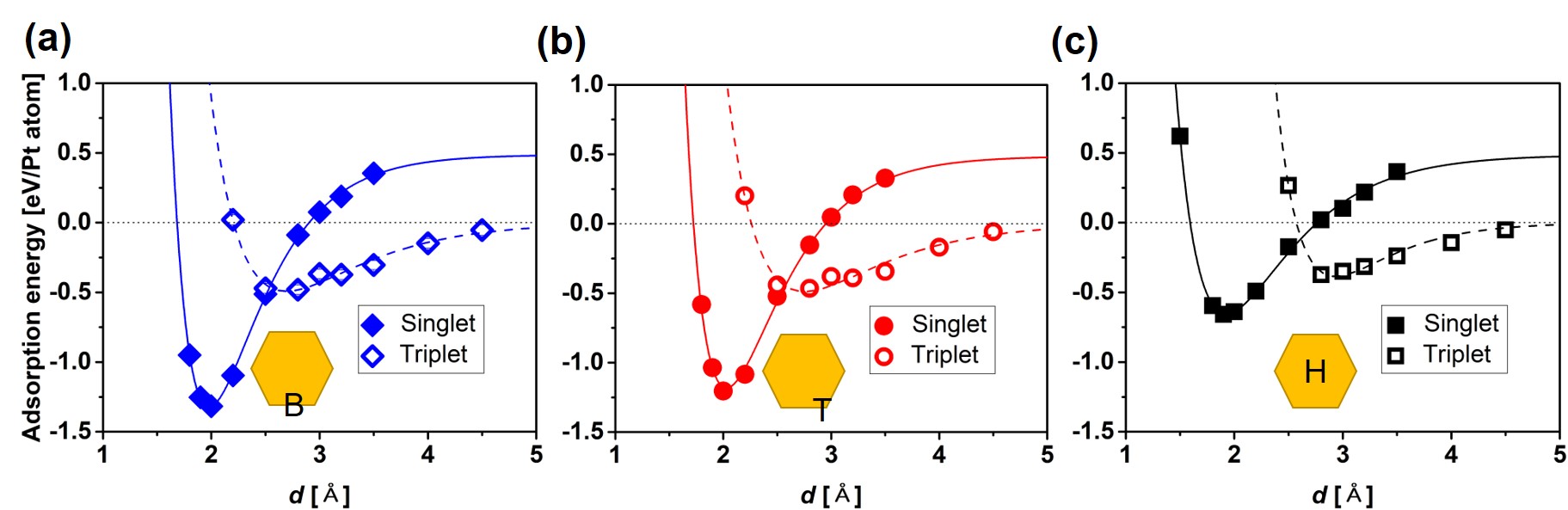}
\par\end{centering}
\caption{DMC adsorption energy curves of a single Pt atom adsorbed at three different sites of (a) a bridge, (b) an on-top, and (c) a hollow site, as a function of the vertical distance from a graphene surface. Here, B, T, and H in the insets denote a bridge, an on-top, and a hollow adsorption site, respectively. The singlet data is represented by solid symbols, while the triplet ones are denoted by open symbols. The solid and the dotted lines represent the Morse potential fits of spin singlet and triplet adsorption energies, respectively. Reproduced with permission from Ahn et al., Phys. Chem. Chem. Phys., 23, 22147-22154 (2021). Copyright 2021 Royal Society of Chemistry.}
\label{fig:Pt_adsorption_qmc}
\end{figure*}

\textbf{Pt clusters on graphene:} A graphene-supported Pt cluster has received a great deal of attention because of its enhanced catalytic properties and long-term stability compared to a conventional Pt catalyst.\cite{yoo2009enhanced,li2010catalytic,wu2011durability} Especially, a single Pt atom catalysis anchored on the graphene surface is considered highly desirable in terms of its reaction efficiency as well as the amount of Pt metal.\cite{sun2013single,cheng2019single} Nevertheless, its experimental realization has not yet been achieved, and thus it is imperative to understand this system theoretically with thorough investigation of the interaction between graphene and a single Pt atom.

Fig.~\ref{fig:Pt_adsorption_qmc} shows our DMC results for the singlet and triplet adsorption energy curves of a single Pt atom adsorbed on three symmetric adsorption sites (a bridge, an on-top, and a hollow site), as a function of the vertical distance from a graphene surface.\cite{D1CP02473F} 
Note that the singlet-state many-body wavefunction can be achieved within the single-reference framework of Eq.~\ref{Slater-jastrow} in the absence of SOC. This is possible because both spin-up and spin-down Slater determinants are constructed from the same Kohn-Sham orbitals obtained from spin-unpolarized DFT calculations.
%\textcolor{red}{Note that the singlet state whose spin part is antisymmetric is realized through spin-unpolarized DFT calculations, which leads to spatially-symmetric Kohn-Sham orbtials to ensure the antisymmetry of the total fermionic many-body wave function as expressed in Eq \ref{Slater-jastrow}.}
The bridge-site adsorption with the spin-singlet state turns out to be the most stable, which is consistent with our DFT benchmarking results.\cite{D1CP02473F} However, it is found that the triplet state becomes energetically preferred over the singlet state for all adsorption sites as the Pt-graphene distance increases, which corresponds to the spin crossing from the chemisorbed singlet state to the physisorbed triplet state.  The DMC calculations also predict the presence of local minima in the triplet region. Comparison of the DMC results for the Pt-benzene, the Pt-coronene, and the Pt-graphene system revealed that additive long-range dispersion forces (induced by carbon atoms outside a carbon ring surrounding the Pt atom) are responsible for the formation of the local minima in the adsorption curves. DFT calculations do not capture this local-minimum feature, indicating the significance of many-body correlations at long distances beyond the spin crossing points. This DMC study provides a comprehensive understanding of the Pt adsorption process on a graphene surface.
%This can be understood from the triplet ground state of an isolated Pt atom. 
%Furthermore, the DMC calculations predict a presence of the local minima in the triplet region, while none of the DFT calculations capture the local-minimum feature, indicating a crucial role of many-body correlation effects at large vertical distance of the physisorption regime beyond the spin crossing points. Finally, the DMC calculations for the Pt-benzene and Pt-coronene systems in comparison with that of the Pt-graphene system shows that the additive dispersive forces at large distances induced by carbon atoms positioned outside a carbon ring surrounding the Pt atom is responsible the formation of local minima, which provides a comprehensive understanding of Pt adsorption process on the graphene surface.

\begin{figure}
\begin{centering}
\includegraphics[width=8.0cm]{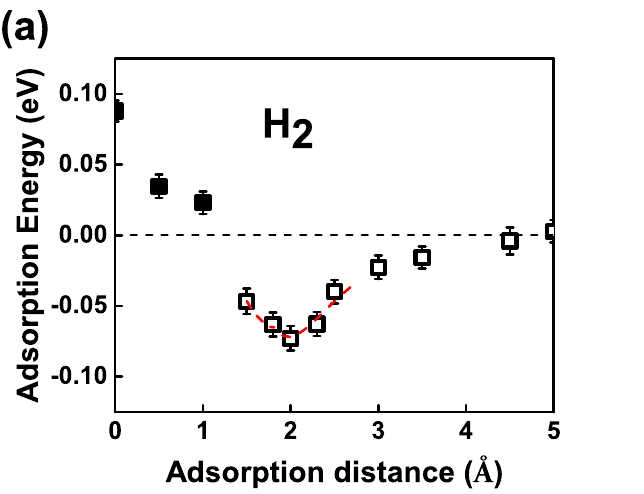}
\includegraphics[width=8.0cm]{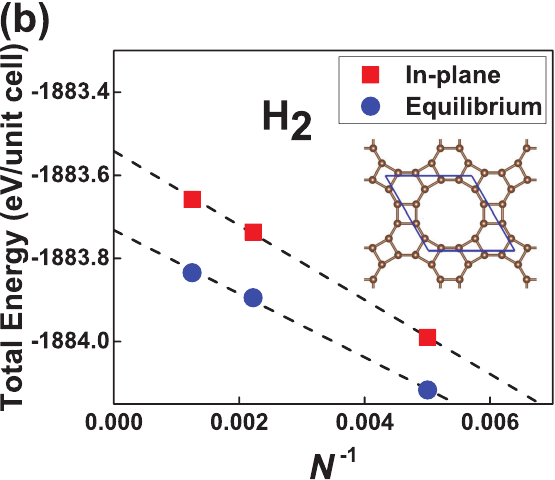}
\par\end{centering}
\caption{(a) DMC adsorption energies of a H$_2$ molecule on graphenylene (GPNL) as a function of the adsorption distance, computed for the  $3\times3\times1$ supercell. Here, the adsorption energies for the H$_2$ orientation modes favored at specific adsorption distances are presented with solid and open symbols corresponding to the V- (vertical to the membrane) and P-modes (parallel to the membrane). The red dashed line represents a Morse potential fit. (b) Twist-averaged DMC supercell energies of a H$_2$-GPNL complex, with the molecule adsorbed at the in-plane or the equilibrium adsorption site, as a function of $N^{-1}$, where $N$ is the number of electrons in a supercell.
Statistical errors of the DMC data are smaller than their symbol sizes.
The dashed lines represent linear regression fits. The inset depicts the top view of the atomic structure of GPNL. Reproduced from Lee et al., J. Chem. Phys. 157, 144703 (2022), with the permission of AIP Publishing.}
\label{fig:H2GPNL}
\end{figure}

\textbf{H$_2$ adsorption on graphenylene:} Using hydrogen as a renewable energy resource is a promising route for future technology. Graphenylene, a newly proposed 2D network of sp$^2$-bonded carbon atoms with large near-circular pores, is a promising membrane for separation of H$_2$ from gas mixtures.\cite{C6TA04456E} Its DMC cohesive energy is estimated to be 6.755(3) eV per atom,\cite{H2GPNL} which is smaller only by $\approx$10 meV per atom than the corresponding energy of $\gamma$-graphyne, the most stable structure in a graphyne family. An experimental report of its successful synthesis \cite{synthesis-gp} is understood to reflect this DMC result. DMC calculations are also performed to estimate the adsorption energies of different gas molecules, including H$_2$, on graphenylene, from which the H$_2$ separation capability of a graphenylene membrane against other gas molecules is estimated.

Fig.~\ref{fig:H2GPNL}(a) presents our DMC adsorption energy of a H$_2$ molecule as a function of the adsorption distance (the vertical distance from a pore center), which was computed for a $3 \times 3 \times 1$ supercell. The equilibrium adsorption distance is determined through the Morse potential fit of the DMC adsorption energies (see the red dotted line in the figure), which produced the equilibrium distances of 1.97(4)~\AA~for H$_2$. The same procedure results in the DMC equilibrium distances of 2.73(3)~\AA~and 2.73(4)~\AA~for N$_2$ and CO molecules, respectively. After establishing the equilibrium adsorption distances for these molecules, DMC calculations are performed to compute their adsorption energies at both in-plane and equilibrium adsorption distances. Fig.~\ref{fig:H2GPNL}(b) shows the DMC total energies of the H$_2$-graphenylene complex computed for three different supercell sizes, where the horizontal axis represents the inverse of the number of electrons in a supercell. The adsorption energies of H$_2$ are estimated from the total energies extrapolated to the thermodynamic limit ($N \rightarrow \infty$). The difference between in-plane and equilibrium adsorption energies determines the diffusion barrier of a gas molecule passing through a graphenylene membrane, whose DMC values are estimated to 0.19(2) eV, 0.87(5) eV, and 0.79(2) eV for H$_2$, N$_2$, and CO, respectively. The large difference in the diffusion barrier between H$_2$ and other molecules results in extremely high values for hydrogen selectivity against a gas of N$_2$ ($\approx$$10^{11}$) or CO ($ \approx$$10^{10}$).\cite{H2GPNL} This suggests that an application of graphenylene as a high-performing hydrogen separator is promising.

\textbf{Atomic H adsorption on graphene:} The chemisorption of atomic hydrogen on graphene is another system of interest because of the tunability of electronic (opening up of the band gap \cite{gap-graphene}) and magnetic (inducing an extended magnetic moment in a graphene sheet \cite{doi:10.1126/science.aad8038,doi:10.1142/S0129183118500924}) properties when hydrogen atoms are chemisorbed. In addition, graphene and graphitic surfaces can be utilized for hydrogen storage and energy applications.\cite{graphene-h} In addition to the interesting applications, the lack of experimental benchmarks (i.e., binding energy) for H chemisorbed on graphene and the variability of standard DFT approaches for atomic adsorption on 2D surfaces make this an excellent system to apply DMC techniques.

\begin{figure}
\begin{center}
\includegraphics[width=8cm]{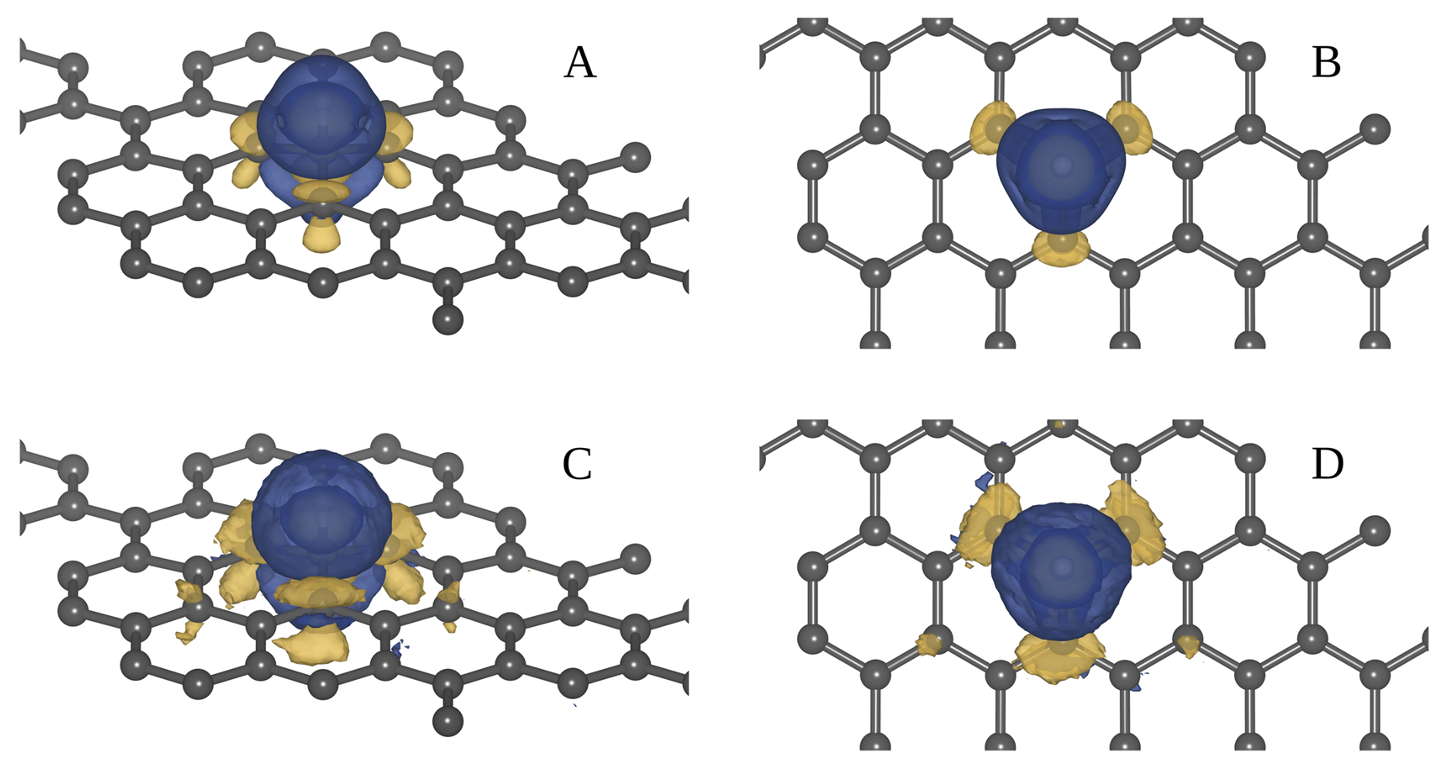}
\caption{The change in electron density associated with the adsorption of H on a distorted graphene sheet. (a) and (b) display the PBE density from different perspectives, and (c) and (d) display the DMC density from different perspectives. Gold indicates a gain of electron density, while blue indicates a loss. Reproduced from Dumi et al., J. Chem. Phys. 156, 144702 (2022), with the permission of AIP Publishing.}
\label{h-graphene}
\end{center}
\end{figure}

In this work, we performed DFT and DMC calculations to obtain the binding energy of a single H atom chemisorbed on the surface of a graphene sheet.\cite{10.1063/5.0085982} With DMC, we find this binding energy to be $-$691 meV $\pm$ 19 meV. We find that PBE (plane-wave) overestimates the binding energy by approximately 20$\%$ compared to DMC. We also find that PBE0 %cite PBE0 in intro 
results in a binding energy close to PBE, but HSE yields a binding energy in closer agreement to DMC ($-$743 meV). We also find significant differences between the DMC and PBE charge densities of graphene and H chemisorbed on graphene (see Fig. \ref{h-graphene}). More details of this work can be found in Ref. \onlinecite{10.1063/5.0085982}.

%Feel free to swap this figure out for another or modify this text above

\begin{figure}
\begin{center}
\includegraphics[width=8cm]{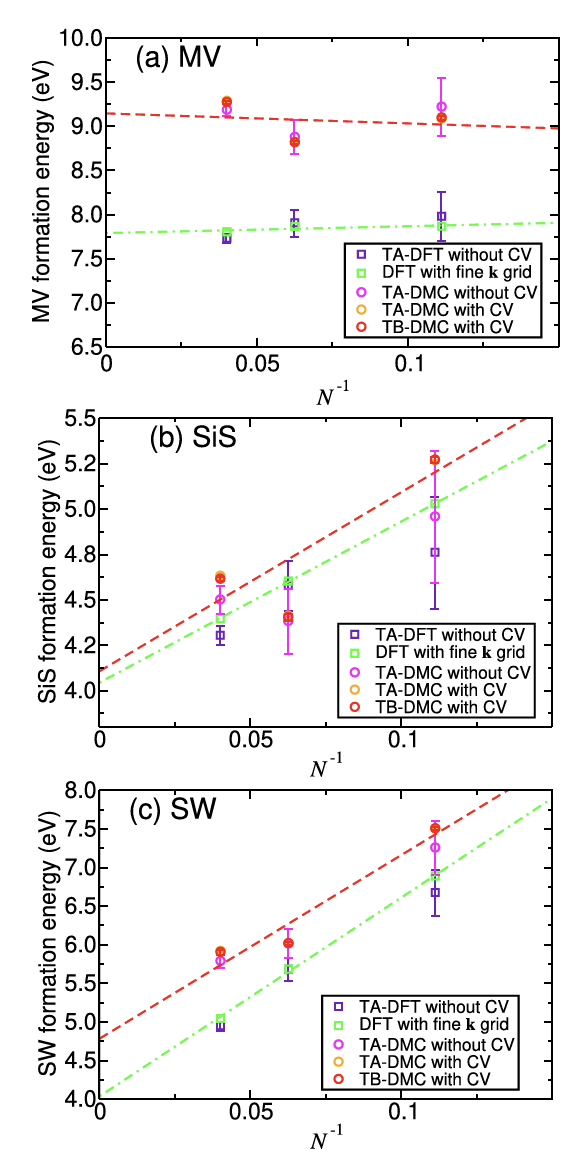}
\caption{Defect formation energy of graphene as a function of supercell size N (calculated with DFT and DMC) for (a) MV, (b) SiS, and (c) SW defects with different methods of addressing one-body FS errors (TA, twist-blocking [TB], and CV). Reproduced with permission from Thomas et al., Phys. Rev. B 105, 184114 (2022). Copyright 2022 American Physical Society.}
\label{graphene-vac}
\end{center}
\end{figure}

\textbf{Point defect formation energy of graphene:} In addition to our work reported in this section that focuses on atomic and molecular adsorption on monolayer surfaces, there have been efforts to utilize DMC to accurately calculate the point defect formation energy in 2D materials.\cite{PhysRevB.105.184114} Thomas et al. \cite{PhysRevB.105.184114} performed detailed DFT and DMC calculations for point defects in graphene. The types of point defects considered were monovacancies (MVs), silicon substitutions (SiSs), and Stone-Wales (SW) defects. MV involves the removal of one carbon atom, SiS involves replacing a single carbon atom with silicon, and an SW defect is created by rotating a single carbon-carbon bond (in-plane) 90$^{\circ}$ about its midpoint. FS errors of the defect calculations were taken into account by TA and performing the simulations at multiple supercell sizes (ranging from 3 $\times$ 3 to 5 $\times$ 5) and extrapolating to the thermodynamic limit. To reduce the error in TA, the DFT energies were used as a control variate (CV) when computing the DMC energies. A summary of the FS extrapolated results for defect formation energy is depicted in Fig. \ref{graphene-vac}. It was found that DFT (PBE) underestimates the defect formation energy on the order of 1 eV (being slightly smaller for SW and SiS defects). In addition, it was found that the vibrational contribution to the defect formation energy was nonnegligible (on the order of 0.5--1 eV). This work highlights the challenges of accurately computing defect properties with QMC, including systematic and quasirandom finite-concentration effects and significant vibrational contributions. More details of this work can be found in Ref. \onlinecite{PhysRevB.105.184114}.

%more details if we need it: The red dashed lines show an unweighted least-squares fit of Eq. (5) tothe TA-DMC data. Both the DFT and DMC calculations used DiracFock pseudopotentials. The carbon and silicon chemical potentials were taken to be the energy per atom of monolayer graphene and bulk silicon, extrapolated to infinite system size; hence the N-dependence shown in this figure only arises from the finite-concentration and finite-size effects in the pure formation energy

\section{Conclusion}\label{conc}

We have demonstrated the successes of applying the many-body DMC approach to a wide variety of 2D material systems. This review article provides detailed summaries of several instances in which DMC can improve the prediction of magnetic, electronic (including excitonic and topological) properties and accurately capture the interlayer interactions and the energetics of cohesion and adsorption. The results reported in this review demonstrate how accuracy beyond standard DFT can realistically be achieved using many-body electronic structure methods such as DMC. We hope that by showcasing the recent advancements in the field of QMC methods being applied to 2D materials, other researchers will be motivated to employ these highly accurate techniques for future work.  

\section{Competing interests}
The authors have no conflicts to disclose.

\section{Data Availability}
Data sharing is not applicable to this article because no new data were created or analyzed in this study. 

\section{Notes}
Please note that certain equipment, instruments, software, or materials are identified in this paper to specify the experimental
procedure adequately. Such identification is not intended to imply the recommendation or endorsement of any product or service by the National Institute of Standards and Technology,
nor is it intended to imply that the materials or equipment identified is necessarily the best available for the purpose.
 
\section{acknowledgments}
%Everybody PLEASE add their funding and acknowledgements 
D.W. acknowledges the National Institute of Standards and Technology for funding and support. J.A., A.B., P.R.C.K, J.T.K., L.M., B.R., and H.S. were supported by the US Department of Energy, Office of Science, Basic Energy Sciences, Materials Sciences and Engineering Division as part of the Computational Materials Sciences Program and the Center for Predictive Simulation of Functional Materials. L.M. also received support (excitonic effects in 2D) from US National Science Foundation grant DMR-2316007. Y.K. was supported by the Basic Science Research Program (2018R1D1A1B07042443) through the National Research Foundation of Korea funded by the Ministry of Education. I.S. acknowledges support by APVV-21-0272, VEGA 2/0133/25, VEGA 2/0131/23 and by the H2020 TREX GA 952165 projects and Funding by the EU NextGenerationEU through the Recovery and Resilience Plan for Slovakia under the project Nos. 09I02-03-V01-00012 and 09I05-03-V02-00055. K.S. and F.A.R. were supported by the US Department of Energy, Office of Science, Basic Energy Sciences, Materials Sciences and Engineering Division. C.A. acknowledges funding from the National Science Foundation under grant NSF DMR-2213398 and the US Department of Energy under grant DE-SC0024236. 
\section*{References}
\bibliography{main}% Produces the bibliography via BibTeX.

% \appendix
%\section{New Capabilities of QMC}\label{cap}

%Possible topics (this can be a short appendix)
%-geometry optimization (surrogate line-search method)

%-geometry optimization highlights, these will be covered in other sections but may be nice to reiterate/cite

%-spin-orbit coupling

%-alloys

%-Fermi surface 

%-high-throughput efforts

%-improved pseudopotentials

%-performance of QMC

%-JARVIS, JARVIS Leaderboard

%\textcolor{blue}{-Surrogate Hessian accelerated structural optimization for stochastic electronic structure theories, Juha Tiihonen,  Paul R. C. Kent, Jaron T. Krogel (2022): https://aip.scitation.org/doi/full/10.1063/5.0079046}

\end{document}